\documentclass[aps,prd,showpacs,twocolumn]{revtex4}

\usepackage{graphicx}
\usepackage{longtable}
\usepackage{morefloats}

\begin{document}

\title{Bottomonium Mesons and Strategies for Their Observation}
\author{ Stephen Godfrey\footnote{Email: godfrey@physics.carleton.ca} and Kenneth Moats}
\affiliation{
Ottawa-Carleton Institute for Physics, 
Department of Physics, Carleton University, Ottawa, Canada K1S 5B6 
}

\date{\today}

\begin{abstract}
The $B$-factories and Large Hadron Collider experiments have demonstrated the ability to observe and measure
the properties of bottomonium 
mesons. In order to discover missing states it is useful to know their properties
to develop a successful
search strategy.  To this end 
we calculate the masses and decay properties of excited bottomonium states. We use the relativized quark model to 
calculate the masses and wavefunctions and the $^3P_0$ quark-pair creation model
to calculate decay widths to open bottom.
We also summarize results for 
radiative transitions, annihilation decays, hadronic transitions 
and production cross sections which are used to develop strategies to find these states. 
We find that the $b\bar{b}$ system has a rich spectroscopy that we expect to 
be substantially extended by the LHC and $e^+e^-$ experiments in the near future.  
Some of the most promising possibilities at the LHC are observing the $\chi_{b(1,2)}(3P)$, 
$\chi_{b(1,2)}(4P)$ and $\eta_b(3S)$ states in $\gamma \mu^+\mu^-$ final states that proceed
via radiative transitions through $\Upsilon(nS)$ intermediate states  
and $1^3D_J$ and $2^3D_J$ into $\gamma\gamma \mu^+\mu^-$ final states proceeding via $1^3P_J\to 1^3S_1$
and $2^3P_J\to 2^3S_1$ intermediate states respectively. Some of the most interesting possibilities
in $e^+e^-$ collisions are studying the $1^3D_J$ states via $4\gamma$ cascades starting with
the $\Upsilon (3S)$ and the $3^3P_J$ states in $\gamma\gamma\mu^+ \mu^-$ final states starting
with the $\Upsilon (4S)$ and proceeding via $\Upsilon (nS)$ intermediate states.  
Completing the bottomonium spectrum is an important validation of lattice QCD calculations
and a 
test of our understanding of bottomonium states in the context of the quark model.
\end{abstract}

\maketitle

\section{Introduction}

The Large Hadron Collider experiments have demonstrated that they can discover some of the missing bottomonium states.
In fact, the first new particle discovered at the LHC was a $3P$ bottomonium state 
\cite{Aad:2011ih,Chisholm:2014sca}.  
The Belle II experiment at SuperKEKB 
will also offer the possibility of studying excited bottomonium states \cite{Drutskoy:2012gt}.
At the same time, 
lattice QCD calculations of bottomonium properties have advanced considerably in recent years 
\cite{Lewis:2012ir,Lewis:2011ti,Lewis:2012bh,Dowdall:2013jqa} so it is important to expand our experimental knowledge of 
bottomonium states to test these calculations.  
With this motivation, we calculate properties of bottomonium mesons to suggest 
experimental strategies to observe missing states. The observation of these states is 
a crucial test of lattice QCD calculations and will also test the various models of hadron properties.
Some recent reviews of bottomonium spectroscopy are 
Ref.~\cite{Brambilla:2010cs,Patrignani:2012an,Eichten:2007qx}.

We use the relativized quark model to calculate the masses and wavefunctions \cite{godfrey85xj}.  
The mass predictions for this model are given in Section II. 
The wavefunctions are used to calculate radiative transitions between states, annihilation decays
and as input for estimating hadronic transitions as described in Sections III-V respectively.  
The strong decay widths to open bottom are described in Section VI and are calculated 
using the $^3P_0$ model \cite{Micu:1968mk,Le Yaouanc:1972ae} with simple harmonic
 oscillator (SHO) wavefunctions with the oscillator parameters, $\beta$, found by fitting the SHO wavefunction 
rms radii to those of the corresponding relativized quark model wavefunctions.  This approach has 
proven to be a useful phenomenological tool for calculating hadron properties which has helped 
to understand the observed spectra 
\cite{Barnes:2005pb,Godfrey:2014,Barnes:2003vb,Godfrey:2003kg,Blundell:1995ev}. 
Additional details of the $^3P_0$ model are given in the appendix, primarily so that the various conventions are
written down explicitly so that the interested reader is able to reproduce our results.

We combine our results for the various decay modes to produce branching ratios (BR) for
each of the bottomonium states we study. 
The purpose of this paper is to suggest strategies to find some of the missing bottomonium 
states in $pp$ collisions at the LHC and in $e^+e^-$ collisions at SuperKEKB.  The final 
missing input is an estimate of production rates for bottomonium states in $pp$ and $e^+e^-$ 
collisions.  This is described in Section VII.  We combine the cross sections with 
the expected integrated luminosities and various BR's to estimate the number of events
expected for the production of bottomonium states with decays to various final states.  
This is the main result of the paper, to identify which of the missing bottomonium states
are most likely to be observed and the most promising signal to find them.  However, there
are many experimental issues that could alter our conclusions  so we hope  
that the interested reader can use the information in this paper as a starting point 
to study other potentially 
useful experimental signatures that we might have missed.  In the final section we summarize 
the most promising signatures.

\section{Spectroscopy}
\label{sec:spectroscopy}

We calculate the bottomonium 
mass spectrum using the relativized quark model \cite{godfrey85xj}. 
This model assumes a relativistic kinetic energy term and the potential
incorporates a Lorentz vector one-gluon-exchange interaction with a
QCD motivated running coupling constant, $\alpha_{s}(r)$, and a  Lorentz scalar linear confining 
interaction.
The details of this model, including the parameters, 
can be found in Ref.~\cite{godfrey85xj} 
(see also Ref.~\cite{Godfrey:1985by,godfrey85b,Godfrey:2004ya,Godfrey:2005ww}). 
This is typical of most such models which 
are based on some variant of the Coulomb plus 
linear potential expected from QCD and often include some relativistic effects.  
The relativized quark 
model has been reasonably successful in describing most known mesons and is a useful template against which to identify newly found states. 
However in recent years, starting with the discovery of the 
$D_{sJ}(2317)$ \cite{Aubert:2003fg,Besson:2003cp,Krokovny:2003zq} and $X(3872)$ states \cite{Choi:2003ue},
an increasing number of states have been observed that do not fit into this picture 
\cite{Godfrey:2008nc,Godfrey:2009qe,Braaten:2013oba,DeFazio:2012sg}  
pointing to the need to include physics which has hitherto been neglected such as 
coupled channel effects \cite{Eichten:2004uh}.  As a consequence of neglecting coupled channel effects
and the crudeness of the relativization procedure we do not 
expect the mass predictions to be accurate to better than $\sim 10-20$~MeV.

\begin{figure*}
\begin{center}
\includegraphics[width=7.5in]{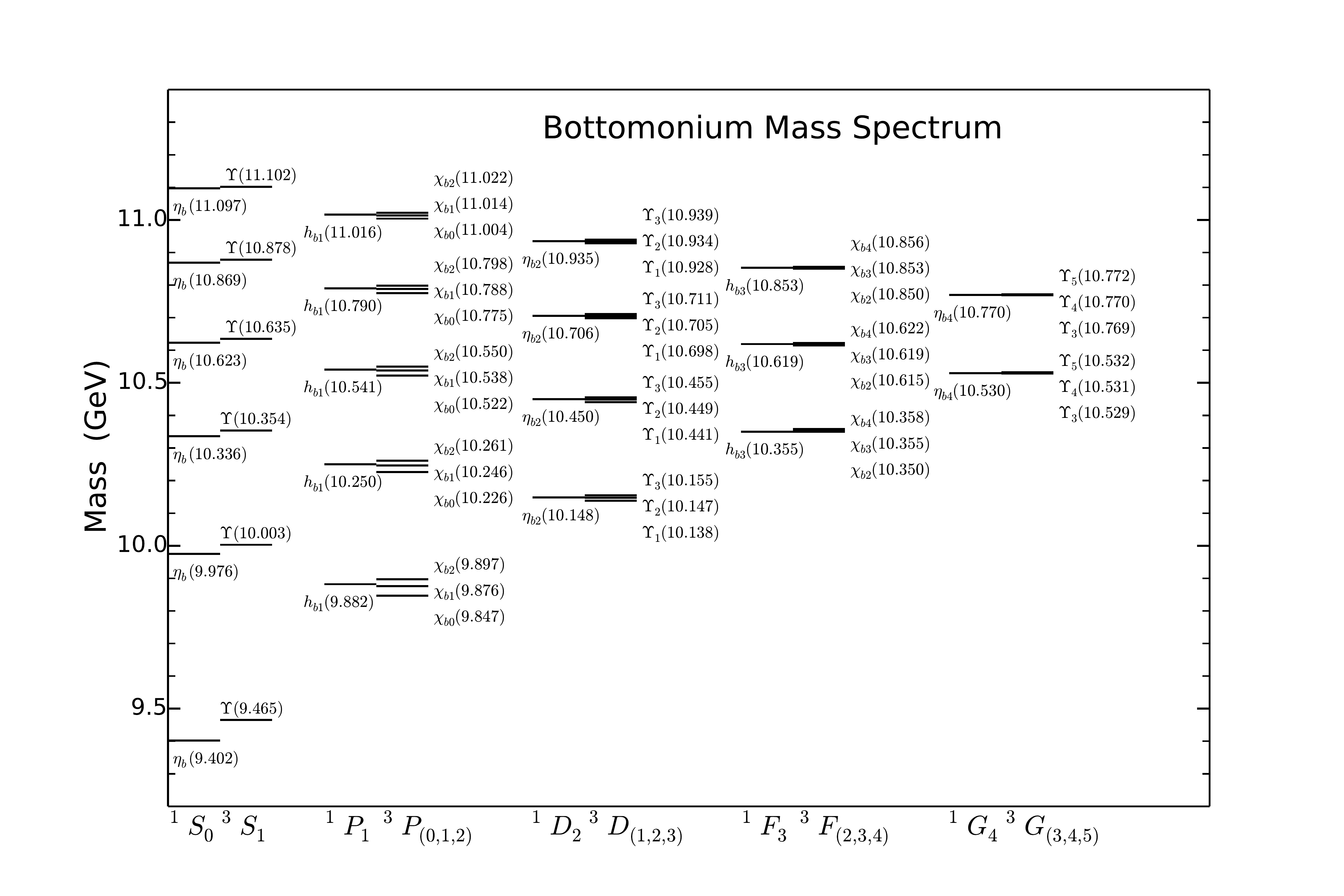}
\end{center}
\caption{The $b\bar{b}$ mass spectrum as predicted by the relativized quark model \cite{godfrey85xj}.
}
\label{Fig1}
\end{figure*}

The bottomonium mass predictions for this model are shown in Fig.~\ref{Fig1}.  These 
are also listed in Tables~\ref{tab:Upsilonparams1}-\ref{tab:Upsilonparams2}
along with known experimental masses and
the effective SHO wavefunction parameters, $\beta$. 
These, along with the masses and effective $\beta$'s for the $B$ meson states, listed
 in Table~\ref{tab:Bparams}, are used in the calculations
of the open bottom strong decay widths as 
described in Sec.~\ref{sec:strongdecays}.
We note that the $1^1P_1$ and $1^3P_1$ $B$ meson states mix to form the physical $1P_1$ and $1P_1^\prime$ states, 
as defined in Table \ref{tab:Bparams}, with a singlet-triplet mixing angle of $\theta_{1P} = -30.3^\circ$ 
for $b\bar{q}$ ordering.

If available, the experimental masses are used as input in our calculations rather 
than the predicted masses.  When the mass of only one meson in a multiplet has been measured, 
we shift our input masses for the remaining states using the measured mass and the 
predicted splittings.  
Specifically, to obtain the $\eta_b(n^1S_0)$ masses (for $n=3,4,5,6$) we subtracted the predicted $n^3S_1 - n^1S_0$ splitting 
from the measured $\Upsilon(n^3S_1)$ mass \cite{Olive:2014kda}.  
For the $\chi_b(3P)$ states, we calculated the predicted mass differences with respect 
to the $\chi_{b1}(3^3P_1)$ state and subtracted them from the observed $\chi_{b1}(3^3P_1)$ 
mass recently measured by LHCb \cite{Aaij:2014hla}.  We used a similar procedure for
 the $\Upsilon(1D)$ mesons  \cite{Olive:2014kda} as well as for the currently unobserved
  $1P$ $B$ mesons \cite{Olive:2014kda} listed in Table~\ref{tab:Bparams}.

\begin{table}[tp]
\begin{center}
\caption{Masses and effective harmonic oscillator parameter values ($\beta$) 
 for  $S$-, $P$- and $D$-wave Bottomonium mesons.}
\begin{tabular}{lccc} \hline \hline
Meson				&	$M_{theo}$ (MeV)	&	$M_{exp}$ (MeV)							& 	$\beta$ (GeV)	\\
\hline
$\Upsilon(1^3S_1)$		&	9465				&	$9460.30\pm0.26$  \footnotemark[1]				&	1.157		\\
$\eta_b(1^1S_0)$		&	9402				&	$9398.0\pm3.2$  \footnotemark[1]				&	1.269		\\
$\Upsilon(2^3S_1)$		&	10003			&	$10023.26\pm0.31$  \footnotemark[1]			&	0.819		\\
$\eta_b(2^1S_0)$		&	9976				&	$9999.0\pm3.5^{+2.8}_{-1.9}$  \footnotemark[1]	&	0.854		\\
$\Upsilon(3^3S_1)$		&	10354			&	$10355.2\pm0.5$  \footnotemark[1]				&	0.698		\\
$\eta_b(3^1S_0)$		&	10336			&	10337 \footnotemark[2]						&	0.719		\\
$\Upsilon(4^3S_1)$		&	10635			&	$10579.4\pm1.2$  \footnotemark[1]				&	0.638		\\
$\eta_b(4^1S_0)$		&	10623			&	10567 \footnotemark[2]						&	0.654		\\
$\Upsilon(5^3S_1)$		&	10878			&	$10876\pm11$  \footnotemark[1]				&	0.600		\\
$\eta_b(5^1S_0)$		&	10869			&	10867 \footnotemark[2]						&	0.615		\\
$\Upsilon(6^3S_1)$		&	11102			&	$11019\pm8$  \footnotemark[1]				&	0.578		\\
$\eta_b(6^1S_0)$		&	11097			&	11014 \footnotemark[2]						&	0.593		\\
\hline
$\chi_{b2}(1^3P_2)$		&	9897				&	$9912.21\pm0.26\pm0.31$  \footnotemark[1]		&	0.858		\\
$\chi_{b1}(1^3P_1)$		&	9876				&	$9892.78\pm0.26\pm0.31$  \footnotemark[1]		&	0.889		\\
$\chi_{b0}(1^3P_0)$		&	9847				&	$9859.44\pm0.42\pm0.31$  \footnotemark[1]		&	0.932		\\
$h_{b}(1^1P_1)$		&	9882				&	$9899.3\pm1.0$  \footnotemark[1]				&	0.880		\\
$\chi_{b2}(2^3P_2)$		&	10261			&	$10268.65\pm0.22\pm0.50$  \footnotemark[1]		&	0.711		\\
$\chi_{b1}(2^3P_1)$		&	10246			&	$10255.46\pm0.22\pm0.50$  \footnotemark[1]		&	0.725		\\
$\chi_{b0}(2^3P_0)$		&	10226			&	$10232.5\pm0.4\pm0.5$  \footnotemark[1]		&	0.742		\\
$h_{b}(2^1P_1)$		&	10250			&	$10259.8\pm0.5\pm1.1$  \footnotemark[1]		&	0.721		\\
$\chi_{b2}(3^3P_2)$		&	10550			&	10528 \footnotemark[2]						&	0.640		\\
$\chi_{b1}(3^3P_1)$		&	10538			&	$10515.7^{+2.2+1.5}_{-3.9-2.1}$ \footnotemark[3]	&	0.649		\\
$\chi_{b0}(3^3P_0)$		&	10522			&	10500 \footnotemark[2]						&	0.660		\\
$h_{b}(3^1P_1)$		&	10541			&	10519 \footnotemark[2]						&	0.649		\\
$\chi_{b2}(4^3P_2)$		&	10798			&	N/A										&	0.598		\\
$\chi_{b1}(4^3P_1)$		&	10788			&	N/A										&	0.605		\\
$\chi_{b0}(4^3P_0)$		&	10775			&	N/A										&	0.613		\\
$h_{b}(4^1P_1)$		&	10790			&	N/A										&	0.603		\\
$\chi_{b2}(5^3P_2)$		&	11022			&	N/A										&	0.570		\\
$\chi_{b1}(5^3P_1)$		&	11014			&	N/A										&	0.576		\\
$\chi_{b0}(5^3P_0)$		&	11004			&	N/A										&	0.585		\\
$h_{b}(5^1P_1)$		&	11016			&	N/A										&	0.575		\\
\hline
$\Upsilon_3(1^3D_3)$	&	10155			&	10172 \footnotemark[2]						&	0.752		\\
$\Upsilon_2(1^3D_2)$	&	10147			&	$10163.7\pm1.4$  \footnotemark[1]				&	0.763		\\
$\Upsilon_1(1^3D_1)$	&	10138			&	10155 \footnotemark[2]						&	0.776		\\
$\eta_{b2}(1^1D_2)$	&	10148			&	10165 \footnotemark[2]						&	0.761		\\
$\Upsilon_3(2^3D_3)$	&	10455			&	N/A										&	0.660		\\
$\Upsilon_2(2^3D_2)$	&	10449			&	N/A										&	0.666		\\
$\Upsilon_1(2^3D_1)$	&	10441			&	N/A										&	0.672		\\
$\eta_{b2}(2^1D_2)$	&	10450			&	N/A										&	0.665		\\
$\Upsilon_3(3^3D_3)$	&	10711			&	N/A										&	0.609		\\
$\Upsilon_2(3^3D_2)$	&	10705			&	N/A										&	0.613		\\
$\Upsilon_1(3^3D_1)$	&	10698			&	N/A										&	0.618		\\
$\eta_{b2}(3^1D_2)$	&	10706			&	N/A										&	0.612		\\
$\Upsilon_3(4^3D_3)$	&	10939			&	N/A										&	0.577		\\
$\Upsilon_2(4^3D_2)$	&	10934			&	N/A										&	0.580		\\
$\Upsilon_1(4^3D_1)$	&	10928			&	N/A										&	0.583		\\
$\eta_{b2}(4^1D_2)$	&	10935			&	N/A										&	0.579		\\
\hline \hline
\end{tabular}
 \footnotetext[1]{Measured mass from Particle Data Group \cite{Olive:2014kda}.}	
 \footnotetext[2]{Using predicted multiplet mass splittings with measured mass as described in Sec.~\ref{sec:spectroscopy}.}	
 \footnotetext[3]{Measured mass from LHCb \cite{Aaij:2014hla}.}	
\label{tab:Upsilonparams1}
\end{center}
\end{table}

\begin{table}[tp]
\begin{center}
\caption{Masses and effective harmonic oscillator parameter values ($\beta$) 
 for  $F$- and $G$-wave Bottomonium mesons.}
\begin{tabular}{lccc} \hline \hline
Meson				&	$M_{theo}$ (MeV)	&	$M_{exp}$ (MeV)	& 	$\beta$ (GeV)	\\
\hline
$\chi_{b4}(1^3F_4)$		&	10358			&	N/A				&	0.693		\\
$\chi_{b3}(1^3F_3)$		&	10355			&	N/A				&	0.698		\\
$\chi_{b2}(1^3F_2)$		&	10350			&	N/A				&	0.704		\\
$h_{b3}(1^1F_3)$		&	10355			&	N/A				&	0.698		\\
$\chi_{b4}(2^3F_4)$		&	10622			&	N/A				&	0.626		\\
$\chi_{b3}(2^3F_3)$		&	10619			&	N/A				&	0.630		\\
$\chi_{b2}(2^3F_2)$		&	10615			&	N/A				&	0.633		\\
$h_{b3}(2^1F_3)$		&	10619			&	N/A				&	0.629		\\
$\chi_{b4}(3^3F_4)$		&	10856			&	N/A				&	0.587		\\
$\chi_{b3}(3^3F_3)$		&	10853			&	N/A				&	0.590		\\
$\chi_{b2}(3^3F_2)$		&	10850			&	N/A				&	0.592		\\
$h_{b3}(3^1F_3)$		&	10853			&	N/A				&	0.589		\\
\hline
$\Upsilon_5(1^3G_5)$	&	10532			&	N/A				&	0.653		\\
$\Upsilon_4(1^3G_4)$	&	10531			&	N/A				&	0.656		\\
$\Upsilon_3(1^3G_3)$	&	10529			&	N/A				&	0.660		\\
$\eta_{b4}(1^1G_4)$	&	10530			&	N/A				&	0.656		\\
$\Upsilon_5(2^3G_5)$	&	10772			&	N/A				&	0.602		\\
$\Upsilon_4(2^3G_4)$	&	10770			&	N/A				&	0.604		\\
$\Upsilon_3(2^3G_3)$	&	10769			&	N/A				&	0.606		\\
$\eta_{b4}(2^1G_4)$	&	10770			&	N/A				&	0.604		\\
\hline \hline
\end{tabular}
\label{tab:Upsilonparams2}
\end{center}
\end{table}

\begin{table*}[tp]
\begin{center}
\caption{Masses and effective $\beta$ values for $B$ mesons used in the calculations of bottomonium strong decay widths.  
The physical $1P_1'$ and $1P_1$ states are mixtures of $1^1P_1$ and $1^3P_1$ with
singlet-triplet mixing angle $\theta_{1P} = -30.3^\circ$ for $b\bar{q}$ ordering.  
Where two values of $\beta$ are listed, the first (second) value is for the singlet (triplet) state.}

\begin{tabular}{lcccc} \hline \hline
Meson			&	State											&	$M_{theo}$ (MeV)	&	$M_{exp}$ (MeV)					& 	$\beta$ (GeV)	\\
\hline
$B^{\pm}$			&	$1^1S_0$										&	5312				&	$5279.26\pm0.17$ \footnotemark[1]		&	0.580		\\ 
$B^{0}$			&	$1^1S_0$										&	5312				&	$5279.58\pm0.17$ \footnotemark[1]		&	0.580		\\ 
$B^*$			&	$1^3S_1$										&	5371				&	$5325.2\pm0.4$ \footnotemark[1]		&	0.542		\\ 
$B(1^3P_0)$		&	$1^3P_0$										&	5756				&	5702 \footnotemark[2]				&	0.536		\\
$B(1P_1)$		&	$\cos\theta_{1P} (1^1P_1) + \sin\theta_{1P} (1^3P_1)$	&	5777				&	$5723.5\pm2.0$ \footnotemark[1]		&	0.499, 0.511	\\
$B(1P_1^\prime)$	&	$-\sin\theta_{1P} (1^1P_1) + \cos\theta_{1P} (1^3P_1)$	&	5784				&	5730 \footnotemark[2]				&	0.499, 0.511	\\
$B(1^3P_2)$		&	$1^3P_2$										&	5797				&	$5743\pm5$ \footnotemark[1]			&	0.472		\\
$B_s$			&	$1^1S_0$										&	5394				&	$5366.77\pm0.24$ \footnotemark[1]		&	0.636		\\ 
$B_s^*$			&	$1^3S_1$										&	5450				&	$5415.4^{+2.4}_{-2.1}$ \footnotemark[1]	&	0.595		\\ 
\hline \hline
\end{tabular}
 \footnotetext[1]{Measured mass from Particle Data Group \cite{Olive:2014kda}.}	
 \footnotetext[2]{Input mass from predicted mass splittings, as described in Sec.~\ref{sec:spectroscopy}.}	
\label{tab:Bparams}
\end{center}
\end{table*}

\begin{table*}[tp]
\caption{Partial widths and branching ratios
for strong and electromagnetic decays and transitions for the $1S$ and $2S$ bottomonium states. The state's mass is given 
in GeV and is listed below the state's name in column 1.  Column 4, labelled $\cal M$ gives the matrix 
element appropriate to the particular decay;  for $S$-wave annihilation decays ${\cal M}$ designates
$\Psi(0)=R(0)/\sqrt{4\pi}$ representing the wavefunction at the origin and 
for radiative transitions the E1 or M1 matrix elements are  $\langle \psi_f | r | \psi_i \rangle$ (GeV$^{-1}$)
and $\langle \psi_f| j_0(kr/2) | \psi_i \rangle$ respectively.
Details of the calculations are given in the text.  
\label{tab:sum_12Swave}}
\begin{center}
\begin{tabular}{l l l l l l l l } \hline \hline
Initial \phantom{www}       & Final 	 				& $M_f$  				& $\cal M$	& \multicolumn{2}{c}{Predicted}				& \multicolumn{2}{c}{Measured} 							\\
 state          			& state 					& (GeV)				&			& Width (keV)  			& \ \ BR (\%) 			& Width (keV)  	& \ \ BR (\%) 		\\ 
 \hline\hline
$\Upsilon (1^3S_1)$		& $\ell^+\ell^-$ 				&  					& $0.793$ 		& 1.44 	&  2.71		& $1.34 \pm 0.04$  	&  	$2.48\pm 0.05$\footnotemark[1]		\\
	9.460 \footnotemark[1]			& $ggg$				&  					& $0.793$ 		& 47.6 	&  89.6		& $44.1 \pm 1.1 $   &  	$81.7\pm 0.7$\footnotemark[1] 	\\
					& $\gamma gg$						&  					& $0.793$ 		&  1.2 				&  2.3			& $1.2 \pm 0.3 $   	&  	$2.2 \pm 0.6$\footnotemark[1]	\\
					& $\gamma\gamma\gamma $				&  					& $0.793$ 		& $1.7\times10^{-5}$	& $3.2\times10^{-5}$	&    		&  				\\
					& $\eta_b(1^1S_0) \gamma $ 	  	 	& 9.398\footnotemark[1] 	& $0.9947$  		& $0.010$	&  0.019				&  									&  	\\
					& Total								& 					&  			& 53.1		&  100				& $54.02 \pm 1.25$ \footnotemark[1]						&  				\\
\hline					
$\eta_b (1^1S_0)$		& $gg $ 		 				&   				& $1.081 $ 		& 16.6 MeV			&  $\sim 100$			&  									&  				\\
9.398 \footnotemark[1]	& $\gamma\gamma $  				&   				& $1.081 $  	& 0.94	 			&  0.0057				&  									&  				\\
					& Total								&  					&  			& 16.6 MeV			&  100				& $10.8^{+6.0}_{-4.2}$ MeV \footnotemark[1]	&				\\ 
\hline	
$\Upsilon (2^3S_1)$		& $\ell^+\ell^-$				&  					& $0.597 $		& 0.73 				&  1.8		& $0.62 \pm 0.06$ 	& $1.93 \pm 0.17$		\\
	10.023 \footnotemark[1]			& $ggg$				&  					& $0.597 $		& 26.3 				&  65.4		& $18.8 \pm 1.6 $   & $58.8 \pm 1.2$		\\
					& $\gamma gg$			  			&  					& $0.597 $		&  0.68 			&  1.7		& $2.81 \pm 0.42$   & $8.8\pm 1.1$			\\
					& $\gamma\gamma\gamma $				&  					& $0.597 $		& $9.8\times10^{-6}$ 	&  $2.4\times10^{-5}$ 	&    		&  				\\
					& $\chi_{b2}(1^3P_2) \gamma$	& 9.912\footnotemark[1] 	& $-1.524 $  		& 1.88	 &  4.67			&  $2.29\pm 0.22 $	& $7.15\pm 0.35$\footnotemark[1]	\\
					& $\chi_{b1}(1^3P_1) \gamma$	& 9.893\footnotemark[1] 	& $-1.440 $ 		& 1.63	 &  4.05			&  $2.21 \pm 0.22$	& $6.9\pm 0.4$\footnotemark[1]	\\
					& $\chi_{b0}(1^3P_0) \gamma$	& 9.859\footnotemark[1] 	& $-1.330 $ 		& 0.91	 &  2.3		&  $1.22 \pm 0.16$	& $3.8\pm 0.4$\footnotemark[1]	\\
					& $\eta_b(2^1S_0) \gamma $ 		& 9.999\footnotemark[1] 	& $0.9924 $ 		& $5.9\times 10^{-4}$	&  $1.5\times10^{-3}$	&  		&   \\
					& $\eta_b(1^1S_0) \gamma $ 		& 9.398\footnotemark[1] 	& $0.0913 $ 		& 0.081	 			&  0.20			& $0.012\pm 0.004$	& $(3.9\pm 1.5)\times 10^{-2}$ \footnotemark[1]	\\
					& $\Upsilon(1^3S_1) \pi \pi $ 	&  					&  			& 8.46\footnotemark[1]	& 21.0		& $8.46 \pm 0.71$ 			& $26.45 \pm 0.48$\footnotemark[1]	\\  
					& Total							&  					&  			& 40.2 				& 100				& $31.98 \pm 2.63$ 			&  				\\
\hline	
$\eta_b (2^1S_0)$		& $gg $ 					&   				& $0.718 $ 		& 7.2 MeV	 			& $\sim 100$\footnotemark[1] 	&   		&  				\\
9.999 \footnotemark[1]	& $\gamma\gamma $ 			&   				& $0.718 $ 		& 0.41	 			&  $5.7\times10^{-3}$	&  									&  				\\
					& $h_{b}(1^1P_1) \gamma $ 		& 9.899\footnotemark[1] 	& $-1.526 $  		& 2.48	 			&  0.034				&  									&  	\\
					& $\Upsilon(1^3S_1) \gamma$		& 9.460 			& $-0.0610 $	& 0.068			&  $9.4\times10^{-4}$	&  									& 				\\
					& $\eta_b(1^1S_0) \pi \pi $  	& 					& 			& 12.4			&  0.17				&  									& 				\\
					& Total							&  					&  			& 7.2 MeV		&  100				& $< 24$ MeV\footnotemark[1] 	&				\\
\hline \hline
\end{tabular}
\end{center}
\footnotetext[1]{PDG Ref.\cite{Olive:2014kda}.}
\end{table*}

\begin{table*}[tp]
\caption{Partial widths and branching ratios
for strong and electromagnetic decays and transitions for the $3S$ states.  
See the caption to Table~\ref{tab:sum_12Swave} for further details.
\label{tab:sum_3S}}
\begin{center}
\begin{tabular}{l l l l l l l l } \hline \hline
Initial \phantom{www}       & Final 	 				& $M_f$  				& $\cal M$	& \multicolumn{2}{c}{Predicted}				& \multicolumn{2}{c}{Measured} 							\\
 state          			& state 					& (GeV)				&			& Width (keV)  			& \ \ BR (\%) 			& Width (keV)  	& \ \ BR (\%) 		\\ 
 \hline\hline
$\Upsilon (3^3S_1)$		& $\ell^+\ell^-$			&  					& $0.523 $	& 0.53 		&  1.8	& $0.44 \pm 0.06 $ 	& $2.18\pm 0.21$		\\
10.355 \footnotemark[1]	& $ggg$						&  					& $0.523 $	& 19.8 		&  67.9	& $7.25 \pm 0.84 $ 	& $35.7\pm 2.6$	\\
					& $\gamma gg$					&  					& $0.523 $	& 0.52		&  1.8	& $0.20 \pm 0.04 $ 	& $0.97 \pm 0.18$ 	\\
					& $\gamma\gamma\gamma $			&  					& $0.523 $ 	& $7.6\times10^{-6}$ &  $2.6\times10^{-5}$	& 	&  \\
					& $\chi_{b2}(2^3P_2) \gamma$	& 10.269\footnotemark[1] & $-2.446 $ & 2.30 	&  7.90	& $2.66 \pm 0.40$ 	& $13.1 \pm 1.6$\footnotemark[1]	\\
					& $\chi_{b1}(2^3P_1) \gamma$	& 10.255\footnotemark[1] & $-2.326 $ & 1.91	&  6.56	& $2.56 \pm 0.34$ 	& $12.6\pm 1.2$\footnotemark[1]	\\
					& $\chi_{b0}(2^3P_0) \gamma$	& 10.232\footnotemark[1] & $-2.169 $ & 1.03	&  3.54	& $1.20\pm 0.16$ 	& $5.9\pm 0.6$\footnotemark[1]	\\
					& $\chi_{b2}(1^3P_2) \gamma$ 	& 9.912\footnotemark[1]  & $0.096 $	& 0.45	&  1.5 	& $0.20 \pm 0.03$ 	& $0.99\pm0.13$\footnotemark[1]	\\
					& $\chi_{b1}(1^3P_1) \gamma$	& 9.893\footnotemark[1]  & $0.039 $	& 0.05	&  0.2	& $0.018\pm 0.010$ 	& $0.09\pm 0.05$\footnotemark[1]	\\
					& $\chi_{b0}(1^3P_0) \gamma$	& 9.859\footnotemark[1]  & $-0.028 $	& 0.01	&  0.03 & $0.055 \pm 0.010$	& $0.27\pm 0.04$\footnotemark[1]	\\
					& $\eta_b(3^1S_0) \gamma $ 		& 10.337\footnotemark[1] & $0.9920 $ & $2.5\times 10^{-4}$	&  $8.6\times 10^{-4}$	&  			&  	\\
					& $\eta_b(2^1S_0) \gamma $ 		& 9.999\footnotemark[1]  & $0.1003 $ & 0.19	&  0.65	& $<0.12$ 			&  $< 0.062$ at 90\% C.L.\footnotemark[1]	\\
					& $\eta_b(1^1S_0) \gamma $ 		& 9.398\footnotemark[1]  & $0.0427 $ & 0.060	&  0.20 & $0.01\pm 0.002$	& $0.051\pm 0.007$\footnotemark[1]	\\
					& $\Upsilon(1^3S_1) \pi \pi $  	& 					& 			& 	 1.34\footnotemark[1]	& 4.60  & $1.335 \pm 0.125$ & $6.57 \pm 0.15$\footnotemark[1]	\\  
					& $\Upsilon(2^3S_1) \pi \pi $  	& 					& 			& 	 0.95\footnotemark[1] 	& 3.3  & $0.949 \pm 0.098$ & $4.67 \pm 0.23$ \footnotemark[1]	\\  
					& Total							& 					&			& 29.1		&  100	& $20.32\pm 1.85$ 	&  				\\
\hline
$\eta_b (3^1S_0)$		& $gg $ 					&   					& $0.601 $	 	& 4.9 MeV	 	&  $\sim 100$			&   									&  	\\
	10.337\footnotemark[2]			& $\gamma\gamma $ 	&  					& $0.601 $ 		& 0.29			&  $5.9\times 10^{-3}$	& 			 				&  	\\
					& $h_{b}(2^1P_1) \gamma$		& 10.260\footnotemark[1]& $-2.461 $ 	& 2.96	 			&  0.060				&  									&  	\\
					& $h_{b}(1^1P_1) \gamma$		& 9.899\footnotemark[1] & $0.1235 $ 	& 1.3	 				&  0.026				&  									&  	\\
					& $\Upsilon(2^3S_1) \gamma$ 	& 10.023\footnotemark[1]& $-0.0484 $ 	& $9.1\times 10^{-3}$	& $1.8\times 10^{-4}$	&  									&  	\\
					& $\Upsilon(1^3S_1) \gamma$ 	& 9.460\footnotemark[1] 	& $-0.031 $ & $0.074$	 			& $1.5\times 10^{-3}$	&  									&  	\\
					& $\eta_b(1^1S_0) \pi \pi $  	& 					& 			& $1.70 \pm 0.12$  		&  $3.5\times 10^{-2}$		& 			& 				\\  
					& $\eta_b(2^1S_0) \pi \pi $  	& 					& 			& $1.16 \pm 0.10$ 		&  $2.4\times 10^{-2}$		& 			& 				\\  
					& Total							&  					&  			& 4.9 MeV				&  100				& 	 				&				\\
\hline \hline
\end{tabular}
\end{center}
\footnotetext[1]{PDG Ref.\cite{Olive:2014kda}.}
\footnotetext[2]{Using predicted $3^3S_1-3^1S_0$ splitting and measured $3^3S_1$ mass.}
\end{table*}

\begin{table*}[tp]
\caption{Partial widths and branching ratios
for strong and electromagnetic decays and transitions and OZI allowed strong decays for the $4S$ and $5S$ 
bottomonium states. 
Details of the OZI allowed decay amplitudes are described in the appendix.  
See the caption to Table~\ref{tab:sum_12Swave} for further details.
\label{tab:sum_45Swave}}
\begin{center}
\begin{tabular}{l l  l l l l l l } \hline \hline
Initial \phantom{www}       & Final 				& $M_f$  			& $\cal M$& \multicolumn{2}{c}{Predicted}			& \multicolumn{2}{c}{Measured}					 			\\
 state          			& state 				&(GeV)				&		& Width (keV)  		& \ \ BR (\%) 			& Width (keV)  			& \ \ BR (\%)							\\ 
 \hline\hline
$\Upsilon (4^3S_1)$		& $\ell^+\ell^-$		 	&  					& $0.459$ 	& 0.39			& $1.8\times 10^{-3}$  	& $0.32 \pm 0.04$\footnotemark[1]	& $(1.57 \pm 0.08)\times 10^{-3}$\footnotemark[1]		\\
10.579\footnotemark[1]	& $ggg$						&  					& $0.459$ 	& 15.1 			&  0.0686				&   			& 									\\
					& $\gamma gg$					&  					& $0.459$ 	&  0.40			& $1.8\times 10^{-3}$ 	&   			&									\\
					& $\gamma\gamma\gamma $			&  					& $0.459$  	& $6.0\times10^{-6}$& $2.7\times 10^{-8}$ 	&    					& 									\\
					& $\chi_{b2}(3^3P_2) \gamma $ 		& 10.528\footnotemark[4]& $-3.223$  	& 0.82	 		& $3.7\times 10^{-3}$ 	&  					&  									\\
					& $\chi_{b1}(3^3P_1) \gamma $		& 10.516\footnotemark[4]& $-3.072$  	& 0.84	 		& $3.8\times 10^{-3}$ 	&  					&  									\\
					& $\chi_{b0}(3^3P_0) \gamma $ 		& 10.500\footnotemark[4]& $-2.869$  	& 0.48	 		& $2.2\times 10^{-3}$ 	&  					&  									\\
					& $\Upsilon(1^3S_1) \pi^+ \pi^- $ 		&		& 		& 	1.66\footnotemark[2] &  $7.54\times 10^{-3}$	& $1.66 \pm 0.24$\footnotemark[2] 		& $(8.1\pm 0.6)\times 10^{-3}$\footnotemark[1] \\  
					& $\Upsilon(2^3S_1) \pi^+ \pi^- $ 		&		& 		&	1.76\footnotemark[2] &  $8.00\times 10^{-3}$	& $1.76 \pm 0.34$\footnotemark[2] 		& $(8.6\pm 1.3)\times 10^{-3}$\footnotemark[1] \\  
					& $BB$							& 					&		& 22.0 MeV		& $\sim 100$			&					& $ > 96 $	 at 95\% C.L. \footnotemark[1] 								\\
					& Total					 		& 					&		& 22.0 MeV		&  100				& $20.5 \pm 2.5$ MeV\footnotemark[1] 	&									\\
\hline
$\eta_b (4^1S_0)$		& $gg $ 				  &   					& $0.500$  	& 3.4 MeV	 		& $\sim 100$	 		&   					&									\\
10.567\footnotemark[3]	& $\gamma\gamma $ 		 &   					& $0.500$  	& 0.20	 		& $5.9\times 10^{-3}$ 	&  					&									\\
					& $h_{b}(3^1P_1) \gamma $ 			& 10.519 \footnotemark[4]	& $-3.238$  	& 1.24	 		& $3.6\times 10^{-2}$  	& 	 				&									\\
					& $\eta_b(1^1S_0) \pi^+ \pi^- $		& 				& 		& $2.03 \pm 0.29$  	&  	$6.0\times 10^{-2}$	&  					&									\\   
					& $\eta_b(2^1S_0) \pi^+ \pi^- $		& 				& 		& $1.90 \pm 0.36$ 	&  	$5.6\times 10^{-2}$					&  					&									\\ 	
					& Total							& 					&		& 3.4 MeV			& 100 				& 				 	&									\\
\hline
$\Upsilon (5^3S_1)$		& $\ell^+\ell^-$				&  					& $0.432$ 	& 0.33	& $1.2\times 10^{-3}$ 	& $0.31 \pm 0.23 $\footnotemark[1] 	& $(5.6\pm 3.1)\times 10^{-4}$\footnotemark[1]		\\
10.876\footnotemark[1]	& $ggg$					 		&  					& $0.432$ 	& 13.1 			& $4.78\times 10^{-2}$	&  	 		&									\\
					& $\chi_{b2}(4^3P_2) \gamma $		& 10.798  				& $-3.908$ 	& 4.3	 			& $1.6\times 10^{-2}$ 	&  					&									\\
					& $\chi_{b1}(4^3P_1) \gamma $		& 10.788  				& $-3.724$ 	& 3.4	 			& $1.2\times 10^{-2}$  	&  					&									\\
					& $\chi_{b0}(4^3P_0) \gamma $		& 10.775  				& $-3.483$ 	& 1.5	 			& $5.5\times 10^{-3}$  	&  					&									\\
					& $\chi_{b2}(3^3P_2) \gamma $		& 10.528 \footnotemark[4] & $0.131$ 	& 0.42	 		& $1.5\times 10^{-3}$ 	& 	 				&									\\
					& $\chi_{b1}(3^3P_1) \gamma $	& 10.516 \footnotemark[4] & $0.0020$ 	& $6.2\times 10^{-5}$& $2.3\times 10^{-7}$ 	&  					&									\\
					& $\chi_{b0}(3^3P_0) \gamma $	& 10.500 \footnotemark[4] & $-0.156$ 	& 0.15	 		& $5.5\times 10^{-4}$ 	&  					&									\\
					& $BB$							& 					& 		& 5.35 MeV		& 19.5				&  $3.0\pm1.6$ MeV\footnotemark[2]		&	$5.5\pm1.0$\footnotemark[1]	\\
					& $BB^*$				 		& 					& 		& 16.6 MeV		& 60.6					&  $7.5\pm3.9$ MeV\footnotemark[2]		&	$13.7\pm1.6$\footnotemark[1]	\\
					& $B^*B^*$						& 					& 		& 2.42 MeV		& 8.83				&  $21\pm11$ MeV\footnotemark[2]		&	$38.1\pm3.4$\footnotemark[1]	\\
					& $B_sB_s$			 			& 					&		& 0.157 MeV		& 0.573				&  $0.3\pm0.3$ MeV\footnotemark[2]		&	$0.5\pm0.5$\footnotemark[1]	\\
					& $B_sB_s^*$			 		& 					&		& 0.833 MeV		& 3.04					&  $0.74\pm0.42$ MeV\footnotemark[2]	&	$1.35\pm0.32$\footnotemark[1]	\\
					& $B_s^*B_s^*$			 		& 					&		& 2.00 MeV		& 7.30				&  $9.7\pm5.1$	MeV\footnotemark[2]	&	$17.6\pm2.7$\footnotemark[1]	\\
					& Total					 		& 					& 		& 27.4 MeV		& 100				& $55 \pm 28$ MeV\footnotemark[1] 		&				\\
\hline
$\eta_b (5^1S_0)$		& $gg $ 					&   					& $0.464$  	& 2.9 MeV	 		& 13 					&   					&									\\
10.867\footnotemark[3]	& $\gamma\gamma $ 		 	&   					& $0.464$  	& 0.17	 		& $7.4\times 10^{-4}$ 	&  					&									\\
					& $h_{b}(4^1P_1) \gamma $ 	  		& 10.790 	 			& $3.925$  	& 7.5	 			& $3.3\times 10^{-2}$ 	&  					&									\\
					& $h_{b}(3^1P_1) \gamma $ 	  		& 10.519 \footnotemark[4]	& $0.162$  	& 1.1	 			&  $4.8\times 10^{-3}$	&  					&									\\
					& $BB^*$					 		& 					& 		& 13.1  MeV		& 57.0				&  					&									\\
					& $B^*B^*$						& 					& 		& 0.914  MeV		& 3.97				&  					&									\\
					& $B_sB_s^*$			 		& 					& 		& 0.559  MeV		& 2.43				&  					&									\\
					& $B_s^*B_s^*$			 		& 					& 		& 5.49  MeV		& 23.9				&  					&									\\
					& Total							& 					&		& 23.0 MeV		& 100				& 				 	&									\\
\hline \hline
\end{tabular}
\end{center}
\footnotetext[1]{From PDG Ref.\cite{Olive:2014kda}.}
\footnotetext[2]{Found by combining the PDG BR with the PDG  total widths for the $\Upsilon(4^3S_1)$ or $\Upsilon(5^3S_1)$ \cite{Olive:2014kda}} as appropriate.
\footnotetext[3]{Using predicted $n^3S_1-n^1S_0$ splitting and measured $n^3S_1$ mass.}
\footnotetext[4]{ $3^3P_1$ from LHCb \cite{Aaij:2014hla} and $3^3P_2$, $3^3P_0$ and $3^1P_1$ using predicted splittings with $3^3P_1$ measured mass.}

\end{table*}

\begin{table*}[tp]
\caption{Partial widths and branching ratios
for strong and electromagnetic decays and transitions and OZI allowed strong decays for the $6S$ states.
Details of the OZI allowed decay amplitudes are described in the appendix.  
See the caption to Table~\ref{tab:sum_12Swave} for further details.
\label{tab:sum_6Swave}}
\begin{center}
\begin{tabular}{l l  l l l l l l } \hline \hline
Initial \phantom{www}       & Final 					&  $M_f$  	& $\cal M$ 	& \multicolumn{2}{c}{Predicted}		& \multicolumn{2}{c}{Measured} 	\\
 state          			& state 					& 		&			& Width (keV)  			& \ \ BR (\%) 			& Width (keV)  		& \ \ BR (\%)  	\\ 
 \hline\hline
$\Upsilon (6^3S_1)$	& $\ell^+\ell^-$		 	&  					& $0.396$ 		& 0.27				& $8.0 \times 10^{-4}$ 	& $0.13 \pm 0.05$\footnotemark[1]		& $(1.6\pm 0.5)\times 10^{-4}$\footnotemark[1]	\\
 11.019\footnotemark[1] & $ggg$					& 		  			& $0.396$ 		& 11.0 				&  0.0324				&  		&  			\\
					& $\chi_{b2}(5^3P_2) \gamma $ 	& 11.022  				&  			& below threshold		&  -					&  				&  			\\
					& $\chi_{b1}(5^3P_1) \gamma $	& 11.014  				& $-4.282$ 		& $8.3\times 10^{-4}$ 	& $2.4 \times 10^{-6}$ 	&  				&  			\\
					& $\chi_{b0}(5^3P_0) \gamma $	& 11.004 				& $-3.992$ 		& $6.4 \times 10^{-3}$	& $1.9 \times 10^{-5}$ 	&  				&  			\\
					& $\chi_{b2}(4^3P_2) \gamma $  	& 10.798  				& $0.116$ 		& $8.5\times 10^{-2}$	& $2.5 \times 10^{-4}$ 	&  				&  			\\
					& $\chi_{b1}(4^3P_1) \gamma $ 	& 10.788  				& $-0.054$ 		& 0.012	 			& $3.5 \times 10^{-5}$ 	&  				&  			\\
					& $\chi_{b0}(4^3P_0) \gamma $  	& 10.775  				& $-0.244$ 		& 0.1	 				& $3 \times 10^{-4}$ 		&  				&  			\\
					& $\chi_{b2}(3^3P_2) \gamma $  	& 10.528 \footnotemark[3] & $0.089$ 		& 0.53		 		& $1.6 \times 10^{-3}$ 	&  				&  			\\
					& $\chi_{b1}(3^3P_1) \gamma $   & 10.516 \footnotemark[3] & $0.100 $ 		& 0.43	 			& $1.3 \times 10^{-3}$ 	&  				&  			\\
					& $\chi_{b0}(3^3P_0) \gamma $  	& 10.500 \footnotemark[3] & $0.115 $ 		& 0.21	 			& $6.2 \times 10^{-4}$ 	&  				&  			\\
					& $BB$					& 					& 			& 1.32 MeV			& 3.89				&  				&  			\\
					& $BB^*$				& 					& 			& 7.59 MeV			& 22.4				&  				&  			\\
					& $BB(1P_1)$			& 					& 			& 7.81 MeV			& 23.0				&  				&  			\\
					& $BB(1P_1^\prime)$		& 					&			& 10.8 MeV			& 31.8				&  				&  			\\
					& $B^*B^*$				& 					& 			& 5.89 MeV			& 17.4				&  				&  			\\
					& $B_sB_s$				& 					& 			& 1.31				& $3.86 \times 10^{-3}$	&  				&  			\\
					& $B_sB_s^*$			& 					& 			& 0.136 MeV			& 0.401				&  				&  			\\
					& $B_s^*B_s^*$			& 					& 			& 0.310 MeV			& 0.914				&  				&  			\\
					& Total					& 					& 			& 33.9 MeV			&  100				& $79 \pm 16$ MeV\footnotemark[1] 	&  			\\
\hline
$\eta_b (6^1S_0)$		& $gg $ 			&   		& $0.410$  		& 2.2 MeV	 			&  16					&   				&  			\\
11.014\footnotemark[2]	& $\gamma\gamma $  	&   		& $0.410$  		& 0.14	 			&  $1.0 \times 10^{-3}$	&  				&  			\\
					& $h_{b}(5^1P_1) \gamma $ 	& 11.016  				&   			& below threshold	 	&  -					&  				&  			\\
					& $h_{b}(4^1P_1) \gamma $ 	& 10.790 	 			& $0.136 $ 		& 0.22	 			&  $1.6 \times 10^{-3}$	&  				&  			\\
					& $h_{b}(3^1P_1) \gamma $ 	& 10.519\footnotemark[3]	& $0.123 $ 		& 1.8	 				& $1.3 \times 10^{-2}$	&	  			&  			\\
					& $BB^*$					& 					& 			& 8.98 MeV			& 66.0				&  				&			\\
					& $BB(1^3P_0)$				& 					& 			& 0.745 MeV			& 5.48				&  				&			\\
					& $B^*B^*$					& 					& 			& 1.14 MeV			& 8.38				&  				&			\\
					& $B_sB_s^*$				& 					& 			& 0.420 MeV			& 3.09				&  				&			\\
					& $B_s^*B_s^*$				& 					& 			& 0.156 MeV			& 1.15				&  				&			\\
					& Total						& 					&			& 13.6 MeV			& 100				& 		 		&  			\\
\hline \hline
\end{tabular}
\end{center}
\footnotetext[1]{From PDG Ref.\cite{Olive:2014kda}.}
\footnotetext[2]{Using predicted $n^3S_1-n^1S_0$ splitting and measured $n^3S_1$ mass.}
\footnotetext[3]{ $3^3P_1$ from LHCb \cite{Aaij:2014hla} and $3^3P_2$, $3^3P_0$ and $3^1P_1$ using predicted splittings with $3^3P_1$ measured mass.}
\end{table*}

\begin{table*}[tp]
\caption{Partial widths and branching ratios
for strong and electromagnetic decays and transitions for the $1P$ states.
For $P$-wave annihilation decays ${\cal M}$ designates
$R'(0)$, the first derivative of the radial wavefunction at the origin. 
See the caption to Table~\ref{tab:sum_12Swave} for further details.
\label{tab:sum_1Pwave}}
\begin{center}
\begin{tabular}{l l  l l l l l  } \hline \hline
Initial \phantom{www}       & Final   	& $M_f$ 	& $\cal M$ & \multicolumn{2}{c}{Predicted}				& Measured				\\
 state          			& state 	&(GeV)	&		& Width (keV)  			& \ \ BR (\%) 			& \ \ BR (\%)  					\\ 
 \hline\hline
$\chi_{b2} (1^3P_2)$	& $gg$ 					& 		& $1.318$	& 147 				& 81.7	  			&  							\\
9.912\footnotemark[1]	& $\gamma\gamma $ 			& 		& $1.318$		& $9.3\times 10^{-3}$ 	& $5.2\times 10^{-3}$	&  							\\
					& $\Upsilon (1^3S_1) \gamma$	& 9.460 	& $1.028$ 	& 32.8 	 			& 18.2				& $19.1\pm 1.2$\footnotemark[1]				\\
					& $ h_{b} (1^1P_1) \gamma$ 		& 9.899 	& $1.000$ 	& $9.6\times 10^{-5}$ 	& $5.3\times 10^{-5}$ 	&  							\\
					& Total							&		&		& 180.				& 100		 		&  							\\
\hline
$\chi_{b1} (1^3P_1)$	& $q\bar{q}+g$ 			 	& 		& $1.700$		& 67 					& 70.				  	&  							\\
9.893\footnotemark[1]	& $\Upsilon (1^3S_1) \gamma$ & 9.460 		& $1.040$ 	& 29.5 	 			& 31					&  $33.9\pm 2.2$\footnotemark[1]				\\
					& Total						& 		&		& 96					& 100				&  							\\
\hline
$\chi_{b0} (1^3P_0)$	& $gg$ 				 	& 		& $2.255$		& 2.6 MeV  			& $\sim 100$		  	&  							\\
9.859\footnotemark[1] 	& $\gamma\gamma $ 			& 		& $2.255$		& 0.15  				& $5.8\times 10^{-3}$	&  							\\
					& $ \Upsilon (1^3S_1) \gamma$	& 9.460 	& $1.050 $	& 23.8 	 			& 0.92				&  $1.76\pm 0.35$\footnotemark[1]				\\
					& Total						&		&		& 2.6 MeV				& 100 				&  							\\
\hline
$h_b (1^1P_1)$		& $ggg$ 						&		& $1.583$		& 37 					& 51			  		&							\\
 9.899\footnotemark[1] 	& $ \eta_b (1^1S_0)  \gamma$ 	& 9.398	& $0.922 $ 	& 35.7 	 			& 49 					&  $49^{+8}_{-7} $\footnotemark[1]	\\
					& $ \chi_{b1} (1^3P_1)  \gamma$ & 9.893 	& $1.000$ 	& $1.0\times 10^{-5}$ 	& $1.4\times 10^{-5}$ 	&  							\\
					& $ \chi_{b0} (1^3P_0)  \gamma$ 	& 9.859 	& $0.998$ 	& $8.9\times 10^{-4}$ 	& $1.2\times 10^{-5}$ 	&  							\\
					& Total							&		&		& 73					& 100 				&  							\\
\hline \hline
\end{tabular}
\end{center}
\footnotetext[1]{From PDG Ref.\cite{Olive:2014kda}.}
\end{table*}

\begin{table*}[tp]
\caption{Partial widths and branching ratios
for strong and electromagnetic decays and transitions for the $2P$ states.
For $P$-wave annihilation decays ${\cal M}$ designates
$R'(0)$, the first derivative of the radial wavefunction at the origin. 
See the caption to Table~\ref{tab:sum_12Swave} for further details.
\label{tab:sum_2Pwave}}
\begin{center}
\begin{tabular}{l l l l l l l } \hline \hline
Initial \phantom{www}       & Final 							& $M_f$  	& $\cal M$  	& \multicolumn{2}{c}{Predicted}				& Measured								\\
 state          			& state 							&(GeV)	&			& Width (keV)  			& \ \ BR (\%) 			& \ \ BR (\%) 								\\ 
 \hline\hline
$\chi_{b2} (2^3P_2)$	& $gg$ 						 	& 		& 	$1.528$		& 207 				& 89.0	  			&  										\\
10.269\footnotemark[1]	& $\gamma\gamma $ 					& 		& 	$1.528$		& $1.2\times 10^{-2}$	& $5.2\times 10^{-3}$	&  										\\
					& $ \Upsilon (2^3S_1) \gamma$		& 10.023 	& $1.667 $		& 14.3 	 			&  6.15				&  $10.6 \pm 2.6 $\footnotemark[1]				\\
					& $ \Upsilon (1^3S_1) \gamma$	 	& 9.460 	& $0.224 $ 		& 8.4 	 			&  3.6				&  $7.0 \pm 0.7 $\footnotemark[1]				\\
					& $ \Upsilon_3 (1^3D_3) \gamma$	 	& 10.172 	& $-1.684$ 		& 1.5 				&  0.65				&   										\\
					& $ \Upsilon_2 (1^3D_2) \gamma$		& 10.164 	& $-1.625$ 		& 0.3 				&  0.1				&  										\\
					& $ \Upsilon_1 (1^3D_1) \gamma$		& 10.154 	& $-1.561$ 		& 0.03 				&  0.01				&  										\\
					& $ h_{b} (2^1P_1)  \gamma$ 		 	& 10.260 	& $1.000 $		& $2.8\times 10^{-5}$ 	&  $1.2\times 10^{-5}$	&  										\\
					& $ h_{b} (1^1P_1)  \gamma$ 		 	&  9.899 	& $-0.011$ 		& $2.4\times 10^{-4}$ 	&  $1.0\times 10^{-4}$	&  										\\
					& $\chi_{b2}(1^3P_2) \pi \pi $ 	  	& 		& 			&  	$0.62\pm0.12$\footnotemark[2]	& 0.27		& $ 0.51\pm 0.09 $\footnotemark[1]	\\
					& $\chi_{b1}(1^3P_1) \pi \pi $ 		& 		& 			& 	0.23	& 0.10	 			&  										\\
					& $\chi_{b0}(1^3P_0) \pi \pi $  	& 		& 			& 	0.10	& 0.043	 			&  										\\ 
					& Total								& 		& 			& 232.5  	& 100		 			&  										\\
\hline
$\chi_{b1} (2^3P_1)$	& $q\bar{q}+g$ 				 	& 		& $1.857$		& 96 					& 82				  	&  										\\
10.255\footnotemark[1]	& $ \Upsilon (2^3S_1) \gamma $  	& 10.023 	& $1.749$ 		& 13.3 	 			& 11.3				&  $19.9 \pm 1.9 $\footnotemark[1]				\\
					& $ \Upsilon (1^3S_1) \gamma $   	& 9.460 	& $0.184$  		& 5.5 	 			& 4.7		 			&  $9.2 \pm 0.8 $\footnotemark[1]				\\
					& $ \Upsilon_2 (1^3D_2) \gamma $  	& 10.164 	& $-1.701$ 		& 1.2 				& 1.0	 				&  										\\
					& $ \Upsilon_1 (1^3D_1) \gamma $   	& 10.154 	& $-1.639$ 		& 0.5 				& 0.4		 			&  										\\
					& $ h_{b} (1^1P_1) \gamma $ 		 	& 9.899 	& $-0.035$ 		& $2.2\times 10^{-3}$ 	& $1.9\times 10^{-3}$  	&  										\\
					& $\chi_{b2}(1^3P_2) \pi \pi $ 		& 		& 			& 0.38				& 0.32		 		&  										\\
					& $\chi_{b1}(1^3P_1) \pi \pi $ 		& 		&  			& $0.57\pm0.09$\footnotemark[2]	&  0.48			& $0.91\pm 0.13 $\footnotemark[1]	\\
					& $\chi_{b0}(1^3P_0) \pi \pi $ 		& 		& 			& $\sim 0$ 			& $\sim 0$		 	& 										\\
					& Total								& 		& 			& 117 			& 100			  	&  										\\ 
\hline
$\chi_{b0} (2^3P_0)$	& $gg$ 						& 		& 	$2.290$		& 2.6 MeV 			& $\sim 100$		  	&  										\\
10.232\footnotemark[1]	& $\gamma\gamma $ 			 	& 		& 	$2.290$		& 0.15 				& $5.7\times 10^{-3}$	&  										\\
					& $ \Upsilon (2^3S_1) \gamma $  		& 10.023 	& $1.842$  		& 10.9	 			& 0.42	&   $4.6\pm 2.1$\footnotemark[1]							\\
					& $ \Upsilon (1^3S_1) \gamma $ 	 	&  9.460 	& $0.130$  		& 2.5 	 			& $9.6\times 10^{-2}$	&  $0.9\pm 0.6$\footnotemark[1]							\\
					& $ \Upsilon_1 (1^3D_1) \gamma $   	& 10.154 	& $-1.731$ 		& 1.0  				& $3.8\times 10^{-2}$	&  										\\
					& $ h_{b} (1^1P_1) \gamma $ 			& 9.899 	& $-0.079$ 		& $9.7\times 10^{-3}$ 	& $3.7\times 10^{-4}$	&  										\\
					& $\chi_{b2}(1^3P_2) \pi \pi $  	& 		& 			& 0.5				& $2\times 10^{-2}$	 	&  										\\
					& $\chi_{b1}(1^3P_1) \pi \pi $ 		& 		& 			& $\sim 0$ 			& $\sim 0$ 	 		&  										\\
					& $\chi_{b0}(1^3P_0) \pi \pi $ 		& 		& 			& 0.44 				& $1.7\times 10^{-2}$	&  										\\
					& Total								&		&			& 2.6 MeV 		& 100 				&  										\\
\hline
$h_b (2^1P_1)$		& $ggg$ 							& 		& 	$1.758$		& 54 					& 64		  			&  										\\
10.260\footnotemark[1]	& $ \eta_b (2^1S_0) \gamma $  	& 9.999 	& $1.510$  		& 14.1 	 			& 17 	&  	$48\pm 13$\footnotemark[1]									\\
					& $ \eta_b (1^1S_0) \gamma $  		& 9.398 	& $0.252$  		& 13.0 	 			& 15	 				&  	$22\pm 5$\footnotemark[1]									\\
					& $ \eta_{b2} (1^1D_2) \gamma $ 		& 10.165 	& $-1.689$  		& 1.7 	 			& 2.0					&  										\\
					& $ \chi_{b2} (1^3P_2) \gamma $  	& 9.912 	& $-0.027$  		& $2.2\times 10^{-3}$  	& $2.6\times 10^{-3}$  	&  										\\
					& $ \chi_{b1} (1^3P_1) \gamma $  	& 9.893 	& $-0.023$  		& $1.1\times 10^{-3}$  	& $1.3\times 10^{-3}$ 	&  										\\
					& $ \chi_{b0} (1^3P_0) \gamma $  	& 9.859 	& $0.019 $ 		& $3.2\times 10^{-4}$  	& $3.8\times 10^{-4}$ 	&  										\\
					& $h_b(1^1P_1) \pi \pi $  			& 			&  				& 0.94	 				& 1.1 				&  										\\
					& Total								&		&			& 84					& 100  				&  										\\
\hline \hline
\end{tabular}
\end{center}
\footnotetext[1]{From PDG Ref.\cite{Olive:2014kda}.}
\footnotetext[2]{Input, see text.}
\end{table*}

\begin{table}[tp]
\caption{Partial widths and branching ratios
for strong and electromagnetic decays and transitions for the $3P$ states.
For $P$-wave annihilation decays ${\cal M}$ designates
$R'(0)$, the first derivative of the radial wavefunction at the origin. 
See the caption to Table~\ref{tab:sum_12Swave} for further details.
\label{tab:sum_3Pwave}}
\vspace{-3mm}
\begin{center}
\begin{tabular}{l l l l l l } \hline \hline
Initial \phantom{www}       & Final 						& $M_f$  	& $\cal M$  	& Width 			& BR 	\\
 state          			& state 						&(GeV)	&			& (keV)  			& (\%) 	\\ 
\hline\hline
$\chi_{b2} (3^3P_2)$	& $gg$ 				 	& 		& $ 1.584$		& 227 				& 91.9							\\
10.528\footnotemark[2]	& $\gamma\gamma $ 			& 		& $1.584$		& $1.3\times 10^{-2}$ 	& $5.3\times 10^{-3}$		  	\\
					& $ \Upsilon (3^3S_1) \gamma $  	& 10.355 	& $2.255$ 		& 9.3 	 			& 3.8  				  		\\
					& $ \Upsilon (2^3S_1) \gamma $  	& 10.023 	& $0.323$  		&  4.5 	 			& 1.8 					  	\\
					& $ \Upsilon (1^3S_1) \gamma $  	&  9.460 	& $0.086$  		&  2.8 	 			& 1.1 					  	\\
					& $ \Upsilon_3 (2^3D_3) \gamma $ & 10.455 	& $-2.568$ 		& 1.5  				& 0.61 				 		\\
					& $ \Upsilon_2 (2^3D_2) \gamma $ 	& 10.449 	& $-2.482 $		& 0.32  				& 0.13 	 		  	\\
					& $ \Upsilon_1 (2^3D_1) \gamma $		& 10.441 	& $-2.389$ 		& 0.027 				& 0.011 		  	\\
					& $ \Upsilon_3 (1^3D_3) \gamma $  	& 10.172 	& $0.042$ 		& 0.046  				& 0.019 		 	\\
					& $\chi_{b2}(1^3P_2) \pi \pi $ 		& 		& 			& 0.68	 	& 0.28 					  	\\
					& $\chi_{b1}(1^3P_1) \pi \pi $ 		& 		& 			& 0.52 		& 0.21 			 		  	\\
					& $\chi_{b0}(1^3P_0) \pi \pi $ 		& 		& 			&  0.24		& 0.10 	 			 		  	\\
					& Total								& 		&  			& 247				& 100  					  	\\
\hline
$\chi_{b1} (3^3P_1)$	& $q\bar{q}+g$ 	& 		& 	$1.814$	& 101 				& 86.3  			 		  	\\
10.516\footnotemark[1] 	& $ \Upsilon (3^3S_1) \gamma $  	& 10.355 	& $2.388 $ 		& 8.4 	 			& 7.2 	 		  	\\
					& $ \Upsilon (2^3S_1) \gamma $  & 10.023 	& $0.278$  		&  3.1 	 			& 2.6 				  	\\
					& $ \Upsilon (1^3S_1) \gamma $  	&  9.460 	& $0.061 $ 		&  1.3 	 			& 1.1 				  	\\
					& $ \Upsilon_2 (2^3D_2) \gamma $  	& 10.449 	& $-2.595 $		& 1.1  				& 0.94 	  		  	\\
					& $ \Upsilon_1 (2^3D_1) \gamma $  	& 10.441 	& $-2.506 $		& 0.47 				& 0.40 			  	\\
					& $ \Upsilon_2 (1^3D_2) \gamma $  & 10.164 	&  $0.060 	$	& 0.080  				& 0.068 		  	\\
					& $ \Upsilon_1 (1^3D_1) \gamma $  	& 10.154 	&  $0.029 	$	& $7.0\times 10^{-3}$ 	& $6.0\times 10^{-3}$ 	  	\\
					& $\chi_{b2}(1^3P_2) \pi \pi $  	& 		& 			& 0.88 	& 0.75	 			 		  	\\
					& $\chi_{b1}(1^3P_1) \pi \pi $ 		& 		& 			& 0.56 	& 0.48 				 		  	\\
					& $\chi_{b0}(1^3P_0) \pi \pi $ 		& 		& 			& $\sim 0$ 			& $\sim 0$ 			 		  	\\
					& Total							& 		&  			& 117				& 100  				  	\\
\hline
$\chi_{b0} (3^3P_0)$	& $gg$ 					& 		& 	$2.104$	& 2.2 MeV 			& $\sim$100  				  	\\
10.500\footnotemark[2]	& $\gamma\gamma $ 			& 		& 	$2.104$	& 0.13 				& $5.9\times 10^{-3}$  		  	\\
					& $ \Upsilon (3^3S_1)  \gamma$	& 10.355 	& $2.537 $ 		& 6.9 	 			& 0.31 		 		  	\\
					& $ \Upsilon (2^3S_1)  \gamma$   & 10.023 	& $0.213  $		&  1.7 	 			& 0.077 	 		  	\\
					& $ \Upsilon (1^3S_1)  \gamma$  	&  9.460 	& $0.029 $ 		&  0.3 	 			& 0.01		 		  	\\
					& $ \Upsilon_1 (2^3D_1)  \gamma$  		& 10.441 	& $-2.637$ 		&  1.0 				& 0.045	 	  	\\
					& $ \Upsilon_1 (1^3D_1)  \gamma$  	& 10.154 	&  $0.084 	$	&  0.20 				& 0.0091	  	\\
					& $\chi_{b2}(1^3P_2) \pi \pi $ 		& 		& 			& 1.2		& 0.054			 		  	\\
					& $\chi_{b1}(1^3P_1) \pi \pi $  	& 		& 			& $\sim 0$ 	& $\sim 0$	 		  	\\
					& $\chi_{b0}(1^3P_0) \pi \pi $  	& 		& 			& 0.27		& 0.012			 		  	\\
					& Total						& 		&  			& 2.2 MeV				& 100  					  	\\
\hline
$h_b (3^1P_1)$		& $ggg$ 							& 		& $1.749$		& 59 					& 71				  	\\
10.519\footnotemark[2]	& $ \eta_b (3^1S_0)  \gamma$  	& 10.337 	& $2.047 $ 		& 8.9 				& 11	 	 	\\
					& $ \eta_b (2^1S_0)  \gamma$  		&  9.999 	& $0.418 $ 		&  8.2 				& 9.9 			 	\\
					& $ \eta_b (1^1S_0)  \gamma$  		&  9.398 	& $0.091 $ 		&  3.6 				& 4.3  			  	\\
					& $ \eta_{b2} (2^1D_2) \gamma $  	& 10.450 	& $-2.579 $ 		& 1.6 	 			& 1.9 			  	\\
					& $ \eta_{b2} (1^1D_2) \gamma $  	& 10.165 	&  $0.052 $ 		& 0.081 	 			& 0.098 			  	\\
					& $ \chi_{b2} (1^3P_2)  \gamma$  	&  9.912 	&  $-0.032 $ 		& $1.4\times 10^{-2}$  	& $1.7\times 10^{-2}$ 		\\
					& $ \chi_{b1} (1^3P_1)  \gamma$  	&  9.893 	&  $-0.031  $		& $9.3\times 10^{-3}$  	& $1.1\times 10^{-2}$ 	 	\\
					& $ \chi_{b0} (1^3P_0)  \gamma$  	&  9.859 	&  $-0.016  $		& $9.8\times 10^{-4}$  	& $1.2\times 10^{-3}$		\\
					& $h_b(1^1P_1) \pi \pi $ 			& 		& 			& 1.4			& 1.7	 					  	\\
					& Total						& 		&  			& 83					& 100  						  	\\
\hline \hline
\end{tabular}
\end{center}
\vspace{-3mm}
\footnotetext[1]{From LHCb  Ref.\cite{Aaij:2014hla}.}
\footnotetext[2]{Using the predicted $3P$ splittings with the measured $3^3P_1$ mass.}
\end{table}

\begin{table}[tp]
\caption{Partial widths and branching ratios
for strong and electromagnetic decays and transitions and OZI allowed strong decays for the $4P$ states.
For $P$-wave annihilation decays ${\cal M}$ designates
$R'(0)$, the first derivative of the radial wavefunction at the origin. 
Details of the OZI allowed decay amplitudes are described in the appendix.  
See the caption to Table~\ref{tab:sum_12Swave} for further details.
\label{tab:sum_4Pwave}}
\begin{center}
\begin{tabular}{l l l  l l l  } \hline \hline
Initial \phantom{www}       & Final 		 	& $M_f$  	& $\cal M$  	& Width 				& BR 				\\
 state          			& state 			&(GeV)	&			& (keV)  				&  (\%)  				\\ 
 \hline\hline
$\chi_{b2} (4^3P_2)$	& $gg$ 				& 		& 	1.646		& 248 				& 0.569				\\
10.798 					& $\gamma\gamma $ 		& 		& 	1.646		& $1.5\times 10^{-2}$	& $3.4\times 10^{-5}$	\\
					& $ \Upsilon (4^3S_1)  \gamma$  	& 10.579 	& 2.765  		& 28.1 	 			& $6.44\times 10^{-2}$ 	\\
					& $ \Upsilon (3^3S_1)  \gamma$  	& 10.355 	& 0.427  		& 5.4 	 			& $1.2\times 10^{-2}$  	\\
					& $ \Upsilon (2^3S_1)  \gamma$  	& 10.023 	& 0.063  		& 0.59 	 			& $1.4\times 10^{-3}$  	\\
					& $ \Upsilon (1^3S_1)  \gamma$  	&  9.460 	& 0.056  		& 2.2 	 			& $5.0\times 10^{-3}$   	\\
					& $ \Upsilon_3 (3^3D_3)  \gamma$  	& 10.711 	& -3.310  		& 4.3 	 			& $9.9\times 10^{-3}$ 	\\
					& $ \Upsilon_2 (3^3D_2)  \gamma$  	& 10.705 	& -3.202  		& 0.88  				& $2.0\times 10^{-3}$ 	\\
					& $ \Upsilon_1 (3^3D_1)  \gamma$  	& 10.698 	& -3.084  		& 0.68  				& $1.6\times 10^{-3}$  	\\
					& $BB$							&		& 			& 8.74 MeV			& 20.0				\\
					& $BB^*$						&		& 			& 28.1 MeV			& 64.4				\\
					& $B^*B^*$						&		&			& 5.05 MeV			& 11.6				\\
					& $B_sB_s$						&		&			& 0.593 MeV			& 1.36				\\
					& $B_sB_s^*$					&		& 			& 0.833 MeV			& 1.91				\\
					& Total						&		&			& 43.6 MeV			& 100   				\\
\hline
$\chi_{b1} (4^3P_1)$	& $q\bar{q}+g$ 				& 		& 	1.849	& 110 				& 0.36				\\
10.788					& $ \Upsilon (4^3S_1) \gamma $  	& 10.579 	& 2.942  		& 27.7 	 			& $9.17\times 10^{-2}$  	\\
					& $ \Upsilon (3^3S_1)  \gamma$  		& 10.355 	& 0.373  		& 3.8 	 			& $1.3\times 10^{-2}$   	\\
					& $ \Upsilon (2^3S_1)  \gamma$  	& 10.023 	& 0.035  		& 0.18 	 			& $6.0\times 10^{-4}$   	\\
					& $ \Upsilon (1^3S_1)  \gamma$  		&  9.460 	& 0.038  		& 1.0 	 			& $3.3\times 10^{-3}$    	\\
					& $ \Upsilon_2 (3^3D_2)  \gamma$ 		& 10.705 	& -3.345  		& 3.4 	 			& $1.1\times 10^{-2}$   	\\
					& $ \Upsilon_1 (3^3D_1)  \gamma$  		& 10.698 	& -3.234  		& 1.4 	 			& $4.6\times 10^{-3}$   	\\
					& $BB^*$							&		&			& 20.6 MeV			& 68.3				\\
					& $B^*B^*$							&		&			& 0.478 MeV			& 1.58				\\
					& $B_sB_s^*$						&		& 			& 8.93 MeV			& 29.6				\\
					& Total								&		&			& 30.2 MeV			& 100			   	\\
\hline
$\chi_{b0} (4^3P_0)$	& $gg$ 						& 		& 	2.079	& 2.1 MeV 			& 6.1					\\
10.775					& $\gamma\gamma $ 			 	& 		& 	2.079	& 0.13 				& $3.8\times 10^{-4}$    	\\
					& $ \Upsilon (4^3S_1)  \gamma$  	& 10.579 	& 3.139  		& 26.0 		 		&  $7.54\times 10^{-2}$ 	\\
					& $ \Upsilon (3^3S_1)  \gamma$  	& 10.355 	& 0.295  		& 2.2 	 			& $6.4\times 10^{-3}$    	\\
					& $ \Upsilon (2^3S_1)  \gamma$  	& 10.023 	& -0.001  		& $9\times 10^{-5}$ 		& $3\times 10^{-7}$   	\\
					& $ \Upsilon (1^3S_1)  \gamma$  		&  9.460 	& 0.017  		& 0.21 	 			& $6.1\times 10^{-4}$    	\\
					& $ \Upsilon_1 (3^3D_1)  \gamma$  	& 10.698 	& -3.397  		& 3.8 	 			& $1.1\times 10^{-2}$   	\\
					& $BB$								&		& 			& 20.0 MeV			& 58.0				\\
					& $B^*B^*$							&		& 			& 12.2 MeV			& 35.4				\\
					& $B_sB_s$							&		& 			& 0.129 MeV			& 0.374				\\
					& Total								&		&			& 34.5 MeV			& 100   				\\
\hline
$h_b (4^1P_1)$		& $ggg$ 						 	& 		& 	1.790		& 64 					& 0.16				\\
10.790					& $ \eta_{b} (4^1S_0)  \gamma$ 	& 10.567 	& 2.808  		& 24.4 	 			& $6.07\times 10^{-2}$  	\\
					& $ \eta_{b} (3^1S_0)  \gamma$  		& 10.337 	& 0.587  		& 10.8 	 			& $2.69\times 10^{-2}$   	\\
					& $ \eta_{b} (2^1S_0)  \gamma$  		&  9.999 	& 0.055  		& 0.48 	 			& $1.2\times 10^{-3}$   	\\
					& $ \eta_{b} (1^1S_0)  \gamma$  		&  9.398 	& 0.052  		& 2.1 	 			& $5.2\times 10^{-3}$   	\\
					& $ \eta_{b2} (3^1D_2) \gamma $ 		& 10.790 	& -3.325  		& 4.7 	 			& $1.2\times 10^{-2}$    	\\
					& $BB^*$							&		&			& 31.8 MeV			& 79.1				\\
					& $B^*B^*$						&		& 			& 4.09 MeV			& 10.2				\\
					& $B_sB_s^*$					&		& 			& 4.18 MeV			& 10.4				\\
					& Total							&		&			& 40.2 MeV			& 100   				\\
\hline \hline
\end{tabular}
\end{center}
\end{table}

\begin{table}[tp]
\caption{Partial widths and branching ratios
for strong and electromagnetic transitions and OZI allowed strong decays for the $5^3P_2$ and
$5^3P_1$ states.
Details of the OZI allowed decay amplitudes are described in the appendix.  
See the caption to Table~\ref{tab:sum_12Swave} for further details.
\label{tab:sum_5Pwave}}
\begin{center}
\begin{tabular}{l l l l l l } \hline \hline
Initial \phantom{www}       	& Final 		  	& $M_f$  	& $\cal M$  	& Width 				& BR  				\\
 state          			& state 				&(GeV)	&			& (keV)  				& (\%) 				\\
\hline \hline
$\chi_{b2} (5^3P_2)$& $ \Upsilon (5^3S_1)  \gamma$  	& 10.876 	& 3.232  		& 11.5 	 			& $2.06\times 10^{-2}$   	\\
11.022				& $ \Upsilon (4^3S_1)  \gamma$  	& 10.579 	& 0.595  		& 10.4 	 			& $1.86\times 10^{-2}$ 	\\
					& $ \Upsilon (3^3S_1)  \gamma$  	& 10.355 	& 0.024  		& 0.06 	 			& $1\times 10^{-4}$   	\\
					& $ \Upsilon (2^3S_1)  \gamma$  	& 10.023 	& 0.067  		& 1.4 	 			& $2.5\times 10^{-3}$    	\\
					& $ \Upsilon (1^3S_1)  \gamma$ 	&  9.460 	& 0.042  		& 1.9 	 			& $3.4\times 10^{-3}$    	\\
					& $ \Upsilon_3 (4^3D_3)  \gamma$  	& 10.939	& -3.955  		& 5.4 	 			& $9.7\times 10^{-3}$   	\\
					& $ \Upsilon_2 (4^3D_2)  \gamma$  	& 10.934 	& -3.828  		& 1.1 	 			& $2.0\times 10^{-3}$   	\\
					& $ \Upsilon_1 (4^3D_1)  \gamma$  	& 10.928 	& -3.685  		& 0.08 	 			& $1\times 10^{-4}$  		\\
					& $BB$							&		& 			& 0.456 MeV			& 0.816				\\
					& $BB^*$						&		& 			& 2.71 MeV			& 4.85				\\
					& $BB(1P_1)$					&		& 			& 4.72 MeV			& 8.44				\\
					& $BB(1P_1^\prime)$				&		& 			& 15.8 MeV			& 28.3				\\
					& $B^*B^*$					&		& 			& 31.3 MeV			& 56.0				\\
					& $B_sB_s$					&		& 			& 0.154 MeV			& 0.275				\\
					& $B_sB_s^*$				&		& 			& 0.130 MeV			& 0.232				\\
					& $B_s^*B_s^*$				&		& 			& 0.618 MeV			& 1.10				\\
					& Total						&		&			& 55.9 MeV			& 100   				\\
\hline	
$\chi_{b1} (5^3P_1)$	& $ \Upsilon (5^3S_1)  \gamma$  		& 10.876 	& 3.439  		& 11.0 		 		&  $1.75\times 10^{-2}$	\\
 11.014 				& $ \Upsilon (4^3S_1)  \gamma$ 	& 10.579 	& 0.534  		&  8.0 	 			&  $1.3\times 10^{-2}$	\\
					& $ \Upsilon (3^3S_1)  \gamma$  		& 10.355 	& -0.006  		& $2.8\times 10^{-3}$ 	&  $4.4\times 10^{-6}$ 	\\
					& $ \Upsilon (2^3S_1)  \gamma$  		& 10.023 	& 0.052  		& 0.83 	 			&  $1.3\times 10^{-3}$   	\\
					& $ \Upsilon (1^3S_1)  \gamma$  	&  9.460 	& 0.029  		& 0.90 	 			&  $1.4\times 10^{-3}$  	\\
					& $ \Upsilon_2 (4^3D_2)  \gamma$  		& 10.934 	& -3.990  		& 4.4 	 			&  $7.0\times 10^{-3}$ 	\\
					& $ \Upsilon_1 (4^3D_1)  \gamma$  		& 10.928 	& -3.857  		& 1.7 	 			&  $2.7\times 10^{-3}$ 	\\
					& $BB^*$							&		& 			& 16.7 MeV			& 26.5				\\
					& $BB(^3P_0)$						&		& 			& 0.306 				&  $4.86\times 10^{-4}$	\\
					& $BB(1P_1)$					&		& 			& 13.5 MeV			& 21.4				\\
					& $BB(1P_1^\prime)$				&		& 			& 6.82 MeV			& 10.8				\\
					& $B^*B^*$					&		& 			& 25.1 MeV			& 39.8				\\
					& $B_sB_s^*$					&		& 			& 21.5				& $3.41\times 10^{-2}$	\\
					& $B_s^*B_s^*$					&		& 			& 0.614 MeV			& 0.975				\\
					& Total							&		&			& 63.0 MeV			& 100    				\\
\hline \hline
\end{tabular}
\end{center}
\end{table}

\begin{table}[tp]
\caption{Partial widths and branching ratios
for strong and electromagnetic transitions and OZI allowed strong decays for 
the $5^3P_0$ and $5^1P_1$ states.
Details of the OZI allowed decay amplitudes are described in the appendix.  
See the caption to Table~\ref{tab:sum_12Swave} for further details.
\label{tab:sum_5Pwave2}}
\begin{center}
\begin{tabular}{l l l l l l } \hline \hline
Initial \phantom{www}       	& Final 		  	& $M_f$  	& $\cal M$  	& Width 				& BR  				\\
 state          			& state 				&(GeV)	&			& (keV)  				& (\%) 				\\
\hline \hline
$\chi_{b0} (5^3P_0)$	& $ \Upsilon (5^3S_1) \gamma $  		& 10.876 	& 3.668  		& 10.0 				& $1.86\times 10^{-2}$			 	\\
11.004				& $ \Upsilon (4^3S_1)  \gamma$  	& 10.579 	& 0.446  		&  5.2 				& $9.6\times 10^{-3}$	\\
					& $ \Upsilon (3^3S_1)  \gamma$  		& 10.355 	& -0.043 		& 0.16  				& $3.0\times 10^{-4}$   	\\
					& $ \Upsilon (2^3S_1)  \gamma$  		& 10.023 	& 0.035  		& 0.36  				& $6.7\times 10^{-4}$   	\\
					& $ \Upsilon (1^3S_1)  \gamma$  	&  9.460 	& 0.014  		& 0.22 	 			& $4.1\times 10^{-4}$    	\\
					& $ \Upsilon_1 (4^3D_1) \gamma $   		& 10.928 	& -4.038  		& 5.1 	 			& $9.5\times 10^{-3}$  	\\
					& $BB$							&		& 			& 4.52 MeV			& 8.38				\\
					& $BB(1P_1)$				&		& 			& 7.16 MeV			& 13.3				\\
					& $B^*B^*$						&		& 			& 40.3 MeV			& 74.8				\\
					& $B_sB_s$						&		& 			& 0.166 MeV			& 0.308				\\
					& $B_s^*B_s^*$					&		& 			& 1.71 MeV			& 3.17				\\
					& Total							&		&			& 53.9 MeV			& 100    				\\
\hline
$h_b (5^1P_1)$		& $ \eta_{b} (5^1S_0)  \gamma$  	& 10.867 	& 2.900  		& 9.8 	 			& $2.1\times 10^{-2}$  	\\
11.016 				& $ \eta_{b} (4^1S_0) \gamma $  	& 10.567 	& 0.824  		& 20.8 	 			& $4.54 \times 10^{-2}$  	\\
					& $ \eta_{b} (3^1S_0) \gamma $  	& 10.337 	& -0.003 		& $9.7\times 10^{-4}$ 	&  $2.1\times 10^{-6}$ 	\\
					& $ \eta_{b} (2^1S_0) \gamma $  	&  9.999 	& 0.030  		& 0.29 	 			&  $6.3\times 10^{-4}$  	\\
					& $ \eta_{b} (1^1S_0)  \gamma$  	&  9.398 	& 0.042  		& 2.2 	 			&  $4.8\times 10^{-3}$  	\\
					& $ \eta_{b2} (4^1D_2) \gamma $  	& 10.790 	& -3.967  		& 6.0 	 			&  $1.3\times 10^{-2}$	\\
					& $BB^*$							&		& 			& 11.7 MeV			& 25.5				\\
					& $BB(1^3P_0)$					&		& 			& 6.81 MeV			& 14.9				\\
					& $BB(1P_1)$					&		& 			&  0.412				& $9.00\times 10^{-4}$	\\
					& $BB(1P_1^\prime)$				&		& 			&  0.689				& $1.50\times 10^{-3}$	\\
					& $B^*B^*$						&		&			& 26.4 MeV			& 57.6				\\
					& $B_sB_s^*$					&		&			& 55.8 				& 0.122				\\
					& $B_s^*B_s^*$					&		& 			& 0.670 MeV			& 1.46				\\
					& Total						&		&			& 45.8 MeV			& 100  				\\
\hline \hline
\end{tabular}
\end{center}
\end{table}

\begin{table*}[tp]
\caption{Partial widths and branching ratios
for strong and electromagnetic decays and transitions for the $1D$ states.
For $D$-wave annihilation decays ${\cal M}$ designates
$R''(0)$, the second derivative of the radial wavefunction at the origin. 
See the caption to Table~\ref{tab:sum_12Swave} for further details.
\label{tab:sum_1Dwave}}
\begin{center}
\begin{tabular}{l l  l l l l  l } \hline \hline
Initial \phantom{www}       & Final 					 	& $M_f$  	& $\cal M$  	& \multicolumn{2}{c}{Predicted}				& Measured \\
 state          			& state 						&(GeV)	&			& Width (keV)  			& \ \ BR (\%) 					& \ \ BR (\%)  	\\ 
\hline\hline
$\Upsilon_3 (1^3D_3)$ 	& $ \chi_{b2} (1^3P_2) \gamma $  	&  9.912 	& 1.830  		& 24.3 	 			&  91.0				  				& 			\\
10.172\footnotemark[2] 				& $ \eta_{b2} (1^1D_2)  \gamma$ 		&  10.165	& 1.000  		& $1.5\times 10^{-5}$ 	& $5.6\times 10^{-5}$  			& 			\\
					& $ggg $  							& 		& 	0.9923	& 2.07	 			&  7.75				& 						\\
					& $\Upsilon(1^3S_1) \pi^+\pi^-$		& 		& 			& 0.197				& 0.738				&    $0.29\pm 0.23$ (or $< 0.62$ at 90\% C.L.)\footnotemark[1]  \\
					& Total								&		&			& 26.7				& 100  				& 							\\
\hline
$\Upsilon_2 (1^3D_2)$	& $ \chi_{b2} (1^3P_2)  \gamma$   			&  9.912 	& 1.835  		& 5.6 	 			&  22		  				& 			\\
10.164\footnotemark[1]	& $ \chi_{b1} (1^3P_1) \gamma $  &  9.893 	& 1.762  		& 19.2 	 			&  74.7				& 			\\
					& $ggg $  								& 		&  	1.149		& 0.69	 			&  2.7								&			\\
					& $\Upsilon(1^3S_1) \pi^+\pi^-$			&		&			& $0.169 \pm 0.045$ \footnotemark[3] &	0.658			& $0.66 \pm 0.16$\footnotemark[1]  	\\  
					& Total								&		&			& 25.7				& 100 	  						&			\\
\hline
$\Upsilon_1 (1^3D_1)$	& $\ell^+\ell^-$ 		& 		&			& 1.38 eV 				& $3.93\times 10^{-3}$	\\
10.155\footnotemark[2]	& $ \chi_{b2} (1^3P_2)  \gamma$  	 &  9.912 	& 1.839  		& 0.56 	 			& 1.6 			 				& 			\\
					& $ \chi_{b1} (1^3P_1) \gamma $  	 &  9.893 	& 1.768  		& 9.7 	 			&  28			 				& 			\\
					& $ \chi_{b0} (1^3P_0) \gamma $  	 &  9.859 	& 1.673  		& 16.5 	 			&  47.1				 				& 			\\
					& $ggg $   							& 	1.356	& 			& 8.11	 			&  23.1								&			\\
					& $\Upsilon(1^3S_1) \pi^+\pi^-$ 	& 		& 		& 0.140		& 0.399		& $0.42 \pm 0.29$ (or $< 0.82$ at 90\% C.L.)\footnotemark[1]   \\ 
					& Total								&		&			& 35.1				& 100			&			\\
\hline
$\eta_{b2} (1^1D_2)$	& $ h_b (1^1P_1)  \gamma$  	 	&  9.899 	& 0.178  		& 24.9 	 			& 91.5	 				& 			\\
10.165\footnotemark[2]				& $\eta_b(1^1S_0) \pi^+\pi^-$			& 		& 			& 0.35				& 1.3					& 			\\
					& $gg$ 							& 		& 	1.130		& 1.8 				& 6.6						&  			\\
					& Total								&		&			& 27.2				& 100						&			\\

\hline
\hline
\end{tabular}
\end{center}
\footnotetext[1]{From BaBar \cite{delAmoSanchez:2010kz}.}
\footnotetext[2]{Using predicted splittings and $1^3D_2$ mass from Ref.~\cite{delAmoSanchez:2010kz}.}
\footnotetext[3]{See Section IV C for the details of how this was obtained.}
\end{table*}

\begin{table}[tp]
\caption{Partial widths and branching ratios
for strong and electromagnetic decays and transitions for the $2D$ states.
For $D$-wave annihilation decays ${\cal M}$ designates
$R''(0)$, the second derivative of the radial wavefunction at the origin. 
See the caption to Table~\ref{tab:sum_12Swave} for further details.
\label{tab:sum_2Dwave}}
\begin{center}
\begin{tabular}{l l l l l l l } \hline \hline
Initial \phantom{www}       & Final 					  	& $M_f$  	& $\cal M$  	& Width 				& BR 				\\
 state          			& state 						&(GeV)	&			&  (keV)  				& (\%)  				\\ 
 \hline\hline
$\Upsilon_3 (2^3D_3)$ 	& $ \chi_{b2} (2^3P_2)  \gamma$   	&  10.269	& 2.445  		& 16.4 	 			&  65.1				\\
10.455				& $ \chi_{b2} (1^3P_2)  \gamma$   	&   9.912 	& 0.200  		&  2.6 	 			&  10.				\\
					& $ \chi_{b4} (1^3F_4)  \gamma$  	 	&  10.358 	& -1.798 		&  1.7 	 			&  6.7				\\
					& $ \chi_{b3} (1^3F_3)  \gamma$		 	&  10.355 	& -1.751 		&  0.16 	 			&  0.63				\\
					& $ \chi_{b2} (1^3F_2)  \gamma$  	 	&  10.350 	& -1.702 		&  $5\times 10^{-3}$ 		& $2\times 10^{-2}$ 		\\
					& $ \eta_{b2} (2^1D_2)  \gamma$ 		&  10.450 	& 0.999  		& $6.5\times 10^{-6}$ 	& $2.6\times 10^{-5}$   	\\
					& $ \eta_{b2} (1^1D_2)  \gamma$   	&  10.165 	& -0.033  		& $1.1\times 10^{-3}$ 	&  $4.4\times 10^{-3}$  	\\					
					& $ggg $  						 	& 		& 	1.389		& 4.3					&  17					\\
					& $\Upsilon_3(1^3D_3) \pi\pi$			& 		& 			& $7.4 \times 10^{-3}$	& $2.9\times 10^{-2}$ 	\\
					& $\Upsilon_2(1^3D_2) \pi\pi$			&		&			& $1.9 \times 10^{-6}$	& $7.5\times 10^{-6}$ 	\\
					& $\Upsilon_1(1^3D_1) \pi\pi$		&		&			& $1.9 \times 10^{-7}$	& $7.5\times 10^{-7}$ 	\\
					& Total						 		&		&			& 25.2				& 100				\\
\hline
$\Upsilon_2 (2^3D_2)$ 	& $ \chi_{b2} (2^3P_2)  \gamma$  	&  10.269	& 2.490  		& 3.8 	 			&  17					\\
10.449 				& $ \chi_{b1} (2^3P_1)  \gamma$  	&  10.255	& 2.359  		& 12.7 	 			&  56.2 				\\
					& $ \chi_{b2} (1^3P_2)  \gamma$  	 	&   9.912	& 0.161  		& 0.4 	 			&  2					\\
					& $ \chi_{b1} (1^3P_1)  \gamma$  		&   9.893 	& 0.224  		& 2.6 	 			&  12					\\
					& $ \chi_{b3} (1^3F_3)  \gamma$  	 	&  10.355 	& -1.806 		&  1.5 	 			&  6.6				\\
					& $ \chi_{b2} (1^3F_2)  \gamma$  	 	&  10.350 	& -1.758 		&  0.21 				& 0.93 				\\
					& $ \eta_{b2} (1^1D_2)  \gamma$   		&  10.165 	& -0.047  		& $2.1\times 10^{-3}$ 	& $9.3\times 10^{-3}$  	\\
					& $ggg $  							& 		& 1.568			& 1.4	 				& 6.2 				\\
					& $\Upsilon_3(1^3D_3) \pi\pi$		& 		& 			& $2.6 \times 10^{-6}$ 	& $1.2\times 10^{-5}$ 	\\
					& $\Upsilon_2(1^3D_2) \pi\pi$		& 		& 			& $7.4 \times 10^{-3}$	& $3.3\times 10^{-2}$ 	\\
					& $\Upsilon_1(1^3D_1) \pi\pi$		& 		& 			& $2.3 \times 10^{-6}$	& $1.0\times 10^{-5}$ 	\\
					& Total							&		&			& 22.6				& 100				\\

\hline
$\Upsilon_1 (2^3D_1)$ 	& $\ell^+\ell^-$ 				& 		&	& 1.99 eV 				& $5.28\times 10^{-3}$	\\
10.441				& $ \chi_{b2} (2^3P_2) \gamma $   	&  10.269	& 2.535  		& 0.4 	 			&  1					\\
	 				& $ \chi_{b1} (2^3P_1) \gamma $  	&  10.255	& 2.409  		& 6.5 	 			&  17					\\
					& $ \chi_{b0} (2^3P_0) \gamma $  		&  10.232	& 2.243  		& 10.6 	 			&  28.1				\\
					& $ \chi_{b2} (1^3P_2) \gamma $  		&   9.912 	& 0.118  		& 0.02 	 			&  0.05				\\
					& $ \chi_{b1} (1^3P_1) \gamma $  		&   9.893 	& 0.184  		& 0.9 	 			&  2					\\
					& $ \chi_{b0} (1^3P_0) \gamma $  		&   9.859 	& 0.260  		& 2.9 	 			&  7.7			 	\\
					& $ \chi_{b2} (1^3F_2) \gamma $  	 	&  10.350	& -1.815 		& 1.6 	 			&  4.2				\\
					& $ \eta_{b2} (1^1D_2)  \gamma$  		&  10.165	& -0.061  		& $3.3\times 10^{-3}$ 	&  $8.8\times 10^{-3}$	\\
					& $ggg $  						 	& 		& 	1.771	& 14.8	 			&  39.2 				\\
					& $\Upsilon_3(1^3D_3) \pi\pi$			& 		& 			& $4.4\times 10^{-7}$ 	& $1.2\times 10^{-6}$	\\
					& $\Upsilon_2(1^3D_2) \pi\pi$			& 		& 			& $3.9 \times 10^{-6}$	& $1.0\times 10^{-5}$	\\
					& $\Upsilon_1(1^3D_1) \pi\pi$		& 		& 			& $7.4 \times 10^{-3}$	& $2.0\times 10^{-2}$	\\
					& Total								&		&			& 37.7				& 100 				\\
\hline
$\eta_{b2} (2^1D_2)$	& $ h_b (2^1P_1) \gamma $  			& 10.260 	& 2.390  		& 16.5 	 			&  67.1				\\
10.450				& $ h_b (1^1P_1) \gamma $  			&  9.899 	& 0.212  		& 3.0 	 			&  12					\\
					& $ h_{b3} (1^1F_3) \gamma $  	 	& 10.355 	& -1.802 	 	& 1.8 	 			&  7.3				\\
					& $ \Upsilon_3 (1^3D_3) \gamma $  	 	&  10.172	& -0.072  		& $6.5\times 10^{-3}$ 	&  $2.6\times 10^{-2}$ 	\\
					& $ \Upsilon_2 (1^3D_2) \gamma $  	 	&  10.164	& -0.048  		& $2.2\times 10^{-3}$	&  $8.9\times 10^{-3}$	\\
					& $ \Upsilon_1 (1^3D_1) \gamma $  	 	&  10.154	& -0.046  		& $1.4\times 10^{-3}$ 	& $5.7\times 10^{-3}$  	\\
					& $gg$ 						 	& 		& 	1.530	& 3.3 				& 13 				\\
					& $\eta_{b2}(1^1D_2) \pi\pi$	 		& 		& 			& $7.4 \times 10^{-3}$	&$3.0\times 10^{-2}$	\\
					& Total								&		&			& 24.6				& 100 				\\
\hline
\hline
\end{tabular}
\end{center}
\end{table}

\begin{table}[tp]
\caption{Partial widths and branching ratios
for strong and electromagnetic decays and  transitions and OZI allowed decays for the $3D$ states.
For $D$-wave annihilation decays ${\cal M}$ designates
$R''(0)$, the second derivative of the radial wavefunction at the origin. 
Details of the OZI allowed decays amplitudes are described in the appendix.  
See the caption to Table~\ref{tab:sum_12Swave} for further details.
\label{tab:sum_3Dwave}}
\vspace{-5mm}
\begin{center}
\begin{tabular}{l l l l l l l } \hline \hline
Initial \phantom{www}       & Final 					  	& $M_f$  	& $\cal M$  	& Width 				& BR 				\\
 state          			& state 						&(GeV)	&			& (keV)  				&  (\%)  				\\ 
 \hline\hline
$\Upsilon_3 (3^3D_3)$ 	& $ \chi_{b2} (3^3P_2)  \gamma$   	&  10.528 	& 3.022  		& 23.6 				& $1.19\times 10^{-2}$	\\
10.711 				& $ \chi_{b2} (2^3P_2)  \gamma$  	&  10.269 	& 0.265  		& 2.5 				& $1.3\times 10^{-3}$  	\\
					& $ \chi_{b2} (1^3P_2)  \gamma$  		&   9.912 	& 0.064  		&  0.82 				& $4.1\times 10^{-4}$  	\\
					& $ \chi_{b4} (2^3F_4)  \gamma$  		&  10.622 	& -2.683 		&  3.0 	 			& $1.5\times 10^{-3}$   	\\
					& $ \chi_{b3} (2^3F_3)  \gamma$  		&  10.619 	& -2.617 		&  0.27 				& $1.4\times 10^{-4}$  	\\
					& $ \chi_{b2} (2^3F_2) \gamma $  		&  10.615 	& -2.545 		&  $8\times 10^{-3}$		& $4\times 10^{-6}$  		\\
					& $ggg $  						  		& 		& 	1.691		& 6.6	 				& $3.3\times 10^{-3}$ 	\\
					& $BB$								&		& 			& 16.3 MeV			& 8.23				\\
					& $BB^*$							&		& 			& 72.9 MeV			& 36.8				\\
					& $B^*B^*$					 		&		& 			& 109 MeV			& 55.0				\\
					& Total						 		&		&  			& 198 MeV			& 100				\\
\hline
$\Upsilon_2 (3^3D_2)$	& $ \chi_{b2} (3^3P_2) \gamma $  	& 10.528 	& 3.098  		& 5.6 	 			& $4.3\times 10^{-3}$ 	\\
10.705				& $ \chi_{b1} (3^3P_1) \gamma $   	& 10.516 	& 2.919  		& 18.2 	 			& $1.41\times 10^{-2}$ 	\\
					& $ \chi_{b2} (2^3P_2)  \gamma$  		& 10.269 	& 0.218  		& 0.40 		 		& $3.1\times 10^{-4}$ 	\\
					& $ \chi_{b1} (2^3P_1)  \gamma$   		& 10.255 	& 0.303  		& 2.5 	 			& $1.9\times 10^{-3}$ 	\\
					& $ \chi_{b2} (1^3P_2)  \gamma$   		&   9.912 	& 0.043  		& 0.09 	 			& $7\times 10^{-5}$  		\\
					& $ \chi_{b1} (1^3P_1)  \gamma$   		&   9.893 	& 0.062  		& 0.59 	 			& $4.6\times 10^{-4}$  	\\
					& $ \chi_{b3} (2^3F_3)  \gamma$    		& 10.619 	& -2.698 		& 2.6 				& $2.0\times 10^{-3}$ 	\\
					& $ \chi_{b2} (2^3F_2)  \gamma$    		& 10.615 	& -2.628 		& 0.36 				& $2.8\times 10^{-4}$ 	\\
					& $ggg $  						  		& 		&  1.875			& 2.0	 				& $1.6\times 10^{-3}$  	\\
					& $BB^*$						 		&		& 			& 52.4 MeV			& 40.6				\\
					& $B^*B^*$					 		&		& 			& 76.5 MeV			& 59.3				\\
					& Total						 		&		&  			& 128.9 MeV			& 100				\\
\hline
$\Upsilon_1 (3^3D_1)$	& $\ell^+\ell^-$ 				 	& 		&			& 2.38 eV 				& $2.30\times 10^{-6}$	\\
10.698				& $ \chi_{b2} (3^3P_2)  \gamma$ 		& 10.528	& 3.174  		& 0.58 			 	& $5.6\times 10^{-4}$   	\\
					& $ \chi_{b1} (3^3P_1)  \gamma$   		& 10.516	& 3.003  		& 9.5 	 			&  $9.2\times 10^{-3}$  	\\
					& $ \chi_{b0} (3^3P_0)  \gamma$  		& 10.500	& 2.775  		& 14.0 	 			&  $1.35\times 10^{-2}$  	\\
					& $ \chi_{b2} (2^3P_2)  \gamma$    		& 10.269	& 0.165  		& 0.02 	 			&  $2\times 10^{-5}$  	\\
					& $ \chi_{b1} (2^3P_1)  \gamma$  		& 10.255 	& 0.256  		& 0.96 	 			&  $9.3\times 10^{-4}$ 	\\
					& $ \chi_{b0} (2^3P_0)  \gamma$  		& 10.233	& 0.354  		& 2.8 	 			&  $2.7\times 10^{-3}$ 	\\
					& $ \chi_{b0} (1^3P_1)  \gamma$  		&  9.893 	& 0.040  		& 0.13 	 			&  $1.3\times 10^{-4}$ 	\\
					& $ \chi_{b0} (1^3P_0)  \gamma$  		&  9.859 	& 0.069  		& 0.59 	 			&  $5.7\times 10^{-4}$ 	\\
					& $ \chi_{b2} (2^3F_2)  \gamma$   		& 10.615 	& -2.712 		& 2.7 	 			&  $2.6\times 10^{-3}$  	\\
					& $ \chi_{b2} (1^3F_2)  \gamma$  		& 10.350 	&  0.039 		& 0.39 				&  $3.8\times 10^{-5}$					\\
					& $ggg $  						 	 	& 			& 	2.081		& 21.2	 			& $2.05\times 10^{-2}$ 	\\
					& $BB$						 		&		& 			& 23.8 MeV			& 23.0				\\
					& $BB^*$					 		&		&			& 0.245 MeV			& 0.236				\\
					& $B^*B^*$					 		&		& 			& 79.5 MeV			& 76.7				\\
					& Total						 		&		&  			& 103.6 MeV			& 100				\\
\hline
$\eta_{b2} (3^1D_2)$	& $ h_{b1} (3^1P_1)  \gamma$  	& 10.519	& 2.956  		& 24.1 				& $1.43\times 10^{-2}$ 	\\
 10.706				& $ h_{b1} (2^1P_1)  \gamma$  	& 10.260	& 0.285  		& 2.9	 			& $1.7\times 10^{-3}$ 	\\
					& $ h_{b1} (1^1P_1)  \gamma$    		&   9.899 	& 0.061  		& 0.76	 			& $4.5\times 10^{-4}$ 	\\
					& $ h_{b3} (2^1F_3)  \gamma$  		& 10.619 	& -2.691 		& 3.1 				& $1.8\times 10^{-3}$ 	\\
					& $ h_{b2} (1^1F_2)  \gamma$  		& 10.355 	&  0.030 		& 0.02 				& $1\times 10^{-5}$ 		\\
					& $gg$  							& 		& 1.839			& 4.7 				& $2.8\times 10^{-3}$   	\\
					& $BB^*$					 		&		& 			& 77.8 MeV			& 46.1				\\
					& $B^*B^*$					 		&		& 			& 90.9 MeV			& 53.9				\\
					& Total								&		&  			& 168.7 MeV			& 100				\\
\hline
\hline
\end{tabular}
\end{center}
\end{table}

\begin{table}[tp]
\caption{Predicted partial widths and branching ratios
for strong and electromagnetic transitions and OZI allowed decays for the $4^3D_3$ 
and $4^3D_2$ states.
Details of the OZI allowed decay amplitudes are described in the appendix.  
See the caption to Table~\ref{tab:sum_12Swave} for further details.
\label{tab:sum_4Dwave}}
\begin{center}
\begin{tabular}{l l l l l l l l l } \hline \hline
Initial \phantom{www}       & Final 					 	& $M_f$  	& $\cal M$  	& Width 				& BR  				\\
 state          			& state 						&(GeV)	&			& (keV)  				& (\%)  				\\ 
 \hline\hline
$\Upsilon_3 (4^3D_3)$ 	& $ \chi_{b2} (4^3P_2)  \gamma$   	&  10.798	& 3.553  		& 15.0 	 			& $2.59\times 10^{-2}$  	\\
 10.939 				& $ \chi_{b2} (3^3P_2)  \gamma$  	&  10.528	& 0.322 		& 2.9 	 			& $5.0\times 10^{-3}$	\\
					& $ \chi_{b2} (2^3P_2)  \gamma$  	 		&  10.269	& 0.073  	& 0.63 	 			& $1.1\times 10^{-3}$  	\\
					& $ \chi_{b2} (1^3P_2)  \gamma$  	 		&   9.912 	& 0.035  		&  0.50 	 			& $8.6\times 10^{-4}$ 	\\
					& $ \chi_{b4} (3^3F_4)  \gamma$  	  		&  10.856	& -3.143 		&  3.9 			 	& $6.7\times 10^{-3}$ 	\\
					& $ \chi_{b3} (3^3F_3)  \gamma$  	  		&  10.853	& -3.330 		&  0.36 	 			& $6.2\times 10^{-4}$ 	\\
					& $ \chi_{b2} (3^3F_2)  \gamma$  	  		&  10.850 	& -3.241 		&  0.01 				& $2\times 10^{-5}$  		\\
					& $BB$						 		&		& 			& 0.726 MeV			& 1.25				\\
					& $BB^*$					 		&		& 			& 2.94 MeV			& 5.07				\\
					& $B^*B^*$					 		&		&			& 51.5 MeV			& 88.8				\\
					& $B_sB_s$					 		&		&			& 0.265 MeV			& 0.457				\\
					& $B_sB_s^*$				 		&		& 			& 0.0827 MeV			& 0.142				\\
					& $B_s^*B_s^*$				 		&		& 			& 2.44 MeV			& 4.21				\\
					& Total						 		&		&  			& 58.0 MeV			& 100				\\
\hline
$\Upsilon_2 (4^3D_2)$	& $ \chi_{b2} (4^3P_2) \gamma $   	&  10.798 	& 3.655  		& 3.6 	 			& $5.6\times 10^{-3}$  	\\
10.934				& $ \chi_{b1} (4^3P_1)  \gamma$  	&  10.788 	& 3.428  		& 11.6 	 			& $1.80\times 10^{-2}$  	\\
					& $ \chi_{b2} (3^3P_2)  \gamma$    		&  10.528 	& 0.270  		& 0.50 	 			& $7.8\times 10^{-4}$ 	\\
					& $ \chi_{b1} (3^3P_1)  \gamma$   		&  10.516 	& 0.384  		& 3.3 	 			& $5.1\times 10^{-3}$  	\\
					& $ \chi_{b2} (2^3P_2)  \gamma$    		&   10.269	& 0.048  		& 0.07 	 			& $1\times 10^{-4}$ 	\\
					& $ \chi_{b1} (2^3P_1)  \gamma$   		&   10.255	& 0.061  		& 0.35 	 			& $5.4\times 10^{-4}$ 	\\
					& $ \chi_{b2} (1^3P_2)  \gamma$    		&   9.912 	& 0.022  		& 0.05 	 			& $8\times 10^{-5}$  	\\
					& $ \chi_{b1} (1^3P_1) \gamma $    		&   9.893 	& 0.046  		& 0.68 	 			& $1.1\times 10^{-3}$  	\\
					& $ \chi_{b3} (3^3F_3)  \gamma$    		&  10.853 	& -3.431 		& 3.6 				& $5.6\times 10^{-3}$  	\\
					& $ \chi_{b2} (3^3F_2)  \gamma$   		&  10.850 	& -3.344 		& 0.47 				& $7.3\times 10^{-4}$ 	\\
					& $BB^*$						 		&		& 			& 25.7 MeV			& 40.0				\\
					& $B^*B^*$					 		&		& 			& 36.4 MeV			& 56.6				\\
					& $B_sB_s^*$				 		&		& 			& 0.357 MeV			& 0.56				\\
					& $B_s^*B_s^*$						&		& 			& 1.80 MeV			& 2.80				\\
					& Total						 		&		&  			& 64.3 MeV			& 100				\\
\hline
\hline
\end{tabular}
\end{center}
\end{table}

\begin{table}[tp]
\caption{Predicted partial widths and branching ratios
for strong and electromagnetic transitions for $4^3D_1$ and $4^1D_2$ states.
Details of the OZI allowed decay amplitudes are described in the appendix.  
See the caption to Table~\ref{tab:sum_12Swave} for further details.
\label{tab:sum_4Dwave2}}
\begin{center}
\begin{tabular}{l l l l l l l l l } \hline \hline
Initial \phantom{www}       & Final 					 	& $M_f$  	& $\cal M$  	& Width 				& BR  				\\
 state          			& state 						&(GeV)	&			& (keV)  				& (\%)  				\\ 
 \hline\hline
$\Upsilon_1 (4^3D_1)$	& $\ell^+\ell^-$ 				 	& 		&			& 2.18 eV 				& $3.04\times 10^{-6}$	\\
10.928				& $ \chi_{b2} (4^3P_2) \gamma $  	&  10.798 	& 3.756  		& 0.36 	 			& $5.0\times 10^{-4}$ 	\\
					& $ \chi_{b1} (4^3P_1) \gamma $    		&  10.788 	& 3.538  		& 6.1 	 			&  $8.5\times 10^{-3}$ 	\\
					& $ \chi_{b0} (4^3P_0) \gamma $    		&  10.775 	& 3.262  		& 9.0 	 			&  $1.2\times 10^{-2}$	\\
					& $ \chi_{b2} (3^3P_2) \gamma $    		&  10.528 	& 0.210 		& 0.03 	 			&  $4\times 10^{-5}$		\\
					& $ \chi_{b1} (3^3P_1) \gamma $   		&  10.516 	& 0.331  		& 1.3 	 			&  $1.8\times 10^{-3}$ 	\\
					& $ \chi_{b0} (3^3P_0) \gamma $    		&  10.500 	& 0.476  		& 4.0 	 			&  $5.6\times 10^{-3}$ 	\\
					& $ \chi_{b2} (2^3P_2) \gamma $    		&   10.269	& 0.022  		& $1.5\times 10^{-3}$ 	&  $2.1\times 10^{-6}$ 	\\
					& $ \chi_{b1} (2^3P_1) \gamma $    		&   10.255	& 0.037  		& 0.07 				&  $1\times 10^{-4}$		\\
					& $ \chi_{b0} (2^3P_0) \gamma $   		&   10.233	& 0.060  		& 0.26 				&  $3.6\times 10^{-4}$ 	\\
					& $ \chi_{b1} (1^3P_1) \gamma $    		&   9.893 	& 0.033  		& 0.19 				&  $2.6\times 10^{-4}$ 	\\
					& $ \chi_{b0} (1^3P_0) \gamma $    		&   9.859 	& 0.054  		& 0.75 				&  $1.0\times 10^{-3}$	\\
					& $ \chi_{b2} (3^3F_2)  \gamma$    		&  10.850 	& -3.448 		& 3.6 				& $5.0\times 10^{-3}$  	\\
					& $BB$					 		&		& 			& 3.85 MeV			& 5.36				\\
					& $BB^*$				 		&		& 			& 14.0 MeV			& 19.5				\\
					& $B^*B^*$						&		& 			& 50.6 MeV			& 70.5				\\
					& $B_sB_s$						&		& 			& 0.101 MeV			& 0.141				\\
					& $B_sB_s^*$					&		& 			& 0.332 MeV			& 0.462				\\
					& $B_s^*B_s^*$					&		& 			& 2.94 MeV			& 4.09				\\
					& Total					 		&		&  			& 71.8 MeV			& 100				\\
\hline
$\eta_{b2} (4^1D_2)$	& $ h_{b} (4^1P_1)  \gamma$   	&  10.790 	& 3.477  & 15.6 			& $2.58\times 10^{-2}$  	\\
10.935				& $ h_{b} (3^1P_1)  \gamma$ 		&  10.519 	& 0.362  & 3.8 				& $6.3\times 10^{-3}$  	\\
					& $ h_{b} (2^1P_1)  \gamma$  		&   10.260	& 0.065  		& 0.51	 	& $8.4\times 10^{-4}$  	\\
					& $ h_{b} (1^1P_1) \gamma $  		&    9.899	& 0.042  		& 0.75	 	& $1.2\times 10^{-3}$  	\\
					& $ h_{b3} (3^1F_3)  \gamma$   		&  10.853	& -3.423 		& 4.1 		& $6.8\times 10^{-3}$  	\\
					& $BB^*$							&		&			& 19.4 MeV			& 32.1				\\
					& $B^*B^*$							&		&			& 38.9 MeV			& 64.3				\\
					& $B_sB_s^*$				 		&		& 			& 0.239 MeV			& 0.395				\\
					& $B_s^*B_s^*$				 		&		& 			& 1.92 MeV			& 3.17				\\
					& Total								&		&  			& 60.5 MeV			& 100 				\\
\hline
\hline
\end{tabular}
\end{center}
\end{table}

\begin{table}[tp]
\caption{Predicted partial widths and branching ratios
for strong and electromagnetic decays and transitions for $1F$-wave states.
For $F$-wave annihilation decays ${\cal M}$ designates
$R'''(0)$, the third derivative of the radial wavefunction at the origin. 
See the caption to Table~\ref{tab:sum_12Swave} for further details.
\label{tab:sum_1Fwave}}
\begin{center}
\begin{tabular}{l l l l l l l  } \hline \hline
Initial \phantom{www}       & Final 				& $M_f$	& $\cal M$  			& Width 				& BR  		\\
 state          			& state 				&(GeV)	&					&  (keV)  				& (\%) 		\\ 
 \hline\hline
$\chi_{b4} (1^3F_4)$	& $ \Upsilon_{3} (1^3D_3) \gamma $	&  10.172 & 2.479  		& 18.0	 			&  $\sim 100$	\\
10.358				& $\chi_{b2}(1^3P_2) \pi \pi $   	& 			&  			& $3.9\times 10^{-3}$	 & $2.2\times 10^{-2}$		\\  
					& $gg$  						 	& 		& 0.868 				& 0.048 				& 0.27  		\\
					& Total								&		& 					& 18.0 				&  100  		\\
\hline
$\chi_{b3} (1^3F_3)$	& $ \Upsilon_{3} (1^3D_3)  \gamma$ 	&  10.172	& 2.482  				& 1.9 	 			& 10.			\\	
10.355				& $ \Upsilon_{2} (1^3D_2)  \gamma$	&  10.164	& 2.442  				& 16.7 	 			& 89.3  		\\
					& $gg$  						 		& 		& 0.974		& 0.060 				&  0.32 		\\
					& $\chi_{b2}(1^3P_2) \pi \pi $    		& 		&			& $1.3\times 10^{-3}$  	&  $7.0\times 10^{-3}$		\\	
					& $\chi_{b1}(1^3P_1) \pi \pi $    		& 		&			& $2.6\times 10^{-3}$ 	&  $1.4\times 10^{-2}$								\\
					& Total						 			&		&  			& 18.7				& 100		\\
\hline
$\chi_{b2} (1^3F_2)$	& $ \Upsilon_{3} (1^3D_3)  \gamma$ 	&  10.172 	& 2.485  	& 0.070	 			&  0.35 		\\	
10.350				& $ \Upsilon_{2} (1^3D_2)  \gamma$	&  10.164 	& 2.446  	& 2.7 	 			&  14			\\
					& $ \Upsilon_{1} (1^3D_1)  \gamma$		&  10.154 	& 2.402  	& 16.4 	 			&  82.4 		\\
					& $gg$  						 		& 			& 1.091		& 0.70 				&  3.5 		\\
					& $\chi_{b2}(1^3P_2) \pi \pi $    		& 			&			& $2.6\times 10^{-4}$ 	& 	$1.3\times 10^{-3}$			\\	
					& $\chi_{b1}(1^3P_1) \pi \pi $   		& 			&			& $1.8\times 10^{-3}$ 	&   $9.0\times 10^{-3}$			\\	
					& $\chi_{b0}(1^3P_0) \pi \pi $    		& 			&			& $1.8\times 10^{-3}$ 	&   $9.0\times 10^{-3}$			\\	
					& Total						 		&		&  					& 19.9				& 100		\\
\hline
$h_{b3} (1^1F_3)$		& $ \eta_{b2} (1^1D_2) \gamma $   	& 10.165 	& 2.449  	& 18.8 				&  $\sim$100   		\\
10.355				& $h_b(1^1P_1) \pi \pi $   			&			& 			& $3.9\times 10^{-3}$	&  $2.1\times 10^{-2}$ 								\\	
					& Total								&		&  					& 18.8				& 100		\\
\hline \hline
\end{tabular}
\end{center}
\end{table}

\begin{table}[tp]
\caption{Predicted partial widths and branching ratios
for strong decays and electromagnetic transitions for the $2F$-wave states.
For $F$-wave annihilation decays ${\cal M}$ designates
$R'''(0)$, the third derivative of the radial wavefunction at the origin. 
Details of the OZI allowed decay amplitudes are described in the appendix.  
See the caption to Table~\ref{tab:sum_12Swave} for further details.
\label{tab:sum_2Fwave}}
\begin{center}
\begin{tabular}{l  l l l l l } \hline \hline
Initial \phantom{www}       & Final 						& $M_f$ 	& $\cal M$  	& Width 				& BR   				\\
 state          			& state 						&(GeV)	&			&  (keV)  				& (\%)  				\\ 
 \hline \hline
$\chi_{b4} (2^3F_4)$	& $ \Upsilon_{3} (2^3D_3)  \gamma$	&  10.455 & 3.053  		& 19.6 	 			& 0.700 				\\
10.622				& $ \Upsilon_{3} (1^3D_3)  \gamma$ 	&  10.172 & 0.191  		&  1.4 	 			& $5.0\times 10^{-2}$  	\\
					& $ \Upsilon_{5} (1^3G_5)  \gamma$	 	&  10.532 & -1.886  		&  1.5 	 			& $5.4\times 10^{-2}$  	\\
					& $ \Upsilon_{4} (1^3G_4) \gamma $		&  10.531 & -1.848  		&  0.08 	 			& $3\times 10^{-3}$  		\\
					& $ \Upsilon_{3} (1^3G_3) \gamma $	 	&  10.529 & -1.808  		&  $1\times 10^{-3}$ 		& $4\times 10^{-5}$		\\
					& $gg$  							& 		& 1.455 		& 0.13 				& $4.6\times 10^{-3}$  	\\
					& $BB$								&		& 			& 2.73 MeV			& 97.5				\\
					& $BB^*$							&		& 			& 0.0462 MeV			& 1.65				\\
					& Total						 		&		&  			& 2.80 MeV			& 100				\\
\hline
$\chi_{b3} (2^3F_3)$	& $ \Upsilon_{3} (2^3D_3)  \gamma$ 	&  10.455 & 3.084  		& 2.1 	 			& $1.4\times 10^{-2}$	\\
10.619				& $ \Upsilon_{2} (2^3D_2)  \gamma$ 	&  10.449 & 3.009  		& 17.9 	 			& 0.116			\\
					& $ \Upsilon_{3} (1^3D_3)  \gamma$	 	&  10.172 & 0.156  		& 0.1 	 			& $6\times 10^{-4}$  		\\
					& $ \Upsilon_{2} (1^3D_2)  \gamma$	 	&  10.164 & 0.199  		& 1.4 	 			& $9.1\times 10^{-3}$  	\\
					& $ \Upsilon_{4} (1^3G_4)  \gamma$	 	&  10.531 & -1.892  		&  1.4 	 			& $9.1\times 10^{-3}$  	\\
					& $ \Upsilon_{3} (1^3G_3)  \gamma$	 	&  10.529 & -1.852  		&  0.10 				& $6.5\times 10^{-4}$  	\\
					& $gg$  						 	& 		& 1.583 		& 0.16 				& $1.0\times 10^{-3}$ 	\\
					& $BB^*$					 		&		&			& 15.4 MeV			& $\sim 100$			\\
					& Total						 		&		&  			& 15.4 MeV			& 100				\\
\hline
$\chi_{b2} (2^3F_2)$	& $ \Upsilon_{3} (2^3D_3) \gamma $	 	&  10.455 & 3.114  		& 0.08 	 			& $9\times 10^{-5}$		\\
10.615				& $ \Upsilon_{2} (2^3D_2) \gamma $	 	&  10.449 & 3.042  		& 3.0 	 			& $3.4\times 10^{-3}$	\\
					& $ \Upsilon_{1} (2^3D_1) \gamma $	 	&  10.441 & 2.961  		& 17.5 	 			& $1.98\times 10^{-2}$ 	\\
					& $ \Upsilon_{3} (1^3D_3) \gamma $		&  10.172 & 0.118  		& $2\times10^{-3}$ 		& $2\times 10^{-6}$ 		\\
					& $ \Upsilon_{2} (1^3D_2) \gamma $	 	&  10.164 & 0.163  		& 0.16 	 			& $1.8\times 10^{-4}$	\\
					& $ \Upsilon_{1} (1^3D_1) \gamma $	 	&  10.154 & 0.210  		& 1.6 	 			& $1.8\times 10^{-3}$ 	\\
					& $ \Upsilon_{3} (1^3G_3) \gamma $	 	&  10.529 & -1.898  		&  1.4 				& $1.6\times 10^{-3}$ 	\\
					& $gg$  						 	& 		& 1.752 		& 1.77 				& $2.00\times 10^{-3}$ 	\\
					& $BB$						 		&		& 			& 83.4 MeV			& 94.1				\\
					& $BB^*$							&		& 			& 5.20 MeV			& 5.987			\\
					& Total						 		&		&  			& 88.6 MeV			& 100				\\
\hline
$h_{b3} (2^1F_3)$		& $ \eta_{b2} (2^1D_2)  \gamma$   	& 10.450 	& 3.019  		& 19.9 				& 0.169			   	\\
10.619				& $ \eta_{b2} (1^1D_2)  \gamma$   	& 10.148 	& 0.196  		& 1.6 				& $1.4\times 10^{-2}$   	\\
					& $ \eta_{b4} (1^1G_4)  \gamma$  		& 10.530 	& -1.890  		& 1.5 				& $1.3\times 10^{-2}$   	\\
					& $BB^*$							&		& 			& 11.8 MeV			& $\sim 100$			\\
					& Total						 		&		&  			& 11.8 MeV			& 100				\\
\hline \hline
\end{tabular}
\end{center}
\end{table}

\begin{table}[tp]
\caption{Partial widths and branching ratios
for electromagnetic transitions and OZI allowed decays for the $3F$-wave states.
Details of the OZI allowed decay amplitudes are described in the appendix.  
See the caption to Table~\ref{tab:sum_12Swave} for further details.
\label{tab:sum_3Fwave}}
\begin{center}
\begin{tabular}{l l l l l l l  } \hline \hline
Initial \phantom{www}       & Final 						& $M_f$  	& $\cal M$  	& Width  				& BR 				\\
 state          			& state 						&(GeV)	&			& (keV)  				& (\%) 				\\ 
 \hline\hline
$\chi_{b4} (3^3F_4)$	& $ \Upsilon_{3} (2^3D_3) \gamma $	&  10.711 & 3.593  		& 17.9 			 	& $1.87\times 10^{-2}$ 	\\
10.856				& $ \Upsilon_{3} (2^3D_3)  \gamma$	&  10.455 & 0.256  		& 1.9 	 			& $2.0\times 10^{-3}$  	\\
					& $ \Upsilon_{3} (1^3D_3)  \gamma$		&  10.172 & 0.050  		&  0.34 	 			& $3.6\times 10^{-4}$ 	\\
					& $ \Upsilon_{5} (2^3G_5)  \gamma$ 		&  10.772 & -2.776  		&  2.6 	 			& $2.7\times 10^{-3}$  	\\
					& $ \Upsilon_{4} (2^3G_4)  \gamma$ 		&  10.771 & -2.722  		&  0.14 	 			& $1.5\times 10^{-4}$ 	\\
					& $ \Upsilon_{3} (2^3G_3)  \gamma$  		&  10.769 & -2.664  		&  $2.2\times 10^{-3}$ 		& $2.3\times 10^{-6}$		\\
					& $BB$							&		&			& 2.84 MeV			& 2.97				\\
					& $BB^*$				 		&		& 			& 0.681 MeV			& 0.713				\\
					& $B^*B^*$						&		& 			& 85.7 MeV			& 89.7				\\
					& $B_sB_s$						&		& 			& 0.733 MeV			& 0.768			\\
					& $B_sB_s^*$					&		& 			& 1.14 MeV			& 1.19				\\
					& $B_s^*B_s^*$					&		& 			& 4.43 MeV			& 4.64				\\
					& Total							&		&  			& 95.5 MeV			& 100				\\
\hline
$\chi_{b3} (3^3F_3)$	& $ \Upsilon_{3} (2^3D_3) \gamma $ 	&  10.711 & 3.646  		& 1.9 	 			& $1.9\times 10^{-3}$  	\\
10.853 				& $ \Upsilon_{2} (2^3D_2) \gamma $	&  10.705 & 3.542  		& 16.4 	 			& $1.62\times 10^{-2}$ 	\\
					& $ \Upsilon_{3} (2^3D_3) \gamma $		&  10.455 & 0.214  		& 0.14 	 			& $1.4\times 10^{-4}$	\\
					& $ \Upsilon_{2} (2^3D_2) \gamma $		&  10.449 & 0.271  		& 1.9 	 			& $1.9\times 10^{-3}$ 	\\
					& $ \Upsilon_{4} (2^3G_4) \gamma $		&  10.771 & -2.785  		&  2.4 	 			& $2.4\times 10^{-3}$  	\\
					& $ \Upsilon_{3} (2^3G_3) \gamma $		&  10.709 & -2.728  		&  0.17 				& $1.7\times 10^{-4}$ 	\\
					& $BB^*$							&		& 			& 43.8 MeV			& 43.2				\\
					& $B^*B^*$							&		& 			& 52.4 MeV			& 51.7				\\
					& $B_sB_s^*$						&		&			& 3.83 MeV			& 3.78				\\
					& $B_s^*B_s^*$				 		&		& 			& 1.30 MeV			& 1.28				\\
					& Total						 		&		&  			& 101.4 MeV			& 100				\\
\hline
$\chi_{b2} (3^3F_2)$	& $ \Upsilon_{3} (3^3D_3) \gamma $ 	&  10.711 & 3.699  		& 0.07 	 			& $6\times 10^{-5}$		\\
10.850				& $ \Upsilon_{2} (3^3D_2) \gamma $  	&  10.705 & 3.598  		& 2.8 	 			& $2.6\times 10^{-3}$ 	\\
					& $ \Upsilon_{1} (3^3D_1) \gamma $		&  10.698 & 3..487  		& 16.3 	 			& $1.50\times 10^{-2}$ 	\\
					& $ \Upsilon_{1} (2^3D_1) \gamma $		&  10.441 & 0.290  		& 2.1 	 			& $1.9\times 10^{-3}$  	\\
					& $ \Upsilon_{3} (2^3G_3) \gamma $		&  10.769 & -2.794  		&  2.5 				& $2.3\times 10^{-3}$  	\\
					& $BB$								&		& 			& 7.85 MeV			& 7.21				\\
					& $BB^*$					 		&		& 			& 32.0 MeV			& 29.4				\\
					& $B^*B^*$					 		&		& 			& 66.0 MeV			& 60.6				\\
					& $B_sB_s$					 		&		& 			& 0.709				& $6.51\times 10^{-4}$	\\
					& $B_sB_s^*$						&		& 			& 2.50 MeV			& 2.30				\\
					& $B_s^*B_s^*$				 		&		& 			& 0.557 MeV			& 0.511				\\
					& Total								&		&  			& 108.9 MeV			& 100				\\
\hline
$h_{b3} (3^1F_3)$		& $ \eta_{b2} (3^1D_2)  \gamma$   	& 10.706 	& 3.555  		& 18.2 				& $1.88\times 10^{-2}$   	\\
10.853 				& $ \eta_{b2} (2^1D_2)  \gamma$  	& 10.450 	& 0.264  		& 2.0 				& $2.1\times 10^{-3}$    	\\
					& $ \eta_{b2} (1^1D_2)  \gamma$  		& 10.165 	& 0.053  		& 0.4 				& $4\times 10^{-4}$    	\\
					& $ \eta_{b4} (2^1G_4)  \gamma$  		& 10.770 	& -2.781  		& 2.7 				& $2.8\times 10^{-3}$   	\\
					& $BB^*$						 		&		& 			& 33.2 MeV			& 34.3				\\
					& $B^*B^*$							&		& 			& 58.2 MeV			& 60.2				\\
					& $B_sB_s^*$				 		&		& 			& 3.32 MeV			& 3.43				\\
					& $B_s^*B_s^*$						&		& 			& 1.93 MeV			& 2.00				\\
					& Total						 		&		&  			& 96.7 MeV			& 100				\\
\hline \hline
\end{tabular}
\end{center}
\end{table}

\begin{table}[tp]
\caption{Predicted partial widths and branching ratios
for strong and electromagnetic transitions for the $1G$-wave states.
For $G$-wave annihilation decays ${\cal M}$ designates
$R^{(iv)}(0)$, the fourth derivative of the radial wavefunction at the origin. 
See the caption to Table~\ref{tab:sum_12Swave} for further details.
\label{tab:sum_1Gwave}}
\begin{center}
\begin{tabular}{l l l l l l  } \hline \hline
Initial \phantom{www}   	& Final 					 	& $M_f$  	& $\cal M$  	& Width 				& BR  				\\
 state          			& state 						&(GeV)	&			& (keV)  				& (\%)   				\\ 
 \hline\hline
$\Upsilon_5 (1^3G_5)$	& $ \chi_{b4} (1^3F_4) \gamma $   	&  10.358 & 3.057  		& 23.1	 			& $\sim 100$ 			\\
10.532				& $\Upsilon_3(1^3D_3) \pi\pi$		& 		& 			& $2.2\times 10^{-4}$	& $9.5\times 10^{-4}$	\\
					& Total									&		&			& 23.1				& 100				\\
\hline
$\Upsilon_4 (1^3G_4)$	& $ \chi_{b4} (1^3F_4) \gamma $  	&  10.358 & 3.059  		& 1.4 				& 6.0 				\\
10.531				& $ \chi_{b3} (1^3F_3) \gamma $  	&  10.355 & 3.032  		& 22.0 				& 94.0 				\\
					& $\Upsilon_3(1^3D_3) \pi\pi$	 		& 		& 			& $4.0\times 10^{-5}$	& $1.7\times 10^{-4}$	\\
					& $\Upsilon_2(1^3D_2) \pi\pi$	 		& 		& 			& $1.8\times 10^{-4}$	& $7.7\times 10^{-4}$	\\
					& Total						 		&		&			& 23.4				& 100				\\
\hline
$\Upsilon_3 (1^3G_3)$	& $ \chi_{b4} (1^3F_4) \gamma $   	&  10.358 & 3.060  		& 0.028 				& 0.12  				\\
10.529				& $ \chi_{b3} (1^3F_3) \gamma $   	&  10.355 & 3.034  		& 1.8 				& 7.5 				\\
					& $ \chi_{b2} (1^3F_2) \gamma $  		&  10.350 & 3.005  		& 22.3 				& 92.4   				\\
					& $\Upsilon_3(1^3D_3) \pi\pi$	 		& 		& 			& $3.1\times 10^{-6}$	& $1.3\times 10^{-5}$	\\
					& $\Upsilon_2(1^3D_2) \pi\pi$	 		& 		& 			& $4.6\times 10^{-5}$	& $1.9\times 10^{-4}$	\\
					& $\Upsilon_1(1^3D_1) \pi\pi$	 		& 		& 			& $1.7\times 10^{-4}$	& $7.0\times 10^{-4}$	\\
					& Total						 		&		&			& 24.1				& 100				\\
\hline
$\eta_{b 4}(1^1G_4)$ 	& $ h_{b3} (1^1F_3)  \gamma$   	& 10.355 	& 3.034  		& 23.1 				& $\sim 100$     		\\
10.530				& $gg$  					 	& 		& 1.005 		& $2.3\times 10^{-3}$ 	& $1.0\times 10^{-2}$   	\\
					& $\eta_{b2}(1^2D_2) \pi\pi$ 		& 		&			& $2.2\times 10^{-4}$	& $9.5\times 10^{-4}$	\\  
					& Total						 		&		&			& 23.1				& 100				\\
\hline\hline
\end{tabular}
\end{center}
\end{table}

\begin{table}[tp]
\caption{Predicted partial widths and branching ratios
for electromagnetic transitions and OZI allowed strong decays for the $2G$-wave states.
Details of the OZI allowed decay amplitudes are described in the appendix.  
See the caption to Table~\ref{tab:sum_12Swave} for further details.
\label{tab:sum_2Gwave}}
\begin{center}
\begin{tabular}{l l l l l l  } \hline \hline
Initial \phantom{www}   	& Final 					& $M_f$  	& $\cal M$  	& Width 			& BR  				\\
 state          			& state 					&(GeV)	&			& (keV) 			& (\%)  				\\ 
 \hline\hline
$\Upsilon_5 (2^3G_5)$	& $ \chi_{b4} (2^3F_4)  \gamma$ 	&  10.622 	& 3.598  		& 20.6	 		& $7.20\times 10^{-3}$ 	\\
10.772 				& $ \chi_{b4} (1^3F_4)  \gamma$  &  10.358 	& 0.186  		& 1.1	 			& $3.8\times 10^{-4}$ 	\\
					& $BB$								&		& 			& 25.9 MeV		& 9.06				\\
					& $BB^*$					 		&		& 			& 42.4 MeV		& 14.8				\\
					& $B^*B^*$					 		&		& 			& 218 MeV		& 76.2				\\
					& $B_sB_s$					 		&		&			& 4.72			& $1.65\times 10^{-3}$	\\
					& Total								&		&  			& 286 MeV		& 100				\\
\hline
$\Upsilon_4 (2^3G_4)$	& $ \chi_{b4} (2^3F_4)  \gamma$ 	&  10.622 	& 3.620  		& 1.3 			& $6.0\times 10^{-4}$ 	\\
10.770 				& $ \chi_{b3} (2^3F_3)  \gamma$  &  10.619 	& 3.570  		& 19.7 			& $9.1\times 10^{-3}$ 	\\
					& $BB^*$							& 		& 			& 116 MeV		& 53.7				\\
					& $B^*B^*$							&		& 			& 100. MeV		& 46.3				\\
					& Total								&		&  			& 216 MeV		& 100				\\
\hline
$\Upsilon_3 (2^3G_3)$	& $ \chi_{b4} (2^3F_4) \gamma $  	&  10.622 	& 3.643  		& 0.025 			& $1.6\times 10^{-5}$ 	\\
10.769				& $ \chi_{b3} (2^3F_3)  \gamma$ 		&  10.619 	& 3.593  		& 1.6 			& $1.0\times 10^{-3}$ 	\\
					& $ \chi_{b2} (2^3F_2)  \gamma$   		&  10.615 	& 3.538  		& 19.8 			& $1.28\times 10^{-2}$ 	\\
					& $BB$								&		& 			& 10.3 MeV		& 6.68				\\
					& $BB^*$					 		&		& 			& 68.3 MeV		& 44.3				\\
					& $B^*B^*$					 		&		& 			& 74.8 MeV		& 48.5				\\
					& $B_sB_s$					 		&		& 			& 0.744 MeV		& 0.482				\\
					& Total						 		&		&  			& 154.2 MeV		& 100				\\
\hline
$\eta_{b 4}(2^1G_4)$ 	& $ h_{b3} (2^1F_3)  \gamma$  	& 10.619 	& 3.573  		& 20.7 			& $9.00\times 10^{-3}$   	\\
 10.770				& $BB^*$						&		&			& 108 MeV		& 47.0				\\
					& $B^*B^*$					 		&		& 			& 122 MeV		& 53.0				\\
					& Total						 		&		&  			& 230. MeV		& 100				\\
\hline\hline
\end{tabular}
\end{center}
\end{table}

\section{Radiative Transitions}

Radiative transitions of excited bottomonium states are of interest for a number of reasons.  
First, they probe the internal structure of the states and provide a strong test of the 
predictions of the various models.  Moreover, for the purposes of this paper they provide a 
means of accessing $b\bar{b}$ states with different quantum numbers.  Observation of the 
photons emitted in radiative transitions between different $b\bar{b}$ states was in fact 
how the $3P$ $b\bar{b}$ state was observed by the ATLAS collaboration \cite{Aad:2011ih,Chisholm:2014sca}
and subsequently by LHCb  \cite{Aaij:2014caa,Aaij:2014hla}. 
 E1 radiative partial widths of bottomonium are typically ${\cal O}(1-10)$~keV  so 
 can represent a significant BR 
 for $b\bar{b}$ states that are relatively narrow.  As we 
will see, a large number of $b\bar{b}$ states 
fall into this category. With the high statistics available at the LHC it should be possible 
to observe some of the missing $b\bar{b}$ states with a well constrained search strategy.  
Likewise, SuperKEKB can provide large event samples of the $\Upsilon (3S)$ and $\Upsilon(4S)$
 and possibly the $\Upsilon(5S)$ and $\Upsilon(6S)$ which 
 could be used to identify radially excited $P$ and $D$-wave and 
other high $L$ states.  $e^+e^-$ collisions at SuperKEKB could also produce the $\Upsilon (1^3D_1)$ and 
$\Upsilon (2^3D_1)$ directly which could be observed by Belle II 
in decay chains involving radiative transitions.

We calculate the $E1$ radiative partial widths using \cite{Kwo88a} 
\begin{eqnarray}
\Gamma(n^{2S+1}L_J  & \to & {n'} ^{2S+1}L'_{J'} + \gamma) \\
& = & {{4\alpha e^2_b k_\gamma^3}\over 3} C_{fi} \delta_{L,L^{\prime}\pm 1} |\langle \psi_f | r | \psi_i \rangle |^2 
\nonumber
\end{eqnarray}
where the angular momentum matrix element is given by 
\begin{equation}
C_{fi} = \hbox{max}(L, L') (2J' + 1)
\left\{ \begin{array}{ccc} J & 1 & J' \\ L' & S & L \end{array}  \right\},  
\end{equation}
and $\{ {\cdots \atop \cdots} \}$ is a 6-$j$ symbol, 
$e_b=-1/3$ is the $b$-quark charge in units of $|e|$, $\alpha$ is the fine-structure constant, 
$k_\gamma$ is the photon energy and $\langle \psi_f | r | \psi_i \rangle $
is the transition matrix element from the initial state $\psi_i$ to the final state 
$\psi_f$. 
For these initial and final states, we use the relativized quark model wavefunctions \cite{godfrey85xj}.
The E1 radiative widths are given in Tables~\ref{tab:sum_12Swave}- \ref{tab:sum_2Gwave} 
along with the matrix elements so that the interested reader can reproduce our results.
The initial and final state masses are also listed in these tables where   
Particle Data Group (PDG) \cite{Olive:2014kda} 
masses are used when the masses are known.  For unobserved states the masses are taken from 
the predicted values in Tables~\ref{tab:Upsilonparams1}-\ref{tab:Upsilonparams2} except
when a member of a multiplet has been observed. In this latter case the mass used
was obtained using the procedure described in Section~\ref{sec:spectroscopy}.

An interesting observation is that the E1 transitions  $3S\to 1P$ are highly suppressed
relative to other E1 transitions \cite{Skwarnicki:2005pq} (see Ref.~\cite{Eichten:2007qx}
for a detailed discussion).  Grant and Rosner \cite{Grant:1992fi} showed this to be 
a general property of E1 transitions, that E1 transitions between states that differ
by 2 radial nodes are highly suppressed relative to the dominant E1 transitions and are in
fact zero for the 3-dimensional harmonic oscillator.  As a consequence, these radiative
transitions are particularly sensitive to relativistic corrections \cite{Eichten:2007qx}.  
We found that this pattern was also apparent for similar transitions of the type
$|n, l \rangle \to | n-2, l \pm 1 \rangle$ such as $5S \to 3P$, $3P \to 1D$, $4P \to 2S$,
$4D \to 2P$, $3D \to 1F$ etc.

M1 transition rates are typically weaker than E1 rates.  Nevertheless they 
have been useful in observing spin-singlet states that are difficult to 
observe in other ways \cite{Godfrey:2001eb,Aubert:2008ba}.  
The M1 radiative partial widths are evaluated using \cite{Nov78}
\begin{eqnarray}
\Gamma(n^{2S+1}L_J & \to & {n'} ^{2S'+1}L_{J'} + \gamma)  \\
& = & {{4 \alpha e^2_b k_\gamma^3}\over {3 m_b^2} } {{2J'+1}\over {2L+1}} \delta_{S,S'\pm 1} 
|\langle \psi_f | j_0(kr/2) | \psi_i \rangle |^2 \nonumber
\end{eqnarray}
where $j_0(x)$ is the spherical Bessel function and the other factors have been defined above.  
As with the E1 transitions, we use the relativized quark model wavefunctions \cite{godfrey85xj} for the initial and final states. 

The partial widths and branching ratios for the M1 radiative transitions are listed in 
Tables~\ref{tab:sum_12Swave}-\ref{tab:sum_2Gwave} as appropriate. 
For comparison, other calculations of $b\bar{b}$ radiative transitions can be found 
in Ref.~\cite{Kwo88a,Daghighian:1987ru,Ferretti:2014xqa,Ebert:2002pp,Wei-Zhao:2013sta,Pineda:2013lta}

\section{Annihilation Decays}
\label{sec:annihilation}

Annihilation decays into gluons and light quarks
make significant contributions to the total widths of some
$b\bar{b}$ resonances. 
In addition, annihilation decays into leptons or photons can be useful for the production
and identification of some bottomonium states.  For example, the vector mesons are
produced in $e^+e^-$ collisions through their couplings to $e^+e^-$.
Annihilation decay rates have
been studied extensively using perturbative QCD (pQCD) methods
\cite{App75,DeR75,Cha75,Bar76a,Bar76b,Nov78,Bar79,Kwo88b,Ack92a,Ack92b,Belanger:1987cg,Ber91,Robinett:1992px,Bradley:1980eh}.  
The relevant formulas for $S$- and $P$-wave states including first-order QCD corrections 
(when they are known)
are summarized in Ref.~\cite{Kwo88b}.  Expressions for $D$- and $F$-wave decays
are given in Refs.~\cite{Belanger:1987cg,Ber91} 
and Refs.~\cite{Ack92b,Robinett:1992px} respectively.  The expression for $^3D_1\to e^+e^-$ 
including the QCD correction comes from Ref.~\cite{Bradley:1980eh}.
Ackleh, Barnes and Close \cite{Ack92a} 
give a general expression for singlet decays to two gluons.  A general property
of annihilation decays is that the decay amplitude for a state with orbital angular momentum 
$l$ goes like $R^{(l)}/m_Q^{2l+2}$ where  $R^{(l)}$ is
the $l$-th derivative of the radial wavefunction. $R^{(l)}$ is typically ${\cal O} (1)$
so for bottom quark masses the magnitude of the annihilation decay widths 
decreases rapidly as the orbital angular momentum of the bottomonium state increases.  
Expressions for the decay widths including first-order QCD corrections when known are
summarized in Table~\ref{tab:annihilation_decays}.  
To obtain our numerical results for these partial widths we take the number of light quarks
to be $n_f=4$, assumed $m_b=4.977$~GeV, 
$\alpha_s\approx 0.18$ (with some weak mass dependence),
and used the wavefunctions found using the model of Ref.~\cite{godfrey85xj}   
as described in Section~\ref{sec:spectroscopy}.

Considerable uncertainties arise in these expressions from the 
model-dependence of the wavefunctions and possible relativistic 
and QCD radiative corrections (see for example the discussion
in Ref.\cite{godfrey85xj}).   One example is that
the logarithm evident in some of these formulas is evaluated at
a rather arbitrarily chosen scale, 
and that the pQCD radiative corrections to these processes
are often found to be large, but are prescription dependent and so are
numerically unreliable. As a consequence, these formulas should be regarded 
as estimates
of the partial widths for these annihilation processes rather than precise 
predictions.   
The numerical results for partial widths for the annihilation 
processes are included in Tables~\ref{tab:sum_12Swave}- \ref{tab:sum_1Gwave}.

\begin{table*}[t]
\caption{Summary of lowest order expressions and first order QCD corrections with $\alpha_s$ 
computed at the mass scale of the decaying state (see Section \ref{sec:annihilation} for references).  
\label{tab:annihilation_decays}}
\begin{tabular}{lcl} \hline \hline
Process & Rate & Correction  \\
\hline 
$^1S_0 \to gg$ & ${{8 \pi \alpha^2_s }\over {3 m_Q^2}} |\Psi(0)|^2 $ & $ (1+{{ 4.4 \alpha_s}\over \pi})$ (for $b\bar{b}$) \\ 
$^1S_0 \to \gamma\gamma$ & ${{12 \pi e_Q^4 \alpha^2 }\over {m_Q^2}} |\Psi(0)|^2 $ & $ (1-{{ 3.4 \alpha_s}\over \pi})$  \\ 
$^3S_1 \to ggg$ & ${{40 (\pi^2 -9)  \alpha^3_s }\over {81 m_Q^2}} |\Psi(0)|^2 $ & $ (1-{{ 4.9 \alpha_s}\over \pi})$ (for $b\bar{b}$) \\ 
$^3S_1 \to \gamma gg$ & ${{32 (\pi^2 -9) e_Q^2 \alpha \alpha^2_s }\over {9 m_Q^2}} |\Psi(0)|^2 $ & $ (1-{{ 7.4 \alpha_s}\over \pi})$ (for $b\bar{b}$) \\ 
$^3S_1 \to \gamma\gamma\gamma$ & ${{16 (\pi^2 -9)  e_Q^6 \alpha^3 }\over {3 m_Q^2}} |\Psi(0)|^2 $ & $ (1-{{ 12.6 \alpha_s}\over \pi})$  \\ 
$^3S_1 \to e^+e^-$ & ${{ 16\pi e_Q^2 \alpha^2 }\over {M^2}} |\Psi(0)|^2 $ & $ (1-{{ 16 \alpha_s}\over{3 \pi}})$ \\ 
$^3P_2\to gg$ & ${{8  \alpha^2_s }\over {5 m_Q^4}} |R_{nP}'(0)|^2 $ & $(1-{{ 0.2\alpha_s}\over \pi})$ ($\chi_{b2}(1P)$) \\
	& 	& $(1+{{ 0.83\alpha_s}\over \pi})$ ($\chi_{b2}(2P)$) \\
	& 	& $(1+{{ 1.47\alpha_s}\over \pi})$ ($\chi_{b2}(3P)$) \\
	& 	& $(1+{{ 1.91\alpha_s}\over \pi})$ ($\chi_{b2}(4P)$) \\
$^3P_2\to \gamma\gamma$ & ${{36 e_Q^4 \alpha^2 }\over {5 m_Q^4}} |R_{nP}'(0)|^2 $ & $(1-{{ 16\alpha_s}\over {3 \pi}})$ \\
$^3P_1\to q\bar{q}+g$ & ${{32  \alpha^3_s }\over {9\pi m_Q^4}} |R_{nP}'(0)|^2 \ln(m_Q\langle R \rangle )$ &  \\
$^3P_0\to gg$ & ${{6  \alpha^2_s }\over { m_Q^4}} |R_{nP}'(0)|^2 $ & $(1+{{ 9.9\alpha_s}\over \pi})$ ($\chi_{b0}(1P)$) \\
	& 	& $(1+{{ 10.2\alpha_s}\over \pi})$ ($\chi_{b0}(2P)$) \\
	& 	& $(1+{{ 10.3\alpha_s}\over \pi})$ ($\chi_{b0}(3P)$) \\
	& 	& $(1+{{ 10.5\alpha_s}\over \pi})$ ($\chi_{b0}(4P)$) \\
$^3P_0\to \gamma\gamma$ & ${{27  e_Q^4\alpha^2 }\over { m_Q^4}} |R_{nP}'(0)|^2 $ & $(1+{{ 0.2\alpha_s}\over \pi})$  \\
$^1P_1\to ggg$ & ${{20  \alpha^3_s }\over {9\pi m_Q^4}} |R_{nP}'(0)|^2 \ln(m_Q\langle R \rangle )$ &  \\
$^3D_3\to ggg$ & ${{40  \alpha^3_s }\over {9\pi m_Q^6}} |R_{nD}''(0)|^2 \ln(4m_Q\langle R \rangle )$ &  \\ 
$^3D_2\to ggg$ & ${{10  \alpha^3_s }\over {9\pi m_Q^6}} |R_{nD}''(0)|^2 \ln(4m_Q\langle R \rangle )$ &  \\ 
$^3D_1\to ggg$ & ${{760  \alpha^3_s }\over {81\pi m_Q^6}} |R_{nD}''(0)|^2 \ln(4m_Q\langle R \rangle )$ &  \\ 
$^3D_1\to e^+e^-$ & ${{200  e_Q^2 \alpha^2 }\over { M^6}} |R_{nD}''(0)|^2 $ & $ (1-{{ 16 \alpha_s}\over{3 \pi}})$ \\ 
$^1D_2\to gg$ & ${{2  \alpha^2_s }\over {3\pi m_Q^6}} |R_{nD}''(0)|^2 $ &  \\ 
$^3F_4\to gg$ & ${{20  \alpha^2_s }\over {27 m_Q^8}} |R_{nF}'''(0)|^2 $ &  \\ 
$^3F_3\to gg$ & ${{20  \alpha^2_s }\over {27 m_Q^8}} |R_{nF}'''(0)|^2 $ &  \\ 
$^3F_2\to gg$ & ${{919  \alpha^2_s }\over {135 m_Q^8}} |R_{nF}'''(0)|^2 $ &  \\ 
$^1G_4\to gg$ & ${{2  \alpha^2_s }\over {3\pi m_Q^{10} }} |R_{nG}^{iv}(0)|^2 $ &  \\ 
\hline
\hline
\end{tabular}
\end{table*}

\section{Hadronic Transitions}

Hadronic transitions between quarkonium levels are needed to
estimate branching ratios and potentially offer useful signatures for 
some   
missing bottomonium states.
There have been numerous theoretical estimates of 
hadronic transitions over the years 
\cite{Yan80,Kua81,Kua88,Kua90,Rosner03,Vol80,Nov81,Iof80,Vol86,Vol03a,Ko94,Mox88,Ko93,Kuang:2006me}.  
In some cases the estimates disagree by orders of 
magnitude \cite{Rosner03}.  
Hadronic transitions are typically described as a two-step 
process in which the gluons are first emitted from the heavy quarks 
and then recombine into light quarks.  
A multipole expansion of the 
colour gauge field is employed to describe the emission process where 
the intermediate colour octet quarkonium state is typically modeled by some sort 
of quarkonium hybrid wavefunction \cite{Kua81,Kuang:2006me}.  An uncertainty in
predictions arises from how the rehadronization step is estimated.  
To some extent this latter uncertainty can be reduced by employing 
the multipole expansion of the colour gauge fields 
developed by Yan and collaborators \cite{Yan80,Kua81,Kua88,Kua90}
together with the Wigner-Eckart theorem 
to estimate the E1-E1 transition rates \cite{Yan80}.  

In addition to E1-E1 transitions 
such as $^3S_1 \to~^3S_1 \pi \pi$,  
there will be other transitions such 
as $^3S_1 \to~^3S_1 + \eta$, which goes via M1-M1 \& E1-M2 multipoles
and spin-flip transitions such as $^3S_1 \to~^1P_1 \pi\pi$, which
goes via E1-M1 \cite{Kua81}.  
These transitions are suppressed by inverse powers 
of the quark masses and are expected to be small compared to the E1-E1 
and electromagnetic transitions. As a consequence, we will neglect them in our 
estimates of branching ratios. We note however, that in certain situations they
have provided a pathway to otherwise difficult to observe states such as the $h_c$ and 
$h_b$ \cite{Godfrey:2002rp,Rosner:2005ry} and have played an important role in these states' discoveries 
\cite{Rubin:2005px,Lees:2011zp}.  
Another example of a higher multipole transition is
$\chi_{b1,b2}(2P) \to \omega \Upsilon(1S)$ \cite{Severini:2003qw}
which proceeds via three E1 gluons although it turns out that
this particular example has a larger branching ratio 
than the $2P\to 1P + \pi \pi$ transition \cite{Olive:2014kda}.

The differential rate for E1-E1 transitions from an initial quarkonium 
state $\Phi'$ to the final quarkonium state $\Phi$, and a system of 
light hadrons, $h$, is given by the expression \cite{Yan80,Kua81}:
\begin{equation}
{{d\Gamma}\over {d{\cal M}^2}} [\Phi' \to \Phi +h] 
= (2J+1)\sum_{k=0}^2
\left\{ { 
\begin{array}{ccc}  k & \ell ' & \ell \\ s & J & J' 
\end{array} 
} \right\}
^2 A_k(\ell' , \, \ell )
\label{eqn:transition}
\end{equation}
where $\ell'$, $\ell$ are the orbital angular momentum  and
$J'$, $J$ are the total angular momentum of the 
initial and final states respectively,  $s$ is the spin of the 
$Q\bar{Q}$ pair, ${\cal M}^2$ is the invariant mass squared of 
the light hadron system, 
and $A_k(\ell' , \, \ell )$ are the reduced matrix elements. 
For the convenience of the reader we give the expressions for the transition rates
in terms of the reduced matrix elements in Table~\ref{tab:hadronic_amplitudes}.
The magnitudes of the $A_k(\ell' , \, \ell )$ are model dependent with a 
large variation in their estimates.  The $A_k(\ell' , \, \ell )$
are a product of a phase space factor, overlap integrals with the intermediate hybrid 
wavefunction and a fitted constant.  
There is a large variation 
in the predicted reduced rates.  For example, for the transition $1^3D_1\to 
1^3S_1 +\pi\pi$, estimates for $A_2(2,0)$ differ by almost three orders 
of magnitude \cite{Rosner03,Kua81,Mox88,Ko93}.  
In an attempt to minimize the theoretical 
uncertainty we estimate the reduced matrix elements by
rescaling measured transition rates by phase space factors and interquark separation 
expectation values. While imperfect, we hope that this approach
captures the essential features of the reduced matrix elements and gives a reasonable 
order of magnitude estimate of the partial widths. 
In the soft-pion limit the 
$A_1$ contributions are suppressed so, as is the usual practice, we 
will take $A_1(\ell' , \, \ell )=0$ \cite{Yan80} (see also Ref.~\cite{Eichten:1994gt})
so that in practice only $A_0(\ell' , \, \ell )$
and/or $A_2(\ell' , \, \ell )$ will contribute to a given transition.  
The $A_0$ and $A_2$ amplitudes have phase space integrals of the form \cite{Kua81}: 
\begin{equation}
G= {3\over 4} { M_f\over M_i} \pi^3 \int dM_{\pi\pi}^2 K  
\left(1 - {{4m_\pi^2}\over {M_{\pi\pi}^2} } \right)^{1/2} 
(M_{\pi\pi}^2 - 2 m_\pi^2)^2
\end{equation}
and 
\begin{widetext}
\begin{equation}
\label{eqn:Htype}
H= {1\over 20} { M_f\over M_i} \pi^3 \int dM_{\pi\pi}^2 K 
\left(1 - {{4m_\pi^2}\over {M_{\pi\pi}^2} } \right)^{1/2} 
\left[(M_{\pi\pi}^2 - 4 m_\pi^2)^2 \left( 1 +{2\over 3}{K^2 \over {M_{\pi\pi}^2}}\right) 
+ {K^2 \over {15 M_{\pi\pi}^4}}
(M_{\pi\pi}^4 + 2 m_\pi^2 M_{\pi\pi}^2  + 6 m_\pi^4 )\right] 
\end{equation}
\end{widetext}
respectively where 
\begin{equation}
K= {1\over{2M_i}} \left[ { (M_i+M_f)^2 - M_{\pi\pi}^2 }\right] ^{1/2}
\left[ { (M_i-M_f)^2 - M_{\pi\pi}^2 }\right] ^{1/2} .
\end{equation}
The amplitudes for E1-E1 transitions depend 
quadratically on the interquark separation so the scaling law between decay rates
for two $b\bar{b}$ states is given by
\cite{Yan80}
\begin{equation}
{{\Gamma(\Phi_1)}\over {\Gamma(\Phi_2)}}=
{{ \langle r^2 (\Phi_1) \rangle^2}\over 
{ \langle r^2 (\Phi_2) \rangle^2}}.
\label{eqn:scaling}
\end{equation}
Because each set of transitions uses different experimental input 
we will give details of how we rescale the $A_k(\ell' , \, \ell )$ 
sector by sector in the following subsections and give
the predicted partial widths in the summary tables.

\begin{table}[ht]
\caption{Expressions for hadronic transitions in terms of the reduced amplitudes 
$A_k(\ell' ,\ell)$.
Note that reduced amplitudes are dependent on the initial and final states.  Because
the $A_1$ contributions are suppressed we follow the usual practice and will take 
$A_1(\ell' , \ell)=0$ although we include them in the table.  The details on how we
obtain numerical estimates for amplitudes are described in the text.
\label{tab:hadronic_amplitudes}}
\begin{tabular}{lc} \hline \hline
Process & Expression  \\
\hline 
$^3S_1 \to ^3S_1 + \pi\pi$ & $A_0(0,0)$ \\
$^1S_0 \to ^1S_0 + \pi\pi$ & $A_0(0,0)$ \\
$^3P_2 \to ^3P_2 + \pi\pi$ & ${1\over 3} A_0(1,1) + {1\over 4} A_1(1,1) + {7\over{60}} A_2(1,1)$ \\
$^3P_2 \to ^3P_1 + \pi\pi$ & $ {1\over {12}} A_1(1,1) + {3\over{20}} A_2(1,1)$ \\
$^3P_2 \to ^3P_0 + \pi\pi$ & $ {1\over{15}} A_2(1,1)$ \\
$^3P_1 \to ^3P_2 + \pi\pi$ & $ {5\over {36}} A_1(1,1) + {1\over{4}} A_2(1,1)$ \\
$^3P_1 \to ^3P_1 + \pi\pi$ & ${1\over 3} A_0(1,1) + {1\over {12}} A_1(1,1) + {1\over{12}} A_2(1,1)$ \\
$^3P_1 \to ^3P_0 + \pi\pi$ & $ {1\over{9}} A_1(1,1)$ \\
$^3P_0 \to ^3P_2 + \pi\pi$ & $ {1\over{3}} A_2(1,1)$ \\
$^3P_0 \to ^3P_1 + \pi\pi$ & $ {1\over{3}} A_1(1,1)$ \\
$^3P_0 \to ^3P_0 + \pi\pi$ & $ {1\over{3}} A_0(1,1)$ \\
$^1P_1 \to ^1P_1 + \pi\pi$ & ${1\over 3} A_0(1,1) + {1\over {3}} A_1(1,1) + {1\over{3}} A_2(1,1)$ \\
$^3D_3 \to ^3D_3 + \pi\pi$ & ${1\over 5} A_0(2,2) + {8\over {45}} A_1(2,2) + {24\over{175}} A_2(2,2)$ \\
$^3D_3 \to ^3D_2 + \pi\pi$ & $ {1\over {45}} A_1(2,2) + {2\over{35}} A_2(2,2)$ \\
$^3D_3 \to ^3D_1 + \pi\pi$ & $ {1\over{175}} A_2(2,2)$ \\
$^3D_2 \to ^3D_3 + \pi\pi$ & $ {7\over {225}} A_1(2,2) + {2\over{25}} A_2(2,2)$ \\
$^3D_2 \to ^3D_2 + \pi\pi$ & ${1\over 5} A_0(2,2) + {5\over {36}} A_1(2,2) + {1\over{20}} A_2(2,2)$ \\
$^3D_2 \to ^3D_1 + \pi\pi$ & $ {3\over {100}} A_1(2,2) + {7\over{100}} A_2(2,2)$ \\
$^3D_1 \to ^3D_3 + \pi\pi$ & $  {1\over{75}} A_2(2,2)$ \\
$^3D_1 \to ^3D_2 + \pi\pi$ & $ {1\over {20}} A_1(2,2) + {7\over{60}} A_2(2,2)$ \\
$^3D_1 \to ^3D_1 + \pi\pi$ & ${1\over 5} A_0(2,2) + {3\over {20}} A_1(2,2) + {7\over{100}} A_2(2,2)$ \\
$^1D_2 \to ^1D_2 + \pi\pi$ & ${1\over 5} A_0(2,2) + {1\over {5}} A_1(2,2) + {1\over{5}} A_2(2,2)$ \\
$^3D_3 \to ^3S_1 + \pi\pi$ & $ {1\over{5}} A_2(2,0)$ \\
$^3D_2 \to ^3S_1 + \pi\pi$ & $ {1\over{5}} A_2(2,0)$ \\
$^3D_1 \to ^3S_1 + \pi\pi$ & $ {1\over{5}} A_2(2,0)$ \\
$^1D_2 \to ^1S_0 + \pi\pi$ & $ {1\over{5}} A_2(2,0)$ \\
$^3F_4 \to ^3P_2 + \pi\pi$ & $ {1\over{7}} A_2(3,1)$ \\
$^3F_3 \to ^3P_2 + \pi\pi$ & $ {1\over{21}} A_2(3,1)$ \\
$^3F_3 \to ^3P_1 + \pi\pi$ & $ {2\over{21}} A_2(3,1)$ \\
$^3F_2 \to ^3P_2 + \pi\pi$ & $ {1\over{105}} A_2(3,1)$ \\
$^3F_2 \to ^3P_1 + \pi\pi$ & $ {1\over{15}} A_2(3,1)$ \\
$^3F_2 \to ^3P_0 + \pi\pi$ & $ {1\over{15}} A_2(3,1)$ \\
$^1F_3 \to ^1P_1 + \pi\pi$ & $ {1\over{7}} A_2(3,1)$ \\
$^3G_5 \to ^3D_3 + \pi\pi$ & $ {1\over{9}} A_2(4,2)$ \\
$^3G_4 \to ^3D_3 + \pi\pi$ & $ {1\over{54}} A_2(4,2)$ \\
$^3G_4 \to ^3D_2 + \pi\pi$ & $ {5\over{54}} A_2(4,2)$ \\
$^3G_3 \to ^3D_3 + \pi\pi$ & $ {1\over{630}} A_2(4,2)$ \\
$^3G_3 \to ^3D_2 + \pi\pi$ & $ {1\over{42}} A_2(4,2)$ \\
$^3G_3 \to ^3D_1 + \pi\pi$ & $ {3\over{35}} A_2(4,2)$ \\
$^1G_4 \to ^1D_2 + \pi\pi$ & $ {1\over{9}} A_2(4,2)$ \\
\hline
\hline
\end{tabular}
\end{table}

\subsection{${n^\prime} {^1}S_0  \to n ^1S_0 + \pi \pi $}

The ${n^\prime} {^1}S_0  \to n ^1S_0 + \pi \pi $ partial widths are found by rescaling
the measured ${n^\prime} {^3}S_1  \to n ^3S_1 + \pi \pi $ partial widths \cite{Olive:2014kda}.
The $n^\prime S \to n S +\pi\pi $ transitions are described by $A_0(0,0)$ amplitudes 
so that  $\Gamma({n^\prime} {^1}S_0  \to n ^1S_0 + \pi \pi ) $ are given by:
\begin{equation}
\Gamma({n^\prime} {^1}S_0)=
{{ \langle r^2 ({n^\prime} {^1}S_0 ) \rangle^2}\over { \langle r^2 ({n^\prime} {^3}S_1) \rangle^2}}
\times {{ G({n^\prime} {^1}S_0 \to n^1 S_0 \pi \pi ) }
\over { G({n^\prime} {^3}S_1 \to n^3 S_1 \pi \pi )  }}
\times \Gamma({n^\prime} {^3}S_1 ).
\end{equation}
The hadronic transition partial widths for the $n^{\prime 1}S_0$ states are given in 
Tables~\ref{tab:sum_12Swave}-\ref{tab:sum_45Swave} for $n^\prime=2,3,4$. 

We do not make predictions for the $\eta_b(5S)$ state as
the measured hadronic transition rates for the $\Upsilon(5S)$ are anomalously large and inconsistent 
with other transitions between $S$-waves \cite{Chen:2008}.   
This has resulted in speculation that the 
$\Upsilon(5S)$ is mixed with a hybrid state leading to its anomalously large hadronic 
transition rates \cite{Segovia:2014mca}, contains a sizable tetraquark component 
\cite{Ali:2009es,Ali:2010pq}
or is the consequence of $B^{(*)}\bar{B}^{(*)}$ rescattering \cite{Chen:2011qx}.  
See also Ref.~\cite{Patrignani:2012an,Drutskoy:2012gt}.
It could also be the result of a large overlap with the intermediate
states.  This subject needs a separate more detailed study which lies outside the present
work.  We also do not include hadronic transitions for the $6S$ states as there are no 
measurements of hadronic transitions originating from the $6^3S_1$ state and in any case,
the total widths for the $6S$ states are quite large so that the BR's for hadronic transitions
would be rather small. 

\subsection{${n^\prime} P_J  \to n P_J  + \pi \pi $}

All $n' P_J \to n P_J +\pi \pi$ transitions can be expressed in terms of
$A_0(1,1)$ and $A_2(1,1)$ where we have taken $A_1(1,1)=0$. The 
expressions relating the various partial widths in terms of these reduced amplitudes 
are summarized in 
Table~\ref{tab:hadronic_amplitudes}.
We can obtain
$A_0(1,1)$ and $A_2(1,1)$ from the measured values for 
 $\Gamma(2^3P_2\to 1^3P_2 +\pi\pi)$ and  $\Gamma(2^3P_1\to 1^3P_1 +\pi\pi)$.  These
 partial widths were obtained by first finding the total widths for the $\chi_{b2}(2P)$
 and $\chi_{b1}(2P)$ using the measured BR's 
 from the PDG \cite{Olive:2014kda} 
 with our predicted partial widths for
 E1 transitions for $\chi_{b2(1)}(2P) \to \gamma \Upsilon (2S) $ and
  $\chi_{b2(1)}(2P) \to \gamma \Upsilon (1S)$.   We obtain $\Gamma(\chi_{b2})=122\pm 11$~keV
  and $\Gamma(\chi_{b1})=62.6\pm 4.0$~keV.  Combining with the measured hadronic BR's we find
  $\Gamma(2^3P_2\to 1^3P_2 +\pi\pi)=0.62 \pm 0.12$~keV and 
$\Gamma(2^3P_1\to 1^3P_1 +\pi\pi)= 0.57 \pm 0.09$~keV leading to
$A_2(1,1)=1.5$~keV and $A_0(1,1)=1.335$~keV.  We neglected the small differences in phase 
space between the $\chi_{b1}$ and $\chi_{b2}$ transitions and remind the reader that
model dependence has been introduced into these results by using model predictions for
the radiative transition partial widths. For example, using the E1 partial width
predictions from Kwong and Rosner \cite{Kwo88a} results in slightly different total widths
and hadronic transition partial widths.   Using these values for $A_0$ and $A_2$ with 
Eq.~\ref{eqn:transition} we obtain the hadronic transition partial widths for $2P\to 1P$ 
transitions given in Table~\ref{tab:sum_2Pwave}. 
A  note of caution is that $A_0$ and $A_2$ 
are sensitive to small variations in the input values of the partial widths so
given the experimental errors on the input values, the predictions should only be regarded
as rough estimates.

There are no measured BR's for $3P$ hadronic transitions that can be used as input
for other $3P$ transitions.
Furthermore, as pointed out by Kuang and Yan \cite{Kua81}, hadronic transitions 
are dependent on intermediate states with complicated cancellations contributing to the
amplitudes so that predictions are rather model dependent.  
To try to take into account the structure dependence of the 
amplitudes we make the assumption 
that once phase space and scaling factors (as in Eq.~\ref{eqn:scaling})
are factored out, the ratios of amplitudes will be approximately the same for transitions
between states with the same number of nodes in the initial and final states. i.e.:
\begin{equation}
{{ A' (3 P \to 1P)}\over {  A' (2 P \to 1P)} } \sim 
{{ A' (3 S \to 1S)}\over {  A' (2 S \to 1S)} } 
\label{eqn:scaling2}
\end{equation}
where the $A'$'s have factored out the phase space and scaling factors in the amplitude.
Thus, we will relate the $3P\to 1P$ partial widths to measured $2P \to 1P$ partial widths
by rescaling the phase space, the $b\bar{b}$ separation factors and 
using the relationship between amplitudes outlined in Eq.~\ref{eqn:scaling2}.
We understand that this is far from rigorous but hope that it captures the gross features
of the transition and will give us an order of magnitude estimate of the transitions that 
can at least tell us if the transition is big or small and how significant its contribution
to the total width will be.  With this prescription we obtain $A_0(3P\to 1P)=0.68$~keV and
$A_2(3P\to 1P)=4.94$~keV.  The resulting  
estimates for the $3P \to 1P$ hadronic transitions are given in 
Table~\ref{tab:sum_3Pwave}. 
To obtain these results we used spin averaged $P$-wave masses 
for the phase space factors given all the other uncertainties in these estimates.
If we don't include the $\langle r^2 \rangle$ rescaling factors, 
the partial widths increase by 45\%, which is another reminder that these estimates should be
regarded as educated guesses that hopefully get the order of magnitude right.

\subsection{$1 D_J  \to 1S  + \pi \pi $ and $2 D_J  \to 1D  + \pi \pi $}
\label{sec:hadronicD}

The BaBar collaboration measured 
$BR(1^3D_2 \to \Upsilon (1S) + \pi^+ \pi^-) = (0.66 \pm 0.16)\%$ \cite{delAmoSanchez:2010kz}. 
It should be noted
that BaBar used the predicted partial widths for $\chi_{bJ'}\to \gamma \Upsilon(1^3D_J)$ 
from 
Ref.~\cite{Kwo88a} as input to obtain this value.  Combining this measured BR with
the remaining $1^3D_2$ partial decay widths given in Table~\ref{tab:sum_1Dwave} 
we 
obtain $\Gamma(1^3D_2 \to \Upsilon (1S) + \pi^+ \pi^-) = 0.169\pm 0.045$~keV.  For comparison,
using the predictions for the $1^3D_2$ decays from Ref.~\cite{Kwo88a} we obtain the 
hadronic width $\Gamma=0.186\pm 0.047$~keV. 
Prior to the
BaBar measurement, predictions for this transition varied from 0.07~keV to 24~keV 
\cite{Rosner03,Kua81,Mox88,Ko93}.  Using the 
partial width value $\Gamma = 0.169 \pm 0.045$~keV as input we obtain 
 $A_2(2,0)=0.845$~keV which we use along with phase space 
(Eq.~\ref{eqn:Htype}) and 
$\langle r^2 \rangle$ rescaling factors (Eq.~\ref{eqn:scaling}) 
to obtain 
the $1D \to 1S \pi \pi $ transitions 
given in Table~\ref{tab:sum_1Dwave}. 
For comparison we include the measurements
for the $1^3D_1$ and $1^3D_3$ transitions from Ref.~\cite{delAmoSanchez:2010kz} which
are less certain than those for the $1^3D_2$ transitions. 

Because we have no data on the $2D$ states we use the same strategy to estimate 
$2D\to 1D $ transitions as we did in estimating the $3P \to 1P$ transitions.  We assume
that rescaling an amplitude with the same number of nodes in the initial and final state 
wavefunctions will capture the gross features of the complicated overlap integrals with
intermediate wavefunctions.  We use the $A_0$ and $A_2$ amplitudes from the $2P \to 1P$ transitions
as input and rescale the amplitudes using the appropriate phase space factors and 
$\langle r^2 \rangle$ rescaling factors.  This gives $A_0(2,2)=3.1\times 10^{-2}$~keV
and  $A_2(2,2)=4.6 \times 10^{-5}$~keV.
As a check we also estimated $A_0$ found by rescaling the 
$A_0$ amplitude obtained from the $2S \to 1S$ transition where only $A_0$ contributes and
found it to be roughly a factor of 40 smaller than
the value obtained from the $2P \to 1P$ transition.  
This should be kept in mind when assessing the 
reliability of our predictions.  In any case, the partial widths obtained for
the $2D \to 1D$ hadronic transitions are sufficiently small 
(see Table~\ref{tab:sum_2Dwave}) that the  large uncertainties
will not change our conclusions regarding the $2D$ states. 

\subsection{$1F \to 1P + \pi \pi$ and $1G \to 1D  + \pi \pi $}

We take the same approach as we used to estimate some of the hadronic transition widths
given above. We take a measured width,  in this case the $1^3D_2 \to \Upsilon (1S) \pi^+ \pi^-$, 
and rescale it using ratios of phase space factors and separation factors to estimate
the $1F \to 1P$ and $1G \to 1D$ transitions.  We obtain $A_2(1F\to 1P) =0.027$~keV
and $A_2(1G\to 1D) =1.94 \times 10^{-3}$~keV.  These small values are primarily due 
to the ratio of phase space factors which roughly go like the mass difference to the 7th 
power.  Given that the $1D-1S$, $1F-1P$ and $1G-1D$ mass splittings are 
$\sim 0.69$, 0.46 and 0.38~MeV respectively resulting in little available phase space 
one can understand why the amplitudes are small.
We include our estimates for these transitions in Tables \ref{tab:sum_1Fwave} and
\ref{tab:sum_1Gwave}.  While the estimates may be crude the point is that we expect 
these partial widths to be quite small.

\section{Strong Decays}
\label{sec:strongdecays}

For states above the $B\bar{B}$ threshold, we calculate OZI allowed strong decay widths using the $^3P_0$ quark 
pair creation model \cite{Micu:1968mk,Le Yaouanc:1972ae,Ackleh:1996yt,Blundell:1995ev,Barnes:2005pb}
which proceeds through the production of a light $q\bar{q}$ pair ($q=u,d,s$) followed
by separation into $B\bar{B}$ mesons. The $q\bar{q}$ pair is assumed to be produced with vacuum quantum 
numbers ($0^{++}$).
There are a number of predictions for $\Upsilon$ strong decay widths in the literature using the $^3P_0$ model 
\cite{Ferretti:2014, Segovia:2012} and other models \cite{Ebert:2014}, but a complete analysis of their 
strong decays had yet to be carried out prior to this work.  
We give details regarding the notation and 
conventions used in our $^3P_0$ model calculations in Appendix~\ref{app:3P0} to make it
more transparent for an interested reader to reproduce our results. 

We use the meson masses listed in Tables~\ref{tab:Upsilonparams1}-\ref{tab:Bparams}.  
If available, the measured value, $M_{exp}$, is used as 
input for calculating the strong decay widths, rather than the predicted value, $M_{theo}$.  When the mass of only one meson 
in a multiplet has been measured, we estimate the input masses for the remaining states following 
the procedure described at the end of Sec.~\ref{sec:spectroscopy}.

We use simple harmonic oscillator wave functions with the effective oscillator parameter, $\beta$, obtained by equating 
 the rms radius of the harmonic oscillator wavefunction for the specified $(n,l)$ quantum numbers
  to the rms radius of the wavefunctions calculated using the relativized quark model of Ref.~\cite{godfrey85xj}.  
  The effective harmonic oscillator wavefunction parameters found in this way are listed in the final column of 
  Tables~\ref{tab:Upsilonparams1}-\ref{tab:Bparams}.  For the constituent quark masses in our 
  calculations of both the meson masses and of the strong decay widths, we use $m_b=4.977$~GeV, 
  $m_s=0.419$~GeV, and $m_q=0.220$~GeV ($q = u, d$).  Finally, we use 
  ``relativistic phase space'' as described in Ref.~\cite{Blundell:1995ev,Ackleh:1996yt} 
  and in Appendix~\ref{app:3P0}.

Typical values of the parameters $\beta$ and $\gamma$ are found from fits to light meson
 decays \cite{Close:2005se,Blundell:1995ev,Blundell:1996as}. The predicted widths are fairly 
 insensitive to the precise values used for $\beta$ provided $\gamma$ is appropriately rescaled.  
 However $\gamma$ can vary as much as 30\% and still give reasonable overall fits of 
 light meson decay widths \cite{Close:2005se,Blundell:1996as}. This can result in factor 
 of two changes to predicted widths, both smaller or larger.  In our calculations of $D_s$ 
 meson strong decay widths in \cite{Godfrey:2014}, we used a value of $\gamma=0.4$, which 
 has also been found to give a good description of strong decays of charmonium 
 \cite{Barnes:2005pb}.  However, we found that this value underestimated 
 the bottomonium strong decay widths when compared to the PDG values for the $\Upsilon(4S)$, 
 $\Upsilon(5S)$ and $\Upsilon(6S)$ widths.  Therefore, we used a value of $\gamma = 0.6$ 
 in our strong decay width calculations in this paper, which was determined by fitting 
 our results to the PDG values in the $\Upsilon$ sector.  This scaling of the value of 
 $\gamma$ in different quarkonia sectors has been studied in \cite{Segovia:2012}.  The 
 resulting strong decay widths are listed in Tables~\ref{tab:sum_45Swave}-\ref{tab:sum_2Gwave}  
in which we use a more concise notation where $BB$ refers to the $B\bar{B}$ decay mode, 
$BB^*$ refers to $B\bar{B}^* + \bar{B}B^*$, etc.
 
 We note that our results differ from the recent work of Ferretti and Santopinto \cite{Ferretti:2014}, 
 in some cases quite substantially.  
  This is primarily due to the values chosen for the 
 harmonic oscillator parameter $\beta$ (with a corresponding change in the pair creation 
 strength $\gamma$).  In our calculations we used a value for $\beta$ found by fitting the
 rms radius of a harmonic oscillator wavefunction to the ``exact'' wavefunction for each
 state while Ferretti and Santopinto used a common value for all states.  
Another reason our results differ is because Ferretti and Santopinto included an additional Gaussian smearing function 
in their momentum-space wavefunction overlap to model the non-point-like nature of the created $q\bar{q}$ pair.  
As a numerical check of 
our programs we reproduced their results using their parameters and including the Gaussian smearing function, 
although we found that the latter had little effect on our results.   
We believe our approach best describes the properties of individual states but this underlines the 
importance of experimental input to test models and improve predictions.

\section{Search Strategies}

An important motivation for this work is to suggest strategies to observe some of the
missing bottomonium mesons.  While there are similarities between searches at hadron colliders
and $e^+e^-$ colliders there are important differences.  As a consequence we will 
consider the two production channels separately.

\subsection{At the Large Hadron Collider}

\subsubsection{Production}

An important ingredient needed in discussing searches for the missing bottomonium states 
at a hadron collider is an estimate of the production rate for the different states
\cite{Brambilla:2010cs,Bodwin:1994jh,Likhoded:2012hw,Han:2014kxa,Ali:2013xba}.  
The production cross sections for the $\Upsilon(nS)$ and $\chi_{bJ}$ states are in good
agreement with predictions of non-relativistic QCD (NRQCD) also referred to as the colour 
octet model.  However we are interested in higher excitations with both higher principle
quantum number and higher orbital angular momentum for which we are not aware of any 
existing calculations.  To estimate production rates we use the NRQCD 
factorization approach to rescale measured event rates.  In the
NRQCD factorization
approach the cross section goes like \cite{Brambilla:2010cs}
\begin{equation}
\sigma(H) = \sum_n \sigma_n(\Lambda) \langle {\cal O}_n^H (\Lambda)\rangle
\end{equation}
for quarkonium state $H$ and where $n$ denotes the colour, spin and angular momentum of the
intermediate $b\bar{b}$ pair, $\sigma_n(\Lambda)$ is the perturbative short distance (parton level) cross section and 
$\langle {\cal O}_n^H (\Lambda)\rangle$ is the long distance matrix element (LDME) 
which includes the colour octet $Q\bar{Q}$ pair that evolves into quarkonium.  
We work with the assumption that the quarkonium state dependence resides primarily
in the LDME which goes very roughly like (see Ref.~\cite{Bodwin:1994jh})
\begin{equation}
\langle {\cal O}(^{2S+1}L_J)\rangle \propto (2J+1) {{ |R_{nL}^{(\ell)}(0)|^2}\over {M^{2L+2}}}
\label{eqn:nrqcd}
\end{equation}
where $R_{nL}^{(\ell)}(0)$ is the $\ell$~th derivative of the wavefunction at the origin
and $M$ is the mass of the state being produced.
There are numerical factors, the operator coefficients of order 1 that have only been computed
for the $S$- and $P$-wave states \cite{Bodwin:1994jh}.  This gives, for example, an additional factor
of 3 in the numerator for $P$-wave states in Eq.~\ref{eqn:nrqcd}.  We note that 
at LO, NRQCD predictions are not in good agreement with experiment but 
at NLO the agreement is much better \cite{Brambilla:2010cs}.  Some of the additional factors that
contribute to the uncertainty in our crude estimates are not calculating
the relative contributions of colour singlet and colour octet contributions,
the neglect of higher order QCD corrections, 
the sensitivity of event rates to the $p_T$ cuts used in the analysis,
and ignoring the dependence of detector efficiencies on photon energies.

We will base our estimates on LHCb expectations but expect similar estimates for the 
collider experiments ATLAS and CMS based on the measured event rates for bottomonium
production by LHCb \cite{Aaij:2014caa}, ATLAS \cite{Aad:2012dlq} and 
CMS \cite{Chatrchyan:2013yna}. However, there are differences between these experiments 
as LHCb covers the low $p_T$ region while ATLAS and CMS extend to higher
$p_T$ so that the production rates are not expected to be  identical \cite{Likhoded:2012hw},
only that the general trends are expected to be similar.

To estimate production rates we start with the 
production rates measured by
LHCb for LHC Run I 
and rescale them using Eq.~\ref{eqn:nrqcd}.  Further, it is expected
that the cross sections will more than double going from 8~TeV to 14~TeV \cite{Ali:2013xba} 
and the total integrated
luminosity is expected to be an order of magnitude larger for Run II
compared to Run I. 
LHCb observed $\sim 1.07 \times 10^6$ $\Upsilon (1S)$'s in the $\mu^+\mu^-$ final state
for the combined 7~TeV and 8~TeV runs \cite{Aaij:2014caa}.  Taking into account the 
$\Upsilon (1S) \to \mu^+\mu^-$ BR, roughly $ 4.3 \times 10^7$ $\Upsilon (1S)$'s were 
produced.  Multiplying this value by the expected doubling of the production cross section
going to 
the current 13~TeV centre-of-mass energy and the factor of 10 in integrated luminosity leads to $8.6\times 10^8$
$\Upsilon (1S)$'s as our starting point.  We rescale this using Eq.~\ref{eqn:nrqcd} 
and use our estimates for the branching ratios for decay chains to estimate event rates. 
Our results can easily be rescaled to correct for the actual integrated luminosity.

We find that this crude approach agrees with the LHCb event rates for $nS$ and $nP$ 
\cite{Aaij:2014caa} within roughly a factor of 2, in some cases too small and in other cases
too large.  Considering the crudeness of these estimates and the many factors listed above
which were not included
we consider this to be acceptable agreement. 
We want to emphasize before proceeding that we only expect our
estimates to be reliable as order of magnitude estimates but this should be sufficient to 
identify the most promising channels to pursue.

In the following subsections we generally focus on $b\bar{b}$ states below $B\bar{B}$ threshold, as the 
BR's for decay chains originating from states above $B\bar{B}$ threshold 
will generally result in too few events to be observable. 
Likewise, the production cross sections for
high $L$ states are suppressed by large powers of masses in the denominator.

We also focus on decay chains involving radiative transitions although hadronic transitions 
with charged pions often have higher detection efficiencies.  However, hadronic transitions
are not nearly as well understood as radiative transitions so we were only able to even attempt
to estimate a limited number of cases that, as discussed, we could relate to measured 
transitions.  In addition, in many cases of interest, hadronic transitions are expected to
be small.  Nevertheless, as demonstrated by the study of the $\Upsilon (1^3D_J$) by the
BaBar collaboration \cite{delAmoSanchez:2010kz} and the $h_b(1P)$ and $h_b(2P)$ by the
Belle collaboration \cite{Adachi:2011ji}, hadronic transitions offer another means to
find and study bottomonium states.  We include a few examples in the tables that follow
but they are by no means an exhaustive compilation and we encourage experimentalists to not neglect this
decay mode.

\subsubsection{The $3S$ and higher excited $S$-wave states}

We start with the $S$-wave states.  Our interest is that they will be produced 
in large quantities and their decay chains include states we are interested in such as excited $P$- 
and $D$-waves
and will therefore add to the statistics for those states.  We therefore focus on decay
chains that include these states.  

The $3S$ decay
chains of interest are given in 
Table~\ref{tab:3S}.  
The estimates for the number of events expected for LHCb are included but we also
include a column with estimates for $e^+e^-$ collisions expected by Belle II which we
discuss in Sec.~\ref{sec:eecolliders}.  
What is relevant is that numerous $1D$ states will be produced in this manner and when
added to those produced directly and via $P$-wave initial states will give rise to 
significant statistics in $2\gamma +\mu^+\mu^-$ final states.  In addition, it might be
possible to observe the $\eta_b(3S)$ in a $\gamma \mu^+ \mu^-$ final state from
the $\eta_b(3S)\to \Upsilon (1S)$ M1 transition.

\begin{table*}[tp]
\caption{The $3S$ Decay chains, branching ratios and event estimates 
for LHCb Run II and Belle II.  
The number of events in the $pp$ column are based on producing 
$3.1\times 10^8$ $\Upsilon(3S)$'s and 
$1.4 \times 10^8$  $\eta_b(3S)$'s as described in the text while those in the 
$e^+e^-$ column are based on $10^9$ $\Upsilon(3S)$'s assuming 250~fb$^{-1}$ 
integrated luminosity.
\label{tab:3S}}
\begin{tabular}{lllrr} \hline \hline
Parent & Decay chain & Combined  & \multicolumn{2}{c}{Events}  \\
		&			&		BR		& $pp$  & $e^+e^-$ \\
\hline 
$3^3S_1$		&  $\stackrel{13.1\%}{\longrightarrow} 2^3P_2 \gamma \; (86.2)
		\stackrel{1.2\%}{\longrightarrow} 1^3D_3 \gamma \; (96.5)
		\stackrel{91.0\%}{\longrightarrow} 1^3P_2 \gamma \; (256.0)
		\stackrel{19.1\%}{\longrightarrow} 1^3S_1 \gamma \; (441.6)
		\stackrel{2.48\%}{\longrightarrow} \mu^+\mu^-$ & $6.8 \times 10^{-6}$ & 2100 & 6800 \\
				&  $\stackrel{13.1\%}{\longrightarrow} 2^3P_2 \gamma \; (86.2)
		\stackrel{0.2\%}{\longrightarrow} 1^3D_2 \gamma \; (104.4)
		\stackrel{22\%}{\longrightarrow} 1^3P_2 \gamma \; (248.4)
		\stackrel{19.1\%}{\longrightarrow} 1^3S_1 \gamma \; (441.6)
		\stackrel{2.48\%}{\longrightarrow} \mu^+\mu^-$ & $2.7 \times 10^{-7}$ & 84 & 270 \\
				&  $\stackrel{13.1\%}{\longrightarrow} 2^3P_2 \gamma \; (86.2)
		\stackrel{0.2\%}{\longrightarrow} 1^3D_2 \gamma \; (104.4)
		\stackrel{74.7\%}{\longrightarrow} 1^3P_1 \gamma \; (267.3)
		\stackrel{33.9\%}{\longrightarrow} 1^3S_1 \gamma \; (423.0)
		\stackrel{2.48\%}{\longrightarrow} \mu^+\mu^-$ & $1.6 \times 10^{-6}$ & 500 & 1600 \\		
  			& $\stackrel{13.1\%}{\longrightarrow} 2^3P_2 \gamma \; (86.2)
		\stackrel{0.02\%}{\longrightarrow} 1^3D_1 \gamma \; (78.0)
		\stackrel{1.6\%}{\longrightarrow} 1^3P_2 \gamma \; (239.1)
		\stackrel{19.1\%}{\longrightarrow} 1^3S_1 \gamma \; (441.6)
		\stackrel{2.48\%}{\longrightarrow} \mu^+\mu^-$ & $2.0 \times 10^{-9}$ & 0.6 & 2 \\
  			& $\stackrel{13.1\%}{\longrightarrow} 2^3P_2 \gamma \; (86.2)
		\stackrel{0.02\%}{\longrightarrow} 1^3D_1 \gamma \; (78.0)
		\stackrel{28\%}{\longrightarrow} 1^3P_1 \gamma \; (258.0)
		\stackrel{33.9\%}{\longrightarrow} 1^3S_1 \gamma \; (423.0)
		\stackrel{2.48\%}{\longrightarrow} \mu^+\mu^-$ & $6.2 \times 10^{-8}$ & 19 & 62 \\		
  			& $\stackrel{13.1\%}{\longrightarrow} 2^3P_2 \gamma \; (86.2)
		\stackrel{0.02\%}{\longrightarrow} 1^3D_1 \gamma \; (78.0)
		\stackrel{47.1\%}{\longrightarrow} 1^3P_0 \gamma \; (290.5)
		\stackrel{1.76\%}{\longrightarrow} 1^3S_1 \gamma \; (391.1)
		\stackrel{2.48\%}{\longrightarrow} \mu^+\mu^-$ & $5.4 \times 10^{-9}$ & 2 & 5 \\
  			& $\stackrel{13.1\%}{\longrightarrow} 2^3P_2 \gamma \; (86.2)
		\stackrel{0.02\%}{\longrightarrow} 1^3D_1 \gamma \; (78.0)
		\stackrel{0.00393\%}{\longrightarrow} \mu^+\mu^-$ & $1.0 \times 10^{-9}$ & 0.3 & 1 \\
			& $\stackrel{12.6\%}{\longrightarrow} 2^3P_1 \gamma \; (99.3)
		 \stackrel{1.9\%}{\longrightarrow} 1^3D_2 \gamma \; (91.3)
		\stackrel{22\%}{\longrightarrow} 1^3P_2 \gamma \; (248.4)
		\stackrel{19.1\%}{\longrightarrow} 1^3S_1 \gamma \; (441.6)
		\stackrel{2.48\%}{\longrightarrow} \mu^+\mu^-$ & $2.5 \times 10^{-6}$ & 780 & 2500 \\
			&  $\stackrel{12.6\%}{\longrightarrow} 2^3P_1 \gamma \; (99.3)
		\stackrel{1.9\%}{\longrightarrow} 1^3D_2 \gamma \; (91.3)
		\stackrel{74.7\%}{\longrightarrow} 1^3P_1 \gamma \; (267.3)
		\stackrel{33.9\%}{\longrightarrow} 1^3S_1 \gamma \; (423.0)
		\stackrel{2.48\%}{\longrightarrow} \mu^+\mu^-$ & $1.5\times 10^{-5}$ & 4650 & 15,000 \\
			&   $\stackrel{12.6\%}{\longrightarrow} 2^3P_1 \gamma \; (99.3)
		\stackrel{0.80\%}{\longrightarrow} 1^3D_1 \gamma \; (100.8)			
		\stackrel{1.6\%}{\longrightarrow} 1^3P_2 \gamma \; (239.1)
		\stackrel{19.1\%}{\longrightarrow} 1^3S_1 \gamma \; (441.6)
		\stackrel{2.48\%}{\longrightarrow} \mu^+\mu^-$ & $7.6\times 10^{-8}$ & 24 & 76 \\	
		&   $\stackrel{12.6\%}{\longrightarrow} 2^3P_1 \gamma \; (99.3)
		\stackrel{0.80\%}{\longrightarrow} 1^3D_1 \gamma \; (100.8)		
		\stackrel{28\%}{\longrightarrow} 1^3P_1 \gamma \; (258.0)
		\stackrel{33.9\%}{\longrightarrow} 1^3S_1 \gamma \; (423.0)
		\stackrel{2.48\%}{\longrightarrow} \mu^+\mu^-$ & $2.4 \times 10^{-6}$ & 740 & 2400 \\
		&   $\stackrel{12.6\%}{\longrightarrow} 2^3P_1 \gamma \; (99.3)
		\stackrel{0.80\%}{\longrightarrow} 1^3D_1 \gamma \; (100.8)		
		\stackrel{47.1\%}{\longrightarrow} 1^3P_0 \gamma \; (290.5)
		\stackrel{1.76\%}{\longrightarrow} 1^3S_1 \gamma \; (391.1)
		\stackrel{2.48\%}{\longrightarrow} \mu^+\mu^-$ & $2.1 \times 10^{-7}$ & 65 & 210 \\
		&   $\stackrel{5.9\%}{\longrightarrow} 2^3P_0 \gamma \; (122.0)
		\stackrel{0.4\%}{\longrightarrow} 1^3D_1 \gamma \; (78.0)
		\stackrel{1.6\%}{\longrightarrow} 1^3P_2 \gamma \; (239.1)
		\stackrel{19.1\%}{\longrightarrow} 1^3S_1 \gamma \; (441.6)
		\stackrel{2.48\%}{\longrightarrow} \mu^+\mu^-$ & $1.8 \times 10^{-8}$ & 6 & 18 \\
		&   $\stackrel{5.9\%}{\longrightarrow} 2^3P_0 \gamma \; (122.0)
		\stackrel{0.4\%}{\longrightarrow} 1^3D_1 \gamma \; (78.0)
		\stackrel{28\%}{\longrightarrow} 1^3P_1 \gamma \; (258.0)
		\stackrel{33.9\%}{\longrightarrow} 1^3S_1 \gamma \; (423.0)
		\stackrel{2.48\%}{\longrightarrow} \mu^+\mu^-$ & $5.6 \times 10^{-7}$ & 170 & 560 \\
		&   $\stackrel{5.9\%}{\longrightarrow} 2^3P_0 \gamma \; (122.0)
		\stackrel{0.4\%}{\longrightarrow} 1^3D_1 \gamma \; (78.0)
		\stackrel{47.1\%}{\longrightarrow} 1^3P_0 \gamma \; (290.5)
		\stackrel{1.76\%}{\longrightarrow} 1^3S_1 \gamma \; (391.1)
		\stackrel{2.48\%}{\longrightarrow} \mu^+\mu^-$ & $4.8 \times 10^{-8}$ & 15 & 48 \\
$3^1S_0$ & $\stackrel{1.8\times 10^{-6}}{\longrightarrow} 2^3S_1 \gamma \; (309.2)
		\stackrel{1.93\%}{\longrightarrow} \mu^+\mu^-$ & $3.4 \times 10^{-8}$ & 5 & NA \\
		 & $\stackrel{1.5\times 10^{-5}}{\longrightarrow} 1^3S_1 \gamma \; (840.0)
		\stackrel{2.48\%}{\longrightarrow} \mu^+\mu^-$ & $3.7 \times 10^{-7}$ & 52 & NA \\
\hline \hline
\end{tabular}
\end{table*}

The $4^3S_1$ state can decay to $3P$ states which can subsequently decay to $2D$ or $1D$ 
states, as shown in Table~\ref{tab:4S}. 
The $4^3S_1$ is above the $B\bar{B}$ threshold so has a much larger total width
than 
the lower mass $S$-waves leading to a much smaller BR for radiative transitions.  
Decay chains to $F$-waves involve too many transitions making them difficult to 
reconstruct so we do not include them
in our tables.   We also include decay chains which might be of interest for $e^+e^-$ 
studies but would result in insufficient 
statistics to be relevant to hadron collider 
studies.  We do not include the $4^1S_0$ state as the decay chains have too small 
combined BR to be observed.  

For the $5^3S_1$, the BR's to the $4^3P_J$ states are ${\cal O}(10^{-4})$ and the BR's of
the $4^3P_J$ are $\lesssim 10^{-4}$ so this product BR is quite small.  When we include 
BR's to interesting states such as the $3^3D_J$ states the product BR's are likely to be 
far too 
small to be observable.    The BR's to the $3^3P_J$ states are comparable to the 
$4S\to 3P$ transitions, ${\cal O} (10^{-5})$ so it might be possible to see $3P$ states 
starting from the $5^3S_1$.  It is not likely that the $2D$ and $1D$ states can be
observed in decay chains originating from the $5^3S_1$.
We arrive at similar conclusions for the $6^3S_1$ and conclude  
that the only possible states that might
be observed are the $3^3P_J$ states.

\begin{table*}[tp]
\caption{$4^3S_1$ Decay chains, branching ratios and event estimates 
for LHCb Run II and Belle II.  
The number of events in the $pp$ column are based on producing 
 $2.3\times 10^8$ $\Upsilon(4S)$'s as described in the text while
 those in the $e^+e^-$ column are based on $10^{10}$ $\Upsilon(4S)$'s assuming 
 10~ab$^{-1}$ integrated luminosity.  The BR's for the hadronic transitions were taken 
 from the PDG \cite{Olive:2014kda}.
\label{tab:4S}}
\begin{tabular}{lllrr} \hline \hline
Parent & Decay chain & Combined  & \multicolumn{2}{c}{Events}  \\
		&			&		BR		& $pp$  & $e^+e^-$ \\
\hline 
$4^3S_1$ & 	$\stackrel{1.57\times 10^{-5}}{\longrightarrow} \mu^+\mu^-$
			& $1.6\times 10^{-5}$ & 3680 & $1.6\times 10^5$ \\		
		& $\stackrel{3.7\times 10^{-5}}{\longrightarrow} 3^3P_2 \gamma \;(50.8)
			\stackrel{3.8\%}{\longrightarrow} 3^3S_1$ $\gamma \;(171.6) 
			\stackrel{2.18\%}{\longrightarrow} \mu^+\mu^-$
			& $3.1\times 10^{-8}$ & 7 & 310 \\		
		& $\stackrel{3.7\times 10^{-5}}{\longrightarrow} 3^3P_2 \gamma \;(50.8)
			\stackrel{1.8\%}{\longrightarrow} 2^3S_1$ $\gamma \;(492.9) 
			\stackrel{1.93\%}{\longrightarrow} \mu^+\mu^-$
			& $1.3\times 10^{-8}$ & 3 & 130 \\
		& $\stackrel{3.7\times 10^{-5}}{\longrightarrow} 3^3P_2 \gamma \;(50.8)
		\stackrel{1.1\%}{\longrightarrow} 1^3S_1$ $\gamma \;(1013.8) 
		\stackrel{2.48\%}{\longrightarrow} \mu^+\mu^-$
			& $1.0\times 10^{-8}$ & 2 & 100\\	
		& $\stackrel{3.7\times 10^{-5}}{\longrightarrow} 3^3P_2 \gamma \;(50.8)
			\stackrel{3.8\%}{\longrightarrow} 3^3S_1$ $\gamma \;(171.6) 
			\stackrel{2.82\%}{\longrightarrow} 2^3S_1 \pi^+\pi^-
			\stackrel{1.93\%}{\longrightarrow} \mu^+\mu^-$
			& $7.7\times 10^{-10}$ & 0.2 & 8 \\	
		& $\stackrel{3.7\times 10^{-5}}{\longrightarrow} 3^3P_2 \gamma \;(50.8)
			\stackrel{3.8\%}{\longrightarrow} 3^3S_1$ $\gamma \;(171.6) 
			\stackrel{4.37\%}{\longrightarrow} 1^3S_1 \pi^+\pi^-
			\stackrel{2.48\%}{\longrightarrow} \mu^+\mu^-$
			& $1.5\times 10^{-9}$ & 0.3 & 15 \\	
		& $\stackrel{3.7\times 10^{-5}}{\longrightarrow} 3^3P_2 \gamma \;(50.8)
			\stackrel{1.8\%}{\longrightarrow} 2^3S_1$ $\gamma \;(492.9) 
			\stackrel{17.85\%}{\longrightarrow} 1^3S_1 \pi^+\pi^-
			\stackrel{2.48\%}{\longrightarrow} \mu^+\mu^-$
			& $2.9\times 10^{-9}$ & 0.7 & 29 \\
		& $\stackrel{3.7\times 10^{-5}}{\longrightarrow} 3^3P_2 \gamma \;(50.8)
		\stackrel{0.61\%}{\longrightarrow} 2^3D_3 $ $\gamma \;(72.7) 
		\stackrel{65.1\%}{\longrightarrow} 2^3P_2 \gamma \; (185.0)
		\stackrel{10.6\%}{\longrightarrow} 2^3S_1 \gamma \; (242.5)
		\stackrel{1.93\%}{\longrightarrow} \mu^+\mu^-$ & $3.0 \times 10^{-10}$ & 0.07 & 3 \\
		& $\stackrel{3.7\times 10^{-5}}{\longrightarrow} 3^3P_2 \gamma \;(50.8)
		\stackrel{0.61\%}{\longrightarrow} 2^3D_3 $ $\gamma \;(72.7) 
		\stackrel{65.1\%}{\longrightarrow} 2^3P_2 \gamma \; (185.0)
		\stackrel{7.0\%}{\longrightarrow} 1^3S_1 \gamma \; (776.5)
		\stackrel{2.48\%}{\longrightarrow} \mu^+\mu^-$ & $2.6 \times 10^{-10}$ & 0.06 & 3\\
		& $\stackrel{3.7\times 10^{-5}}{\longrightarrow} 3^3P_2 \gamma \;(50.8)
		\stackrel{0.61\%}{\longrightarrow} 2^3D_3 $ $\gamma \;(72.7) 
		\stackrel{10\%}{\longrightarrow} 1^3P_2 \gamma \; (529.0)
		\stackrel{19.1\%}{\longrightarrow} 1^3S_1 \gamma \; (441.6)
		\stackrel{2.48\%}{\longrightarrow} \mu^+\mu^-$ & $1.1 \times 10^{-10}$ & 0.03 & 1 \\
		& $\stackrel{3.7\times 10^{-5}}{\longrightarrow} 3^3P_2 \gamma \;(50.8)
		\stackrel{0.61\%}{\longrightarrow} 2^3D_3 $ $\gamma \;(72.7) 
		\stackrel{6.7\%}{\longrightarrow} 1^3F_4 \gamma \; (96.7)$ & & \\
		&  $\qquad \qquad
			\stackrel{100\%}{\longrightarrow} 1^3D_3 \gamma \; (200.9)
		\stackrel{91.0\%}{\longrightarrow} 1^3P_2 \gamma \; (256.0)
		\stackrel{19.1\%}{\longrightarrow} 1^3S_1 \gamma \; (441.6)
		\stackrel{2.48\%}{\longrightarrow} \mu^+\mu^-$ & $6.5 \times 10^{-11}$ & 0.01 & 0.6 \\
		& $\stackrel{3.7\times 10^{-5}}{\longrightarrow} 3^3P_2 \gamma \;(50.8)
		\stackrel{0.019\%}{\longrightarrow} 1^3D_3 \gamma \; (350.0)
		\stackrel{91.0\%}{\longrightarrow} 1^3P_2 \gamma \; (256.0)
		\stackrel{19.1\%}{\longrightarrow} 1^3S_1 \gamma \; (441.6)
		\stackrel{2.48\%}{\longrightarrow} \mu^+\mu^-$ & $3.0 \times 10^{-11}$ & - & 0.3  \\
		& $\stackrel{3.7\times 10^{-5}}{\longrightarrow} 3^3P_2 \gamma \;(50.8)
		\stackrel{0.13\%}{\longrightarrow} 2^3D_2 $ $\gamma \;(78.7) 
		\stackrel{56.2\%}{\longrightarrow} 2^3P_1 \gamma \; (191.6)
		\stackrel{19.9\%}{\longrightarrow} 2^3S_1 \gamma \; (229.6)
		\stackrel{1.93\%}{\longrightarrow} \mu^+\mu^-$ & $1.0 \times 10^{-10}$ & 0.02 & 1  \\
		 & $\stackrel{3.8\times 10^{-5}}{\longrightarrow} 3^3P_1 \gamma \;(62.8)
			\stackrel{7.2\%}{\longrightarrow} 3^3S_1$ $\gamma \;(159.8) 
			\stackrel{2.18\%}{\longrightarrow} \mu^+\mu^-$
			& $6.1\times 10^{-8}$ & 15 & 610 \\		
		& $\stackrel{3.8\times 10^{-5}}{\longrightarrow} 3^3P_1 \gamma \;(62.8)
			\stackrel{2.6\%}{\longrightarrow} 2^3S_1$ $\gamma \;(481.4) 
			\stackrel{1.93\%}{\longrightarrow} \mu^+\mu^-$
			& $1.9\times 10^{-8}$ & 4 & 190\\
		 & $\stackrel{3.8\times 10^{-5}}{\longrightarrow} 3^3P_1 \gamma \;(62.8)
			\stackrel{1.1\%}{\longrightarrow} 1^3S_1$ $\gamma \;(1003.0) 
			\stackrel{2.48\%}{\longrightarrow} \mu^+\mu^-$
			& $1.0\times 10^{-8}$ & 2 & 100 \\
		 & $\stackrel{3.8\times 10^{-5}}{\longrightarrow} 3^3P_1 \gamma \;(62.8)
			\stackrel{7.2\%}{\longrightarrow} 3^3S_1$ $\gamma \;(159.8) 
			\stackrel{2.82\%}{\longrightarrow} 2^3S_1 \pi^+\pi^-			
			\stackrel{1.93\%}{\longrightarrow} \mu^+\mu^-$
			& $1.5\times 10^{-9}$ & 0.3 & 15 \\		
		 & $\stackrel{3.8\times 10^{-5}}{\longrightarrow} 3^3P_1 \gamma \;(62.8)
			\stackrel{7.2\%}{\longrightarrow} 3^3S_1$ $\gamma \;(159.8) 
			\stackrel{4.37\%}{\longrightarrow} 1^3S_1 \pi^+\pi^-			
			\stackrel{2.48\%}{\longrightarrow} \mu^+\mu^-$
			& $3.0\times 10^{-9}$ & 0.7 & 30 \\		
		& $\stackrel{3.8\times 10^{-5}}{\longrightarrow} 3^3P_1 \gamma \;(62.8)
			\stackrel{2.6\%}{\longrightarrow} 2^3S_1$ $\gamma \;(481.4) 
			\stackrel{17.85\%}{\longrightarrow} 1^3S_1 \pi^+\pi^-			
			\stackrel{2.48\%}{\longrightarrow} \mu^+\mu^-$
			& $4.4\times 10^{-9}$ & 1 & 44 \\					
		 & $\stackrel{3.8\times 10^{-5}}{\longrightarrow} 3^3P_1 \gamma \;(62.8)
		\stackrel{0.94\%}{\longrightarrow} 2^3D_2 $ $\gamma \;(66.8) 
		\stackrel{17\%}{\longrightarrow} 2^3P_2 \gamma \; (178.6)
		\stackrel{10.6\%}{\longrightarrow} 2^3S_1 \gamma \; (242.5)
		\stackrel{1.93\%}{\longrightarrow} \mu^+\mu^-$ & $1.2 \times 10^{-10}$ & 0.03 & 1 \\
		 & $\stackrel{3.8\times 10^{-5}}{\longrightarrow} 3^3P_1 \gamma \;(62.8)
		\stackrel{0.94\%}{\longrightarrow} 2^3D_2 $ $\gamma \;(66.8) 
		\stackrel{17\%}{\longrightarrow} 2^3P_2 \gamma \; (178.6)
		\stackrel{7.0\%}{\longrightarrow} 1^3S_1 \gamma \; (776.5)
		\stackrel{2.48\%}{\longrightarrow} \mu^+\mu^-$ & $1.0 \times 10^{-10}$ & 0.02 & 1 \\
		 & $\stackrel{3.8\times 10^{-5}}{\longrightarrow} 3^3P_1 \gamma \;(62.8)
		\stackrel{0.94\%}{\longrightarrow} 2^3D_2 $ $\gamma \;(66.8) 
		\stackrel{56.2\%}{\longrightarrow} 2^3P_1 \gamma \; (191.6)
		\stackrel{19.9\%}{\longrightarrow} 2^3S_1 \gamma \; (229.6)
		\stackrel{1.93\%}{\longrightarrow} \mu^+\mu^-$ & $7.7 \times 10^{-10}$ & 0.2 & 8 \\
		 & $\stackrel{3.8\times 10^{-5}}{\longrightarrow} 3^3P_1 \gamma \;(62.8)
		\stackrel{0.94\%}{\longrightarrow} 2^3D_2 $ $\gamma \;(66.8) 
		\stackrel{56.2\%}{\longrightarrow} 2^3P_1 \gamma \; (191.6)
		\stackrel{9.2\%}{\longrightarrow} 1^3S_1 \gamma \; (764.3)
		\stackrel{2.48\%}{\longrightarrow} \mu^+\mu^-$ & $4.6 \times 10^{-10}$ & 0.1 & 5 \\
		 & $\stackrel{3.8\times 10^{-5}}{\longrightarrow} 3^3P_1 \gamma \;(62.8)
		\stackrel{0.94\%}{\longrightarrow} 2^3D_2 $ $\gamma \;(66.8) 
		\stackrel{12\%}{\longrightarrow} 1^3P_1 \gamma \; (541.2)
		\stackrel{33.9\%}{\longrightarrow} 1^3S_1 \gamma \; (423.0)
		\stackrel{2.48\%}{\longrightarrow} \mu^+\mu^-$ & $3.6 \times 10^{-10}$ & 0.08 & 4 \\
		 & $\stackrel{3.8\times 10^{-5}}{\longrightarrow} 3^3P_1 \gamma \;(62.8)
		\stackrel{0.40\%}{\longrightarrow} 2^3D_1 $ $\gamma \;(74.7) 
		\stackrel{17\%}{\longrightarrow} 2^3P_1 \gamma \; (183.9)
		\stackrel{19.9\%}{\longrightarrow} 2^3S_1 \gamma \; (229.6)
		\stackrel{1.93\%}{\longrightarrow} \mu^+\mu^-$ & $9.9 \times 10^{-11}$ & 0.02 & 1 \\
		 & $\stackrel{3.8\times 10^{-5}}{\longrightarrow} 3^3P_1 \gamma \;(62.8)
		\stackrel{0.40\%}{\longrightarrow} 2^3D_1 $ $\gamma \;(74.7) 
		\stackrel{28.1\%}{\longrightarrow} 2^3P_0 \gamma \; (206.4)
		\stackrel{0.9\%}{\longrightarrow} 1^3S_1 \gamma \; (743.1)
		\stackrel{2.48\%}{\longrightarrow} \mu^+\mu^-$ & $9.5 \times 10^{-12}$ & - & 0.1 \\
		 & $\stackrel{2.2\times 10^{-5}}{\longrightarrow} 3^3P_0 \gamma \;(78.7)
			\stackrel{0.31\%}{\longrightarrow} 3^3S_1$ $\gamma \;(144.0) 
			\stackrel{2.18\%}{\longrightarrow} \mu^+\mu^-$
			& $1.5\times 10^{-9}$ & 0.3 & 15 \\		
		 & $\stackrel{2.2\times 10^{-5}}{\longrightarrow} 3^3P_0 \gamma \;(78.7)
			\stackrel{0.077\%}{\longrightarrow} 2^3S_1$ $\gamma \;(466.2) 
			\stackrel{1.93\%}{\longrightarrow} \mu^+\mu^-$
			& $3.3\times 10^{-10}$ & 0.08 & 3 \\
		 & $\stackrel{2.2\times 10^{-5}}{\longrightarrow} 3^3P_0 \gamma \;(78.7)
			\stackrel{0.01\%}{\longrightarrow} 1^3S_1$ $\gamma \;(1003.0) 
			\stackrel{2.48\%}{\longrightarrow} \mu^+\mu^-$
			& $5.5\times 10^{-11}$ & 0.01 & 0.6 \\
		 & $\stackrel{2.2\times 10^{-5}}{\longrightarrow} 3^3P_0 \gamma \;(78.7)
			\stackrel{0.31\%}{\longrightarrow} 3^3S_1$ $\gamma \;(144.0) 
			\stackrel{2.82\%}{\longrightarrow} 2^3S_1 \pi^+\pi^-			
			\stackrel{1.93\%}{\longrightarrow} \mu^+\mu^-$
			& $3.7\times 10^{-11}$ & - & 0.4 \\		
		 & $\stackrel{2.2\times 10^{-5}}{\longrightarrow} 3^3P_0 \gamma \;(78.7)
			\stackrel{0.31\%}{\longrightarrow} 3^3S_1$ $\gamma \;(144.0) 
			\stackrel{4.37\%}{\longrightarrow} 1^3S_1 \pi^+\pi^-			
			\stackrel{2.48\%}{\longrightarrow} \mu^+\mu^-$
			& $7.4\times 10^{-11}$ & 0.02 & 1 \\		
		 & $\stackrel{2.2\times 10^{-5}}{\longrightarrow} 3^3P_0 \gamma \;(78.7)
			\stackrel{0.077\%}{\longrightarrow} 2^3S_1$ $\gamma \;(466.2) 
			\stackrel{17.85\%}{\longrightarrow} 1^3S_1 \pi^+\pi^-			
			\stackrel{2.48\%}{\longrightarrow} \mu^+\mu^-$
			& $7.5\times 10^{-11}$ & 0.02 & 1 \\		
		 & $\stackrel{2.2\times 10^{-5}}{\longrightarrow} 3^3P_0 \gamma \;(78.7)
		\stackrel{0.045\%}{\longrightarrow} 2^3D_1  \gamma \;(58.8) 
		\stackrel{17\%}{\longrightarrow} 2^3P_1 \gamma \; (183.9)
		\stackrel{19.9\%}{\longrightarrow} 2^3S_1 \gamma \; (229.6)
		\stackrel{1.93\%}{\longrightarrow} \mu^+\mu^-$ & $6.5 \times 10^{-12}$ & - & 0.06 \\
		 & $\stackrel{2.2\times 10^{-5}}{\longrightarrow} 3^3P_0 \gamma \;(78.7)
		\stackrel{0.0091\%}{\longrightarrow} 1^3D_1  \gamma \;(340.2) 
		\stackrel{1.6\%}{\longrightarrow} 1^3P_2 \gamma \; (239.1)
		\stackrel{19.1\%}{\longrightarrow} 1^3S_1 \gamma \; (441.6)
		\stackrel{2.48\%}{\longrightarrow} \mu^+\mu^-$ & $1.5 \times 10^{-13}$ & - & - \\
		 & $\stackrel{2.2\times 10^{-5}}{\longrightarrow} 3^3P_0 \gamma \;(78.7)
		\stackrel{0.0091\%}{\longrightarrow} 1^3D_1  \gamma \;(340.2) 
		\stackrel{28\%}{\longrightarrow} 1^3P_1 \gamma \; (258.0)
		\stackrel{33.9\%}{\longrightarrow} 1^3S_1 \gamma \; (423.0)
		\stackrel{2.48\%}{\longrightarrow} \mu^+\mu^-$ & $4.7 \times 10^{-12}$ & - & - \\
		 & $\stackrel{2.2\times 10^{-5}}{\longrightarrow} 3^3P_0 \gamma \;(78.7)
		\stackrel{0.0091\%}{\longrightarrow} 1^3D_1  \gamma \;(340.2) 
		\stackrel{47.1\%}{\longrightarrow} 1^3P_0 \gamma \; (290.5)
		\stackrel{1.76\%}{\longrightarrow} 1^3S_1 \gamma \; (391.1)
		\stackrel{2.48\%}{\longrightarrow} \mu^+\mu^-$ & $4.1 \times 10^{-13}$ & - & - \\
\hline \hline
\end{tabular}
\end{table*}

\subsubsection{The $2P$ states}

The $2P$ states are of course well known.  We include them as they can decay to the $1D$ 
states and hence contribute to the $1D$ event rates which was the discovery
channel for the $1^3D_2$ state \cite{Bonvicini:2004yj} 
(see also Ref.~\cite{Godfrey:2001vc}).   
We only include event chains relevant
to these final states which we list in Table~\ref{tab:2P}.
In an attempt to reduce the theoretical uncertainties in the $2P\to 1D$ BR's, 
rather than using the predictions for total widths and BR's in Table~\ref{tab:sum_2Pwave} 
we estimate the $\chi_{bJ}(2P)$ total widths 
using the 
PDG values \cite{Olive:2014kda} for the BR's for 
the  $2P\to 2S +\gamma$ and $2P \to 1S + \gamma$ transitions together with our predictions 
for these partial widths as was described in Section~VB.  The values for $\Gamma(2^3P_2)$ 
and $\Gamma(2^3P_1)$ were given there.  Similarly we find $\Gamma(2^3P_0)=247\pm 93$~keV.
Combining these total widths with the partial widths for $2P\to 1D$ transitions given in  
Table~\ref{tab:sum_2Pwave},  $\Gamma(2^3P_2\to 1^3D_{3,2,1})=$~1.5, 0.3 and 0.03~keV
respectively, we obtain the corresponding BR's of 1.2\%, 0.2\% and 0.02\%.  
Likewise
$\Gamma(2^3P_1\to 1^3D_{2,1})=$~1.2 and 0.5~keV give the corresponding BR's 1.9\% and 0.80\%,
and $\Gamma(2^3P_0\to 1^3D_{1})=$~1.0~keV  has BR~=~0.4\%. 
There is 
uncertainty in these estimates as can be seen by comparing the PDG values \cite{Olive:2014kda}
we used in our estimates to recent BaBar measurements \cite{Lees:2014qea} 
 and by comparing our 
predictions for the partial widths to those of Kwong and Rosner \cite{Kwo88a}.  
Nevertheless these estimates are sufficient for estimating the 
$2P\to 1D +\gamma$ BR's and the resulting event rates for the purposes of identifying
promising channels.
The important point is that in all cases, significant numbers of bottomonium $D$-waves will be produced 
from $P$-wave production and decay
which will improve the statistics from those that are produced directly.

\begin{table*}[tp]
\caption{The $2P$ Decay chains, branching ratios and event estimates 
for LHCb Run II. 
These are based on producing 
$1.0\times 10^7$ $\chi_{b2}(2P)$'s, $9.1\times 10^{6}$ $\chi_{b1}(2P)$'s,
$4.7\times 10^6$ $\chi_{b0}(2P)$'s
 and 
$7.4 \times 10^6$  $h_b(2P)$'s as described in the text.
\label{tab:2P}}
\begin{tabular}{lllc} \hline \hline
Parent & Decay chain & Combined BR & Events  \\
\hline 
$2^3P_2$		&  $\stackrel{1.2\%}{\longrightarrow} 1^3D_3 \gamma \; (96.5)
		\stackrel{91.0\%}{\longrightarrow} 1^3P_2 \gamma \; (256.0)
		\stackrel{19.1\%}{\longrightarrow} 1^3S_1 \gamma \; (441.6)
		\stackrel{2.48\%}{\longrightarrow} \mu^+\mu^-$ & $5.2 \times 10^{-5}$ & 517 \\
				&  $\stackrel{0.2\%}{\longrightarrow} 1^3D_2 \gamma \; (104.4)
		\stackrel{22\%}{\longrightarrow} 1^3P_2 \gamma \; (248.4)
		\stackrel{19.1\%}{\longrightarrow} 1^3S_1 \gamma \; (441.6)
		\stackrel{2.48\%}{\longrightarrow} \mu^+\mu^-$ & $2.1 \times 10^{-6}$ & 21 \\
				&  $\stackrel{0.2\%}{\longrightarrow} 1^3D_2 \gamma \; (104.4)
		\stackrel{74.7\%}{\longrightarrow} 1^3P_1 \gamma \; (267.3)
		\stackrel{33.9\%}{\longrightarrow} 1^3S_1 \gamma \; (423.0)
		\stackrel{2.48\%}{\longrightarrow} \mu^+\mu^-$ & $1.2 \times 10^{-5}$ & 124 \\
$2^3P_1$		&  $\stackrel{1.9\%}{\longrightarrow} 1^3D_2 \gamma \; (91.3)
		\stackrel{22\%}{\longrightarrow} 1^3P_2 \gamma \; (248.4)
		\stackrel{19.1\%}{\longrightarrow} 1^3S_1 \gamma \; (441.6)
		\stackrel{2.48\%}{\longrightarrow} \mu^+\mu^-$ & $2.0 \times 10^{-5}$ & 180 \\
			&  $\stackrel{1.9\%}{\longrightarrow} 1^3D_2 \gamma \; (91.3)
		\stackrel{74.7\%}{\longrightarrow} 1^3P_1 \gamma \; (267.3)
		\stackrel{33.9\%}{\longrightarrow} 1^3S_1 \gamma \; (423.0)
		\stackrel{2.48\%}{\longrightarrow} \mu^+\mu^-$ & $1.2\times 10^{-4}$ & 1100 \\
			&  $\stackrel{0.80\%}{\longrightarrow} 1^3D_1 \gamma \; (100.8)			
		\stackrel{1.6\%}{\longrightarrow} 1^3P_2 \gamma \; (239.1)
		\stackrel{19.1\%}{\longrightarrow} 1^3S_1 \gamma \; (441.6)
		\stackrel{2.48\%}{\longrightarrow} \mu^+\mu^-$ & $6.1\times 10^{-7}$ & 6 \\	
		&  $\stackrel{0.80\%}{\longrightarrow} 1^3D_1 \gamma \; (100.8)		
		\stackrel{28\%}{\longrightarrow} 1^3P_1 \gamma \; (258.0)
		\stackrel{33.9\%}{\longrightarrow} 1^3S_1 \gamma \; (423.0)
		\stackrel{2.48\%}{\longrightarrow} \mu^+\mu^-$ & $1.9 \times 10^{-5}$ & 174 \\
		&  $\stackrel{0.80\%}{\longrightarrow} 1^3D_1 \gamma \; (100.8)		
		\stackrel{47.1\%}{\longrightarrow} 1^3P_0 \gamma \; (290.5)
		\stackrel{1.76\%}{\longrightarrow} 1^3S_1 \gamma \; (391.1)
		\stackrel{2.48\%}{\longrightarrow} \mu^+\mu^-$ & $1.6 \times 10^{-6}$ & 15 \\
$2^3P_0$		&  $\stackrel{0.4\%}{\longrightarrow} 1^3D_1 \gamma \; (78.0)
		\stackrel{1.6\%}{\longrightarrow} 1^3P_2 \gamma \; (239.1)
		\stackrel{19.1\%}{\longrightarrow} 1^3S_1 \gamma \; (441.6)
		\stackrel{2.48\%}{\longrightarrow} \mu^+\mu^-$ & $3.0 \times 10^{-7}$ & 1 \\
				&  $\stackrel{0.4\%}{\longrightarrow} 1^3D_1 \gamma \; (78.0)
		\stackrel{28\%}{\longrightarrow} 1^3P_1 \gamma \; (258.0)
		\stackrel{33.9\%}{\longrightarrow} 1^3S_1 \gamma \; (423.0)
		\stackrel{2.48\%}{\longrightarrow} \mu^+\mu^-$ & $9.4 \times 10^{-6}$ & 44 \\
$2^1P_1$ &  $\stackrel{2.0\%}{\longrightarrow} 1^1D_2 \gamma \; (111.4)
		\stackrel{91.5\%}{\longrightarrow} 1^1P_1 \gamma \; (262.5)
		\stackrel{49\%}{\longrightarrow} 1^1S_0 \gamma \; (488.3)
		\stackrel{100\%}{\longrightarrow} gg $ & $9.0 \times 10^{-3}$ & $6.6\times 10^4$ \\
		& $\stackrel{48\%}{\longrightarrow} 2^1S_0 $ $\gamma \;(257.7)
		\stackrel{100\%}{\longrightarrow} gg$ & 0.48 & $3.6\times 10^6$ \\
		& $\stackrel{22\%}{\longrightarrow} 1^1S_0 $ $\gamma \;(825.8)
		\stackrel{100\%}{\longrightarrow} gg$ & 0.22 & $1.6\times 10^6$ \\
\hline \hline
\end{tabular}
\end{table*}

\subsubsection{The $3P$ states}

The observation of the $\chi_b(3P)$ by the ATLAS collaboration  \cite{Aad:2011ih}
 through its radiative transitions to $\Upsilon (1S)$ and 
$\Upsilon (2S) $ with $\Upsilon(1S, \, 2S) \to \mu^+\mu^-$
was the first new particle discovered at the LHC.  This decay chain represents a clean
experimental signature with the two final state muons a clean signal to trigger on. 
The $\chi_b(3P)$ was confirmed by the D0 collaboration \cite{Abazov:2012gh} and
by the LHCb collaboration \cite{Aaij:2014caa,Aaij:2014hla}. Further,  LHCb identified the
state as the $\chi_{b1}(3P)$ with mass 
$10515.7^{+2.2}_{-3.9}({\rm stat})^{+1.5}_{-2.1}({\rm syst})$~MeV/$c^2$ \cite{Aaij:2014hla}. 
Using the approach outlined above we calculate for Run I  $\sim 243$ events for the decay chain 
$\chi(3P)\to \Upsilon(1S)\to \mu^+\mu^-$, $\sim 371$ for the $\Upsilon(2S)$ decay chain, and
$\sim 1030$ for the   $\Upsilon(3S)$ decay chain 
compared to the observed numbers of events of $329\pm 59$, $121\pm 31$ and $182 \pm 23$
respectively \cite{Aaij:2014caa}.  
The agreement is reasonable for the $\Upsilon (1S)$ chain but becomes
decreasingly so for the $\Upsilon(2S)$ and $\Upsilon(3S)$ chains.  We assume this is 
due to decreasing photon detection efficiency as the photon energy goes down.  
Overall the agreement is not unreasonable considering the
approximations used to obtain these values   
and our neglect of detector efficiencies and 
helps the reader judge the general reliability of our predictions for event rates.

In Table~\ref{tab:3P} we summarize the event rates expected for the most promising
decay chains.  The $\chi_{b0}(3P)$ state is much broader and decays 
predominantly to light hadrons via gluon
intermediate states with a small BR to $\gamma \Upsilon (1S)$ ($\sim 10^{-4}$) so it
would be quite challenging to observe although it might be possible to observe 
via hadronic transitions to the $1^3P_2$ state which subsequently undergoes a 
radiative transition to the $1^3S_1$ state.

For both the $\chi_{b2}(3P)$ and $\chi_{b1}(3P)$ the decay chains to 
$\Upsilon(nS)\to\mu^+\mu^-$ where $n=1$, 2, 3 give rise to the largest event rates and are
the simplest to reconstruct so it is not surprising that these were the discovery channels.
Run II should provide sufficient statistics to separately fit the 
$\chi_{b2}(3P)$ and $\chi_{b1}(3P)$ using these decay chains.  Other decay chains are
potentially interesting as they involve the undiscovered $2^3D_3$, $2^3D_2$, $2^3D_1$,
$1^3D_3$, $1^3F_4$ and $1^3F_3$ $b\bar{b}$ states. However they generally have 
multiple photons in the final state making it difficult to reconstruct the initial particle.
In addition, some of the photons in the decay chains are relatively low energy so could
have low detection efficiencies.  The one exception is $3^3P_1\to 2^3D_1 \gamma \to \mu^+\mu^-$.
This final state 
is relatively clean but has a low combined BR.  As we noted, we would
not be surprised if our production rates are off by an order of magnitude so we do not
rule out the possibility that the $2^3D_1$ could be observed in this process.    

We do not include decay chains for the $h_{b}(3P)$ state, with two exceptions,
as we believe it would
be very difficult to reconstruct in a hadronic environment for the following reasons.  It 
decays predominantly to hadronic final states ($\sim 70\%$ of the time) which would be
difficult to identify in hadron collisions.  The next largest BR's are to the
$\eta_b(3S)$, $\eta_b(2S)$ and $\eta_b(1S)$.  All of these decay almost 100\% of the time
to hadronic final states.  The $\eta_b(3S)$ and   $\eta_b(2S)$ have small BR's to 
$h_b(nP)$ and very small BR's to $\Upsilon(1S)$.  While there are several hundred $h_b$'s
produced they would be difficult to see and the combined BR to $\Upsilon(1S)\to \mu^+\mu^-$ 
is too small to produce 
sufficient numbers to observe.  We include the decay chain
$h_b(3P)\to \eta_b(1S) \gamma \to gg$  as it might be possible to reconstruct the $h_b(3P)$
using the one final state photon and possibly simple $\eta_b$ hadronic final states
and likewise the decay chain $h_b(3P)\to h_b(1P) + \pi\pi \to \eta_b(1S) \gamma \to gg$ 
for the same reasons.

\begin{table*}[tp]
\caption{$3^3P_J$ Decay chains, branching ratios and event estimates 
for LHCb Run II. 
These are based on producing $9.9\times 10^6$ $\chi_{b2}(3P)$'s, 
$7.9\times 10^6$  $\chi_{b1}(3P)$'s, $3.5\times 10^6$  $\chi_{b0}(3P)$'s 
and $7.3\times 10^6$  $h_{b}(3P)$'s as described in the text.
\label{tab:3P}}
\begin{tabular}{lllc} \hline \hline
Parent & Decay chain & Combined BR & Events  \\
\hline 
$3^3P_2$ & $\stackrel{3.8\%}{\longrightarrow} 3^3S_1$ $\gamma \;(171.6) \stackrel{2.18\%}{\longrightarrow} \mu^+\mu^-$
			& $8.3\times 10^{-4}$ & 8,240 \\
		& $\stackrel{1.8\%}{\longrightarrow} 2^3S_1$ $\gamma \;(492.9) \stackrel{1.93\%}{\longrightarrow} \mu^+\mu^-$
			& $3.5\times 10^{-4}$ & 3,460 \\
		& $\stackrel{1.1\%}{\longrightarrow} 1^3S_1$ $\gamma \;(1013.8) \stackrel{2.48\%}{\longrightarrow} \mu^+\mu^-$
			& $2.7\times 10^{-4}$ & 2,710 \\
		& $\stackrel{0.28\%}{\longrightarrow}  1^3P_2 \pi \pi 
		\stackrel{19.1\%}{\longrightarrow} 1^3S_1 \gamma \; (441.6)
		\stackrel{2.48\%}{\longrightarrow} \mu^+\mu^-$ & $1.3 \times 10^{-5}$ & 132 \\
		& $\stackrel{0.21\%}{\longrightarrow}  1^3P_1 \pi \pi 
		\stackrel{33.9\%}{\longrightarrow} 1^3S_1 \gamma \; (423.0)
		\stackrel{2.48\%}{\longrightarrow} \mu^+\mu^-$ & $1.8 \times 10^{-5}$ & 173 \\
		& $\stackrel{0.61\%}{\longrightarrow} 2^3D_3 $ $\gamma \;(72.7) 
		\stackrel{65.1\%}{\longrightarrow} 2^3P_2 \gamma \; (185.0)
		\stackrel{10.6\%}{\longrightarrow} 2^3S_1 \gamma \; (242.5)
		\stackrel{1.93\%}{\longrightarrow} \mu^+\mu^-$ & $8.1 \times 10^{-6}$ & 80 \\
		& $\stackrel{0.61\%}{\longrightarrow} 2^3D_3 $ $\gamma \;(72.7) 
		\stackrel{65.1\%}{\longrightarrow} 2^3P_2 \gamma \; (185.0)
		\stackrel{7.0\%}{\longrightarrow} 1^3S_1 \gamma \; (776.5)
		\stackrel{2.48\%}{\longrightarrow} \mu^+\mu^-$ & $6.9 \times 10^{-6}$ & 68 \\
		& $\stackrel{0.61\%}{\longrightarrow} 2^3D_3 $ $\gamma \;(72.7) 
		\stackrel{10\%}{\longrightarrow} 1^3P_2 \gamma \; (529.0)
		\stackrel{19.1\%}{\longrightarrow} 1^3S_1 \gamma \; (441.6)
		\stackrel{2.48\%}{\longrightarrow} \mu^+\mu^-$ & $2.9 \times 10^{-6}$ & 29 \\
		& $\stackrel{0.61\%}{\longrightarrow} 2^3D_3 $ $\gamma \;(72.7) 
		\stackrel{6.7\%}{\longrightarrow} 1^3F_4 \gamma \; (96.7)
		\stackrel{100\%}{\longrightarrow} 1^3D_3 \gamma \; (200.9)
		\stackrel{91.0\%}{\longrightarrow} 1^3P_2 \gamma \; (256.0)$ & & \\
		& $\hspace{8cm} 
		\stackrel{19.1\%}{\longrightarrow} 1^3S_1 \gamma \; (441.6)
		\stackrel{2.48\%}{\longrightarrow} \mu^+\mu^-$ & $1.8 \times 10^{-6}$ & 18 \\
		& $\stackrel{0.019\%}{\longrightarrow} 1^3D_3 \gamma \; (350.0)
		\stackrel{91.0\%}{\longrightarrow} 1^3P_2 \gamma \; (256.0)
		\stackrel{19.1\%}{\longrightarrow} 1^3S_1 \gamma \; (441.6)
		\stackrel{2.48\%}{\longrightarrow} \mu^+\mu^-$ & $8.2 \times 10^{-7}$ & 8 \\
		& $\stackrel{0.13\%}{\longrightarrow} 2^3D_2 $ $\gamma \;(78.7) 
		\stackrel{17\%}{\longrightarrow} 2^3P_2 \gamma \; (178.6)
		\stackrel{10.6\%}{\longrightarrow} 2^3S_1 \gamma \; (242.5)
		\stackrel{1.93\%}{\longrightarrow} \mu^+\mu^-$ & $4.5 \times 10^{-7}$ & 4 \\
		& $\stackrel{0.13\%}{\longrightarrow} 2^3D_2 $ $\gamma \;(78.7) 
		\stackrel{17\%}{\longrightarrow} 2^3P_2 \gamma \; (178.6)
		\stackrel{7.0\%}{\longrightarrow} 1^3S_1 \gamma \; (776.5)
		\stackrel{2.48\%}{\longrightarrow} \mu^+\mu^-$ & $3.8 \times 10^{-7}$ & 4 \\
		& $\stackrel{0.13\%}{\longrightarrow} 2^3D_2 $ $\gamma \;(78.7) 
		\stackrel{56.2\%}{\longrightarrow} 2^3P_1 \gamma \; (191.6)
		\stackrel{19.9\%}{\longrightarrow} 2^3S_1 \gamma \; (229.6)
		\stackrel{1.93\%}{\longrightarrow} \mu^+\mu^-$ & $2.8 \times 10^{-6}$ & 28 \\
		& $\stackrel{0.13\%}{\longrightarrow} 2^3D_2 $ $\gamma \;(78.7) 
		\stackrel{56.2\%}{\longrightarrow} 2^3P_1 \gamma \; (191.6)
		\stackrel{9.2\%}{\longrightarrow} 1^3S_1 \gamma \; (764.3)
		\stackrel{2.48\%}{\longrightarrow} \mu^+\mu^-$ & $1.7 \times 10^{-6}$ & 17 \\
		& $\stackrel{0.13\%}{\longrightarrow} 2^3D_2 $ $\gamma \;(78.7) 
		\stackrel{12\%}{\longrightarrow} 1^3P_1 \gamma \; (541.2)
		\stackrel{33.9\%}{\longrightarrow} 1^3S_1 \gamma \; (423.0)
		\stackrel{2.48\%}{\longrightarrow} \mu^+\mu^-$ & $1.3 \times 10^{-6}$ & 13 \\
		& $\stackrel{0.13\%}{\longrightarrow} 2^3D_2 $ $\gamma \;(78.7) 
		\stackrel{6.6\%}{\longrightarrow} 1^3F_3 \gamma \; (93.7)
		\stackrel{89.3\%}{\longrightarrow} 1^3D_2 \gamma \; (189.2) 
		\stackrel{74.7\%}{\longrightarrow} 1^3P_1 \gamma \; (267.3)$& & \\
		& $\hspace{8cm} 
		\stackrel{33.9\%}{\longrightarrow} 1^3S_1 \gamma \; (423.0)
		\stackrel{2.48\%}{\longrightarrow} \mu^+\mu^-$ & $4.8 \times 10^{-7}$ & 5 \\
$3^3P_1$ & $\stackrel{7.2\%}{\longrightarrow} 3^3S_1$ $\gamma \;(159.8) \stackrel{2.18\%}{\longrightarrow} \mu^+\mu^-$
			& $1.6\times 10^{-3}$ & 12,360 \\
		& $\stackrel{2.6\%}{\longrightarrow} 2^3S_1$ $\gamma \;(481.4) \stackrel{1.93\%}{\longrightarrow} \mu^+\mu^-$
			& $5.0\times 10^{-4}$ & 3,950 \\
		& $\stackrel{1.1\%}{\longrightarrow} 1^3S_1$ $\gamma \;(1003.0) \stackrel{2.48\%}{\longrightarrow} \mu^+\mu^-$
			& $2.7\times 10^{-4}$ & 2,150 \\
		& $\stackrel{0.75\%}{\longrightarrow}  1^3P_2 \pi \pi 
		\stackrel{19.1\%}{\longrightarrow} 1^3S_1 \gamma \; (441.6)
		\stackrel{2.48\%}{\longrightarrow} \mu^+\mu^-$ & $3.5 \times 10^{-5}$ & 278 \\
		& $\stackrel{0.48\%}{\longrightarrow}  1^3P_1 \pi \pi 
		\stackrel{33.9\%}{\longrightarrow} 1^3S_1 \gamma \; (423.0)
		\stackrel{2.48\%}{\longrightarrow} \mu^+\mu^-$ & $4.0 \times 10^{-5}$ & 317 \\
		& $\stackrel{0.40\%}{\longrightarrow} 2^3D_1$ $\gamma \;(74.7) 
		\stackrel{5.3\times 10^{-3}\%}{\longrightarrow} \mu^+\mu^-$
			& $2.1\times 10^{-7}$ & 2 \\
		& $\stackrel{0.94\%}{\longrightarrow} 2^3D_2 $ $\gamma \;(66.8) 
		\stackrel{17\%}{\longrightarrow} 2^3P_2 \gamma \; (178.6)
		\stackrel{10.6\%}{\longrightarrow} 2^3S_1 \gamma \; (242.5)
		\stackrel{1.93\%}{\longrightarrow} \mu^+\mu^-$ & $3.3 \times 10^{-6}$ & 26 \\
		& $\stackrel{0.94\%}{\longrightarrow} 2^3D_2 $ $\gamma \;(66.8) 
		\stackrel{17\%}{\longrightarrow} 2^3P_2 \gamma \; (178.6)
		\stackrel{7.0\%}{\longrightarrow} 1^3S_1 \gamma \; (776.5)
		\stackrel{2.48\%}{\longrightarrow} \mu^+\mu^-$ & $2.8 \times 10^{-6}$ & 22 \\
		& $\stackrel{0.94\%}{\longrightarrow} 2^3D_2 $ $\gamma \;(66.8) 
		\stackrel{56.2\%}{\longrightarrow} 2^3P_1 \gamma \; (191.6)
		\stackrel{19.9\%}{\longrightarrow} 2^3S_1 \gamma \; (229.6)
		\stackrel{1.93\%}{\longrightarrow} \mu^+\mu^-$ & $2.0 \times 10^{-5}$ & 158 \\
		& $\stackrel{0.94\%}{\longrightarrow} 2^3D_2 $ $\gamma \;(66.8) 
		\stackrel{56.2\%}{\longrightarrow} 2^3P_1 \gamma \; (191.6)
		\stackrel{9.2\%}{\longrightarrow} 1^3S_1 \gamma \; (764.3)
		\stackrel{2.48\%}{\longrightarrow} \mu^+\mu^-$ & $1.2 \times 10^{-5}$ & 95 \\
		& $\stackrel{0.94\%}{\longrightarrow} 2^3D_2 $ $\gamma \;(66.8) 
		\stackrel{2\%}{\longrightarrow} 1^3P_2 \gamma \; (522.8)
		\stackrel{19.1\%}{\longrightarrow} 1^3S_1 \gamma \; (441.6)
		\stackrel{2.48\%}{\longrightarrow} \mu^+\mu^-$ & $8.9 \times 10^{-7}$ & 7 \\
		& $\stackrel{0.94\%}{\longrightarrow} 2^3D_2 $ $\gamma \;(66.8) 
		\stackrel{12\%}{\longrightarrow} 1^3P_1 \gamma \; (541.2)
		\stackrel{33.9\%}{\longrightarrow} 1^3S_1 \gamma \; (423.0)
		\stackrel{2.48\%}{\longrightarrow} \mu^+\mu^-$ & $1.1 \times 10^{-5}$ & 87 \\
		& $\stackrel{0.94\%}{\longrightarrow} 2^3D_2 $ $\gamma \;(66.8) 
		\stackrel{6.6\%}{\longrightarrow} 1^3F_3 \gamma \; (93.7)
		\stackrel{89.3\%}{\longrightarrow} 1^3D_2 \gamma \; (189.2)
		\stackrel{74.7\%}{\longrightarrow} 1^3P_1 \gamma \; (267.3)$& & \\
		& $\hspace{8cm} 
		\stackrel{33.9\%}{\longrightarrow} 1^3S_1 \gamma \; (423.0)
		\stackrel{2.48\%}{\longrightarrow} \mu^+\mu^-$ & $3.5 \times 10^{-6}$ & 28 \\
		& $\stackrel{0.40\%}{\longrightarrow} 2^3D_1 $ $\gamma \;(74.7) 
		\stackrel{17\%}{\longrightarrow} 2^3P_1 \gamma \; (183.9)
		\stackrel{19.9\%}{\longrightarrow} 2^3S_1 \gamma \; (229.6)
		\stackrel{1.93\%}{\longrightarrow} \mu^+\mu^-$ & $2.6 \times 10^{-6}$ & 21 \\
		& $\stackrel{0.40\%}{\longrightarrow} 2^3D_1 $ $\gamma \;(74.7) 
		\stackrel{17\%}{\longrightarrow} 2^3P_1 \gamma \; (183.9)
		\stackrel{9.2\%}{\longrightarrow} 1^3S_1 \gamma \; (764.3)
		\stackrel{2.48\%}{\longrightarrow} \mu^+\mu^-$ & $1.6 \times 10^{-6}$ & 12 \\
		& $\stackrel{0.40\%}{\longrightarrow} 2^3D_1 $ $\gamma \;(74.7) 
		\stackrel{28.1\%}{\longrightarrow} 2^3P_0 \gamma \; (206.4)
		\stackrel{4.6\%}{\longrightarrow} 2^3S_1 \gamma \; (207.1)
		\stackrel{1.93\%}{\longrightarrow} \mu^+\mu^-$ & $1.0 \times 10^{-6}$ & 8 \\
		& $\stackrel{0.40\%}{\longrightarrow} 2^3D_1 $ $\gamma \;(74.7) 
		\stackrel{28.1\%}{\longrightarrow} 2^3P_0 \gamma \; (206.4)
		\stackrel{0.9\%}{\longrightarrow} 1^3S_1 \gamma \; (743.1)
		\stackrel{2.48\%}{\longrightarrow} \mu^+\mu^-$ & $2.5 \times 10^{-7}$ & 2 \\
		& $\stackrel{0.40\%}{\longrightarrow} 2^3D_1 $ $\gamma \;(74.7) 
		\stackrel{4.2\%}{\longrightarrow} 1^3F_2 \gamma \; (90.5)
		\stackrel{13.6\%}{\longrightarrow} 1^3D_2 \gamma \; (184.3) 
		\stackrel{74.7\%}{\longrightarrow} 1^3P_1 \gamma \; (267.3$& & \\
		& $\hspace{8cm} 
		\stackrel{33.9\%}{\longrightarrow} 1^3S_1 \gamma \; (423.0)
		\stackrel{2.48\%}{\longrightarrow} \mu^+\mu^-$ & $1.4 \times 10^{-7}$ & 1 \\
$3^3P_0$ & $\stackrel{0.054\%}{\longrightarrow}  1^3P_2 \pi \pi 
		\stackrel{19.1\%}{\longrightarrow} 1^3S_1 \gamma \; (441.6)
		\stackrel{2.48\%}{\longrightarrow} \mu^+\mu^-$ & $2.6 \times 10^{-6}$ & 9 \\

$3^1P_1$ & $\stackrel{4.3\%}{\longrightarrow} 1^1S_0 $ $\gamma \;(1081.0)
		\stackrel{100\%}{\longrightarrow} gg$ & $4.3 \times 10^{-2}$ & $3.1\times 10^5$ \\
		& $\stackrel{1.7\%}{\longrightarrow}  1^1P_1 \pi \pi 
		\stackrel{49\%}{\longrightarrow} 1^1S_0 $ $\gamma \;(488.3)
		\stackrel{100\%}{\longrightarrow} gg$ & $8.3 \times 10^{-3}$ & $6.0\times 10^4$ \\		
\hline \hline
\end{tabular}
\end{table*}

\subsubsection{The $4P$ and $5P$ states}

The $4P$ and $5P$ states are above $B\bar{B}$ threshold and have total widths 
of the order of tens of MeV so that BR's for radiative transitions are relatively small, 
${\cal O}(10^{-4})$.  However they undergo radiative transitions to $\Upsilon(nS)$ 
states which decay to $\mu^+ \mu^-$ offering a clean final state to study.  Our estimates 
for the expected number of events from these decay chains 
are given in Table~\ref{tab:4P}.  While the number of expected events is small it is possible
that our estimates are off by an order of magnitude.  Also, our estimates are based on
using LHCb event numbers and ATLAS and CMS with their different capabilities might be
able to observe more events.  
Thus, for completeness we include estimates for both the
$4P$ and $5P$ states.  The most important message is that the higher energy and luminosity
of LHC Run II could potentially observe some of the $4P$ and $5P$ states which would be an
important test of models and Lattice QCD results.

\begin{table*}[tp]
\caption{$4^3P_J$ and $5^3P_J$ Decay chains, branching ratios and event estimates 
for LHCb Run II. 
These are based on producing 
$9.7\times 10^6$ $\chi_{b2}(4P)$'s,
$7.4\times 10^6$  $\chi_{b1}(4P)$'s, $3.1\times 10^6$  $\chi_{b0}(4P)$'s,
$8.2\times 10^6$ $\chi_{b2}(5P)$'s,
$5.7 \times 10^6$  $\chi_{b1}(5P)$'s and $2.2\times 10^6$  $\chi_{b0}(5P)$'s
 as described in the text.
\label{tab:4P}}
\begin{tabular}{lllc} \hline \hline
Parent & Decay chain & Combined BR & Events  \\
\hline 
$4^3P_2$ & $\stackrel{0.012\%}{\longrightarrow} 3^3S_1$ $\gamma \;(433.9) \stackrel{2.18\%}{\longrightarrow} \mu^+\mu^-$
			& $2.6\times 10^{-6}$ & 25 \\
		& $\stackrel{1.4\times 10^{-3}\%}{\longrightarrow} 2^3S_1$ $\gamma \;(747.2) \stackrel{1.93\%}{\longrightarrow} \mu^+\mu^-$
			& $2.6\times 10^{-7}$ & 3 \\
		& $\stackrel{5.0\times 10^{-3}\%}{\longrightarrow} 1^3S_1$ $\gamma \;(1255.1) \stackrel{2.48\%}{\longrightarrow} \mu^+\mu^-$
			& $1.3\times 10^{-6}$ & 12 \\
$4^3P_1$ & $\stackrel{0.013\%}{\longrightarrow} 3^3S_1$ $\gamma \;(424.3) \stackrel{2.18\%}{\longrightarrow} \mu^+\mu^-$
			& $2.7\times 10^{-6}$ & 20 \\
		& $\stackrel{6.0\times 10^{-4}\%}{\longrightarrow} 2^3S_1$ $\gamma \;(737.9) \stackrel{1.93\%}{\longrightarrow} \mu^+\mu^-$
			& $1.1\times 10^{-7}$ & 1 \\
		& $\stackrel{3.3\times 10^{-3}\%}{\longrightarrow} 1^3S_1$ $\gamma \;(1003.0) \stackrel{2.48\%}{\longrightarrow} \mu^+\mu^-$
			& $8.2\times 10^{-7}$ & 6 \\
$4^3P_0$ & $\stackrel{6.4\times 10^{-3}\%}{\longrightarrow} 3^3S_1$ $\gamma \;(411.8) \stackrel{2.18\%}{\longrightarrow} \mu^+\mu^-$
			& $1.4\times 10^{-6}$ & 4 \\
		& $\stackrel{6.1\times 10^{-4}\%}{\longrightarrow} 1^3S_1$ $\gamma \;(1234.8) \stackrel{2.48\%}{\longrightarrow} \mu^+\mu^-$
			& $1.5\times 10^{-7}$ & 0.5 \\
$5^3P_2$ & $\stackrel{1\times 10^{-4}\%}{\longrightarrow} 3^3S_1$ $\gamma \;(646.8) \stackrel{2.18\%}{\longrightarrow} \mu^+\mu^-$
			& $2.1\times 10^{-8}$ & 0.2 \\
		& $\stackrel{2.5\times 10^{-3}\%}{\longrightarrow} 2^3S_1$ $\gamma \;(953.7) \stackrel{1.93\%}{\longrightarrow} \mu^+\mu^-$
			& $4.8\times 10^{-7}$ & 4 \\
		& $\stackrel{3.4\times 10^{-3}\%}{\longrightarrow} 1^3S_1$ $\gamma \;(1451.3) \stackrel{2.48\%}{\longrightarrow} \mu^+\mu^-$
			& $8.4\times 10^{-7}$ & 7 \\
$5^3P_1$ & $\stackrel{1.3\times 10^{-3}\%}{\longrightarrow} 2^3S_1$ $\gamma \;(946.4) \stackrel{1.93\%}{\longrightarrow} \mu^+\mu^-$
			& $2.5\times 10^{-7}$ & 1 \\
		& $\stackrel{1.4\times 10^{-3}\%}{\longrightarrow} 1^3S_1$ $\gamma \;(1444.4) \stackrel{2.48\%}{\longrightarrow} \mu^+\mu^-$
			& $3.5\times 10^{-7}$ & 2 \\
$5^3P_0$ & $\stackrel{3.0\times 10^{-4}\%}{\longrightarrow} 3^3S_1$ $\gamma \;(629.9) \stackrel{2.18\%}{\longrightarrow} \mu^+\mu^-$
			& $6.5\times 10^{-8}$ & 0.1 \\
		& $\stackrel{6.7\times 10^{-4}\%}{\longrightarrow} 2^3S_1$ $\gamma \;(937.3) \stackrel{1.93\%}{\longrightarrow} \mu^+\mu^-$
			& $1.3\times 10^{-7}$ & 0.3 \\
		& $\stackrel{4.1\times 10^{-4}\%}{\longrightarrow} 1^3S_1$ $\gamma \;(1435.7) \stackrel{2.48\%}{\longrightarrow} \mu^+\mu^-$
			& $1.0\times 10^{-7}$ & 0.2 \\
\hline \hline
\end{tabular}
\end{table*}

\subsubsection{The $1D$ states}

The production rate for the $D$-waves is significantly lower than that of $P$-waves.  This
is a consequence of the LDMEs 
going like $|R^{(l)}(0)|^2/M^{2l+2}$ so the
production cross section is significantly suppressed by the mass factor in
the denominator.  The mass factor
is simply a consequence of the dimensionality of the $l$th derivative
of the wavefunction and will increasingly suppress the cross sections going to higher
$l$ multiplets.

Nevertheless, it is expected that the $1D$ states can be produced in sufficient quantity
to be observed. There are three sources of the $D$ states; direct production and from
decay chains originating with the $3^3S_1$ and $2^3P_J$ states.  
The decay chains and estimated number of events from direct production are given in 
Table~\ref{tab:1D}.  We estimate that direct production of the $D$-waves will
yield  $\sim 100$ $\Upsilon_3(1D)$, $\sim 150$ 
$\Upsilon_2(1D)$ and $\sim 50$ $\Upsilon_1(1D)$ events. 
This is
roughly comparable to the number of $3P$ events observed by LHCb.  There are two 
differences.  The $3P$ events were comprised of one photon and a $\mu^+\mu^-$ pair
while the $1D$ events are generally comprised of two photons and a $\mu^+\mu^-$ pair making
them more difficult to reconstruct.  On the other hand, we can
estimate the photon energies fairly accurately because 
the $\Upsilon_{2}(1D)$ mass has been measured.  An exception is the $\Upsilon_1(1D)$ which can
decay directly to a $\mu^+\mu^-$ final state.  Unfortunately the BR appears to be too small
to produce a sufficient number of events to find this state in this channel.  On the other hand, 
as we have pointed out a number of times, our estimates can easily be off by an order
of magnitude.  In addition to direct production the $1D$ states will also be produced via 
transitions originating with $3^3S_1$ and $2^3P_J$.  In fact, cascades originating from the
$3^3S_1$ contribute
the largest number of events to the $2\gamma \mu^+\mu^-$ signal with roughly 
another 20\% originating from $2^3P_J$ production.
Specifically we expect $\sim 2700$ events in 
$1^3D_3 \to \gamma 1^3P_2\to \gamma \gamma 1^3S_1 \to \gamma\gamma \mu^+\mu^-$, 
$\sim 6500$  in $1^3D_2 \to \gamma 1^3P_1\to \gamma \gamma 1^3S_1 \to \gamma\gamma \mu^+\mu^-$
and $\sim 1200$ in
$1^3D_1 \to \gamma 1^3P_1\to \gamma \gamma 1^3S_1 \to \gamma\gamma \mu^+\mu^-$.  Other
decay chains will contribute to $1^3D_J$ production but they will have different
photon energies so we only give estimates for the decay chains with the largest statistics.

We have also included the decay chains $1^3D_J \to 1^3S_1 + \pi^+\pi^- \to \mu^+\mu^-$ 
and $1^1D_2 \to 1^1S_0 + \pi^+ \pi^- \to gg$ despite yielding few events as they offer 
a signal complementary to those involving photons.

Finally, we mention that the $\eta_{b2}$ will decay predominantly
via $\eta_{b2}(1D) \to \gamma h_b \to \gamma \gamma \eta_b(1S)$ with the $\eta_b(1S)$
decaying to hadrons.  We do not see how the $\eta_{b2}(1D)$ can be reconstructed 
given the hadronic final state but perhaps experimentalists will come up with a clever
approach that we have not considered.

\begin{table*}[tp]
\caption{The $1D$ Decay chains, branching ratios and event estimates 
for LHCb Run II and Belle II.  
For the $pp$ column these are based on producing 
$2.0\times 10^4$ $\Upsilon_{3}(1D)$'s and $\Upsilon_{2}(1D)$'s,
$1.6\times 10^4$ $\Upsilon_{1}(1D)$'s
 and 
$1.9\times 10^4$  $\eta_{b2}(1D)$'s as described in the text.  For the $e^+e^-$ 
column these are based on  $1.3\times 10^6$ $\Upsilon_{1}(1D)$'s produced assuming $\sigma =13$~pb
and 100~fb$^{-1}$ integrated luminosity.
\label{tab:1D}}
\begin{tabular}{lllrr} \hline \hline
Parent & Decay chain & Combined  & \multicolumn{2}{c}{Events}  \\
		&			&		BR		& $pp$  & $e^+e^-$ \\
\hline 
$1^3D_3$		&  $	\stackrel{91.0\%}{\longrightarrow} 1^3P_2 \gamma \; (256.0)
		\stackrel{19.1\%}{\longrightarrow} 1^3S_1 \gamma \; (441.6)
		\stackrel{2.48\%}{\longrightarrow} \mu^+\mu^-$ & $4.3 \times 10^{-3}$ & 87 & NA \\
				& $	\stackrel{0.74\%}{\longrightarrow} 1^3S_1 \pi^+ \pi^-
		\stackrel{2.48\%}{\longrightarrow} \mu^+\mu^-$ & $1.8 \times 10^{-4}$ & 4 & NA \\
$1^3D_2$ 	&	$\stackrel{22\%}{\longrightarrow} 1^3P_2 \gamma \; (248.4)
		\stackrel{19.1\%}{\longrightarrow} 1^3S_1 \gamma \; (441.6)
		\stackrel{2.48\%}{\longrightarrow} \mu^+\mu^-$ & $1.0 \times 10^{-3}$ & 20 & NA \\
			&	$\stackrel{74.7\%}{\longrightarrow} 1^3P_1 \gamma \; (267.3)
		\stackrel{33.9\%}{\longrightarrow} 1^3S_1 \gamma \; (423.0)
		\stackrel{2.48\%}{\longrightarrow} \mu^+\mu^-$ & $6.3 \times 10^{-3}$ & 126 & NA \\
			& $	\stackrel{0.66\%}{\longrightarrow} 1^3S_1 \pi^+ \pi^-
		\stackrel{2.48\%}{\longrightarrow} \mu^+\mu^-$ & $1.6 \times 10^{-4}$ & 3 & NA \\
$1^3D_1$ 	&	$	\stackrel{1.6\%}{\longrightarrow} 1^3P_2 \gamma \; (239.1)
		\stackrel{19.1\%}{\longrightarrow} 1^3S_1 \gamma \; (441.6)
		\stackrel{2.48\%}{\longrightarrow} \mu^+\mu^-$ & $7.6\times 10^{-5}$ & 1 & 99\\
			&	$\stackrel{28\%}{\longrightarrow} 1^3P_1 \gamma \; (258.0)
		\stackrel{33.9\%}{\longrightarrow} 1^3S_1 \gamma \; (423.0)
		\stackrel{2.48\%}{\longrightarrow} \mu^+\mu^-$ & $2.4 \times 10^{-3}$ & 38 & 3120 \\
			&	$\stackrel{47.1\%}{\longrightarrow} 1^3P_0 \gamma \; (290.5)
		\stackrel{1.76\%}{\longrightarrow} 1^3S_1 \gamma \; (391.1)
		\stackrel{2.48\%}{\longrightarrow} \mu^+\mu^-$ & $2.0 \times 10^{-4}$ & 3 & 260 \\
			& $	\stackrel{0.40\%}{\longrightarrow} 1^3S_1 \pi^+ \pi^-
		\stackrel{2.48\%}{\longrightarrow} \mu^+\mu^-$ & $9.9 \times 10^{-5}$ & 2 & 129 \\
		& $\stackrel{3.93\times 10^{-3}\%}{\longrightarrow} \mu^+\mu^-$
			& $3.9\times 10^{-5}$ & 1 & 50 \\
$1^1D_2$ 	&  	$\stackrel{91.5\%}{\longrightarrow} 1^1P_1 \gamma \; (262.5)
		\stackrel{49\%}{\longrightarrow} 1^1S_0 \gamma \; (488.3) \stackrel{100\%}{\longrightarrow} gg$
			 & 0.45 & 8550 & NA \\
			& $	\stackrel{1.3\%}{\longrightarrow} 1^1S_0 \pi^+ \pi^- \stackrel{100\%}{\longrightarrow} gg$ 
					& $1.3 \times 10^{-2}$ & 247 & NA \\

\hline \hline
\end{tabular}
\end{table*}

\subsubsection{The $2D$ and higher $D$-wave states}

For the $2D$ states in Table~\ref{tab:2D}, we focus on the simplest decay chains with 2 photons and a $\mu^+\mu^-$ pair
in the final state which are most likely to be reconstructed.  
To estimate the number of expected events, we include 
contributions to the $\gamma\gamma \mu^+\mu^-$ final state from direct 
production and from decay chains originating from $3^3P_J$ production.  There are a number
of possible decay chains but the ones with the largest expected statistics are
$\sim 125$ events for $2^3D_3 \to \gamma 2^3P_2 \to \gamma \gamma 2^3S_1 \to \gamma\gamma \mu^+\mu^-$,
$\sim 252$ events for $2^3D_2 \to \gamma 2^3P_1 \to \gamma \gamma 2^3S_1 \to \gamma\gamma \mu^+\mu^-$  
and 
$\sim 37$ events for $2^3D_1 \to \gamma 2^3P_1 \to \gamma \gamma 2^3S_1 \to \gamma\gamma \mu^+\mu^-$.
In all cases more events are expected from $3^3P_J$ production than from direct $2D$ 
production.

For the $3D$ and $4D$ states we expect that each member of both multiplets will have 
of the order of $10^4$ produced.  However the predicted widths are ${\cal O}$(100 MeV)
so the BR's for radiative transitions will be small.  Thus,  we only expect that
they can be observed in $B\bar{B}$, $B\bar{B}^*$ or $B^*\bar{B}^*$ final states if they
can be reconstructed with high enough efficiencies and separated from backgrounds.

\begin{table*}[tp]
\caption{The $2D$ Decay chains, branching ratios and event estimates 
for LHCb Run II.  
These are based on producing 
$3.4\times 10^4$ $\Upsilon_{3}(2D)$'s, $3.1\times 10^4$ $\Upsilon_{2}(2D)$'s,
$2.4\times 10^4$ $\Upsilon_{1}(2D)$'s
 and 
$2.9\times 10^4$  $\eta_{b2}(2D)$'s as described in the text.
\label{tab:2D}}
\begin{tabular}{lllc} \hline \hline
Parent & Decay chain & Combined BR & Events  \\
\hline 
$2^3D_3$		&  $\stackrel{65.1\%}{\longrightarrow} 2^3P_2 \gamma \; (185.0)
		\stackrel{10.6\%}{\longrightarrow} 2^3S_1 \gamma \; (242.5)
		\stackrel{1.93\%}{\longrightarrow} \mu^+\mu^-$ & $1.3 \times 10^{-3}$ & 45 \\
				&  $\stackrel{65.1\%}{\longrightarrow} 2^3P_2 \gamma \; (185.0)
		\stackrel{7.0\%}{\longrightarrow} 1^3S_1 \gamma \; (776.5)
		\stackrel{2.48\%}{\longrightarrow} \mu^+\mu^-$ & $1.1 \times 10^{-3}$ & 38 \\
				&  $\stackrel{10.3\%}{\longrightarrow} 1^3P_2 \gamma \; (529.0)
		\stackrel{19.1\%}{\longrightarrow} 1^3S_1 \gamma \; (441.6)
		\stackrel{2.48\%}{\longrightarrow} \mu^+\mu^-$ & $4.9 \times 10^{-4}$ & 16 \\
				&$\stackrel{6.7\%}{\longrightarrow} 1^3F_4 \gamma \; (96.7)
		\stackrel{100\%}{\longrightarrow} 1^3D_3 \gamma \; (200.9)
		\stackrel{91.0\%}{\longrightarrow} 1^3P_2 \gamma \; (256.0)
		\stackrel{19.1\%}{\longrightarrow} 1^3S_1 \gamma \; (441.6)
		\stackrel{2.48\%}{\longrightarrow} \mu^+\mu^-$ & $2.9 \times 10^{-4}$ & 10 \\
$2^3D_2$ 	&	$\stackrel{17\%}{\longrightarrow} 2^3P_2 \gamma \; (178.6)
		\stackrel{10.6\%}{\longrightarrow} 2^3S_1 \gamma \; (242.5)
		\stackrel{1.93\%}{\longrightarrow} \mu^+\mu^-$ & $3.5 \times 10^{-4}$ & 11 \\		
			& $\stackrel{17\%}{\longrightarrow} 2^3P_2 \gamma \; (178.6)
		\stackrel{7.0\%}{\longrightarrow} 1^3S_1 \gamma \; (776.5)
		\stackrel{2.48\%}{\longrightarrow} \mu^+\mu^-$ & $3.0 \times 10^{-4}$ & 9 \\
			&$\stackrel{56.2\%}{\longrightarrow} 2^3P_1 \gamma \; (191.6)
		\stackrel{19.9\%}{\longrightarrow} 2^3S_1 \gamma \; (229.6)
		\stackrel{1.93\%}{\longrightarrow} \mu^+\mu^-$ & $2.1 \times 10^{-3}$ & 66 \\
			& $	\stackrel{56.2\%}{\longrightarrow} 2^3P_1 \gamma \; (191.6)
		\stackrel{9.2\%}{\longrightarrow} 1^3S_1 \gamma \; (764.3)
		\stackrel{2.48\%}{\longrightarrow} \mu^+\mu^-$ & $1.3 \times 10^{-3}$ & 39 \\
			&  $\stackrel{12\%}{\longrightarrow} 1^3P_1 \gamma \; (541.2)
		\stackrel{33.9\%}{\longrightarrow} 1^3S_1 \gamma \; (423.0)
		\stackrel{2.48\%}{\longrightarrow} \mu^+\mu^-$ & $1.0 \times 10^{-3}$ & 31 \\
			&  $\stackrel{6.6\%}{\longrightarrow} 1^3F_3 \gamma \; (93.7)
		\stackrel{89.3\%}{\longrightarrow} 1^3D_2 \gamma \; (189.2)
		\stackrel{74.7\%}{\longrightarrow} 1^3P_1 \gamma \; (267.3)
		\stackrel{33.9\%}{\longrightarrow} 1^3S_1 \gamma \; (423.0)
		\stackrel{2.48\%}{\longrightarrow} \mu^+\mu^-$ & $3.7 \times 10^{-4}$ & 11 \\
$2^3D_1$	&	$\stackrel{5.28\times 10^{-3}\%}{\longrightarrow} \mu^+\mu^-$
			& $5.3\times 10^{-5}$ & 1 \\
			&  $\stackrel{17\%}{\longrightarrow} 2^3P_1 \gamma \; (183.9)
		\stackrel{19.9\%}{\longrightarrow} 2^3S_1 \gamma \; (229.6)
		\stackrel{1.93\%}{\longrightarrow} \mu^+\mu^-$ & $6.5 \times 10^{-4}$ & 16 \\
			& $\stackrel{28.1\%}{\longrightarrow} 2^3P_0 \gamma \; (206.4)
		\stackrel{4.6\%}{\longrightarrow} 2^3S_1 \gamma \; (207.1)
		\stackrel{1.93\%}{\longrightarrow} \mu^+\mu^-$ & $2.5 \times 10^{-4}$ & 6 \\
		& $\stackrel{28.1\%}{\longrightarrow} 2^3P_0 \gamma \; (206.4)
		\stackrel{0.9\%}{\longrightarrow} 1^3S_1 \gamma \; (743.1)
		\stackrel{2.48\%}{\longrightarrow} \mu^+\mu^-$ & $6.3 \times 10^{-5}$ & 2 \\
		& $\stackrel{4.2\%}{\longrightarrow} 1^3F_2 \gamma \; (90.5)
		\stackrel{13.6\%}{\longrightarrow} 1^3D_2 \gamma \; (184.3)
		\stackrel{74.7\%}{\longrightarrow} 1^3P_1 \gamma \; (267.3)
		\stackrel{33.9\%}{\longrightarrow} 1^3S_1 \gamma \; (423.0)
		\stackrel{2.48\%}{\longrightarrow} \mu^+\mu^-$ & $3.6 \times 10^{-5}$ & 1 \\
$2^1D_2$ 	&  	$\stackrel{67.1\%}{\longrightarrow} 2^1P_1 \gamma \; (188.3)
		\stackrel{48\%}{\longrightarrow} 2^1S_0 \gamma \; (257.7) \stackrel{100\%}{\longrightarrow} gg$
			 & 0.32 & 9280 \\
			&  	$\stackrel{67.1\%}{\longrightarrow} 2^1P_1 \gamma \; (188.3)
		\stackrel{22\%}{\longrightarrow} 1^1S_0 \gamma \; (825.8) \stackrel{100\%}{\longrightarrow} gg$
			 & 0.15 & 4350 \\
			&  	$\stackrel{12\%}{\longrightarrow} 1^1P_1 \gamma \; (536.5)
		\stackrel{49\%}{\longrightarrow} 1^1S_0 \gamma \; (488.3) \stackrel{100\%}{\longrightarrow} gg$
			 & 0.059 & 1710 \\
\hline \hline
\end{tabular}
\end{table*}

\subsubsection{The $nF$ states}

The production rate decreases quite dramatically for states with larger $L$ as a 
consequence of our estimates which use Eq.~\ref{eqn:nrqcd} where the NR approximation
of the cross section goes like the $l$th derivative of the wavefunction at the origin 
with the corresponding mass in the denominator needed for dimensional reasons.  We only expect
that $\sim 200$ for each of the $1F$ states will be produced.  Once the BR's for the decay
chains are included we expect that only 1 or less events will result. Considering experimental
challenges in making these measurements we do not expect that the $1F$ states will
be observed from direct production.  Nevertheless, we include the dominant decay chains in
Table~\ref{tab:1F} for completeness.

We do expect a small number of $1F$ states; $\sim 28$ $1^3F_4$'s, $\sim 44$ $1^3F_3$'s and $\sim 2$ $1^3F_2$,  
to be produced via radiative transitions 
originating with $3^3P_J$ and $2^3D_J$ states.  Since there are 3 $\gamma$'s in the final state
it is unlikely that the $1F$ states will be observed in hadron production. 

We expect the $2F$ and $3F$ multiplets to be even more challenging to observe using 
radiative transition decay chains primarily because they are above $B\bar{B}$ threshold 
and are therefore broader, ranging from 2.8~MeV for the $2^3F_4$ state to 88.6~MeV for
the $2^3F_2$ state.  These states are very close to $B\bar{B}$ and $B\bar{B}^*$ threshold
and therefore are very sensitive to available phase space.  If our mass predictions are too high
it is possible that the total widths could be significantly smaller leading to significantly
larger BR's to the decay chains we have been focusing on.  Nevertheless, given our 
expectations for the $1F$ states we do not feel it is likely that they would be discovered
using radiative transitions.  If the $B$ and $B^*$ mesons can be observed with high 
efficiencies the excited $F$-wave states might be observed in $B\bar{B}$ and $B\bar{B}^*$ 
final states.

The $3F$ multiplets are sufficiently above $B\bar{B}$ and $B\bar{B}^*$ threshold that
they are much broader.  The only possibility that they might be observed would be in
$B\bar{B}$, $B\bar{B}^*$ or $B^*\bar{B}^*$ final states.

\begin{table*}[tp]
\caption{The $1F$ Decay chains, branching ratios and event estimates 
for LHCb Run II.  
These are based on producing 166 $\chi_{b4}(1F)$'s,
163 $\chi_{b3}(1F)$'s, 147 $\chi_{b2}(1F)$'s and
158 $h_{b3}(1F)$'s as described in the text.
\label{tab:1F}}
\begin{tabular}{lllc} \hline \hline
Parent & Decay chain & Combined BR & Events  \\
\hline 
$1^3F_4$	&  $\stackrel{100\%}{\longrightarrow} 1^3D_3 \gamma \; (184.9)
		\stackrel{91.0\%}{\longrightarrow} 1^3P_2 \gamma \; (256.0)
		\stackrel{19.1\%}{\longrightarrow} 1^3S_1 \gamma \; (441.6)
		\stackrel{2.48\%}{\longrightarrow} \mu^+\mu^-$ & $4.3 \times 10^{-3}$ & 0.7 \\
$1^3F_3$ 	&	$\stackrel{89.3\%}{\longrightarrow} 1^3D_2 \gamma \; (189.2)
		\stackrel{74.7\%}{\longrightarrow} 1^3P_1 \gamma \; (267.3)
		\stackrel{33.9\%}{\longrightarrow} 1^3S_1 \gamma \; (423.0)
		\stackrel{2.48\%}{\longrightarrow} \mu^+\mu^-$ & $5.6 \times 10^{-3}$ & 1 \\
$1^3F_2$ 	&	$\stackrel{82.4\%}{\longrightarrow} 1^3D_1 \gamma \; (194.1)
		\stackrel{28\%}{\longrightarrow} 1^3P_1 \gamma \; (258.0)
		\stackrel{33.9\%}{\longrightarrow} 1^3S_1 \gamma \; (423.0)
		\stackrel{2.48\%}{\longrightarrow} \mu^+\mu^-$ & $1.9 \times 10^{-3}$ & 0.3 \\
$1^1F_3$ 	&  	$\stackrel{100\%}{\longrightarrow} 1^1D_2 \gamma \; (189.2)
		\stackrel{91.5\%}{\longrightarrow} 1^1P_1 \gamma \; (262.5)
		\stackrel{49\%}{\longrightarrow} 1^1S_0 \gamma \; (488.3) \stackrel{100\%}{\longrightarrow} gg$
			 & 0.45 & 71 \\
\hline \hline
\end{tabular}
\end{table*}

\subsubsection{The $nG$ states}

Even more so than the $F$-waves, the $G$-wave production cross sections are highly
suppressed so we expect only ${\cal O}$(1) meson will be produced for each of the states in 
the $G$-wave multiplets in Run II which is far too small a number to have any hope
of being seen.

\subsection{At $e^+e^-$ Colliders}
\label{sec:eecolliders}

The question we wish to address in this section is whether previously unobserved states
can be observed in $e^+e^-$ collisions and outline how to do so.  We will focus 
on $1^{--}$ states, the $n^3S_1$ and $n^3D_1$ states, since only $1^{--}$ states can
be produced directly in $e^+e^-$ collisions.  To estimate the number of events 
requires a detailed Monte Carlo study that includes beam spread and initial state radiation
which is beyond the scope of this work.  Instead we will use cross sections derived from 
Belle and BaBar events along with integrated luminosities suggested for Belle II.  

We start with the $\Upsilon (3S)$ where we use the cross section of 4~nb based
on Belle and BaBar measurements \cite{Bevan:2014iga}. 
An integrated luminosity of 250~fb$^{-1}$ would yield 
$10^9$ $\Upsilon (3S)$'s, about a factor of 7 1/2 times the combined Belle-BaBar dataset.  
We give estimates of number of events in Table~\ref{tab:3S}  
based on this but the numbers
can be easily rescaled for other integrated luminosities.  We expect that ${\cal O}(10^3)$
$1^3D_J$'s will be produced via radiative transitions for each $J$ value, which should be 
sufficient to identify each
of the $1^3D_J$ states and measure and compare their masses to theoretical 
predictions.  

The expectations are to accumulate several ab$^{-1}$ of integrated luminosity 
at the $\Upsilon (4S)$.  The cross section of $\sim 1$~nb would yield several billion
$\Upsilon (4S)$'s depending on the eventual integrated luminosity.  We estimate
the number of events given in Table~\ref{tab:4S} assuming that
$10^{10}$ $\Upsilon (4S)$'s will be produced based on 10~ab$^{-1}$ of integrated luminosity
but these numbers can
easily be rescaled assuming different values of integrated luminosities.  We expect 
sufficient events in the chains that proceed via the $3^3P_2$ and $3^3P_1$ so that these
states should be observed in radiative decays of the $\Upsilon (4S)$.  We don't expect that
the $3^3P_0$ will be observed in this manner.  Another interesting possibility for studying
the $3P$ states via radiative transitions from the $\Upsilon (4S)$ utilizes hadronic transitions
from the $\Upsilon(3S)$ and $\Upsilon(2S)$ to $\Upsilon (2S)$ or $\Upsilon (1S)$ in the decay 
chain.  This is experimentally very clean and would yield some tens of events for $3^3P_2$ and
$3^3P_1$ intermediate states but only ${\cal O} (1)$ event for the $3^3P_0$.  This
might be sufficient to resolve some of the $J$ states.
It would have been interesting to be
able to observe the $2D$ states in radiative transitions originating with the $\Upsilon (4S)$
but this does not appear to be likely.  Of all the $2^3D_J$ states the decay chain
proceeding via the $2^3D_1$ state will have the largest statistics although not sufficient
to be observed considering the 4 photons in the decay chain.  

It is unlikely that sitting on the $\Upsilon (5S)$ or $\Upsilon (6S)$ would produce enough
bottomonium states via radiative transitions to be seen.
This stems from the $\Upsilon (5S)$ and $\Upsilon (6S)$ having large total widths 
which leads to small BR's to $e^+e^-$ and subsequently smaller production cross sections
than the lower mass $^3S_1$ states.  This also results in small BR's for photon decays.  
For the  $\Upsilon (5S)$ the Belle collaboration measures its $e^+e^-$ production 
cross section to be 0.3~nb \cite{Bevan:2014iga}.  The radiative transition with the
largest BR is to $4^3P_2$ with BR~$=1.6 \times 10^{-4}$ yielding 
$\sim 5 \times 10^{3}$ $4^3P_2$'s per 100~fb$^{-1}$.  From Table~\ref{tab:4P} the combined BR
for representative $4^3P_2$ decay chains are ${\cal O}(10^{-6})$ so it is unlikely that 
$b\bar{b}$ states could be seen in decays originating from the $\Upsilon (5S)$ in $e^+e^-$ 
collisions.  We repeat the exercise for the $\Upsilon (6S)$.  We assume an $e^+e^-$ production 
cross section similar to that of the $\Upsilon (5S)$ although in fact it should be smaller
due to its smaller BR to $e^+e^-$.  For the $\Upsilon (6S)$ the largest BR is expected to be
to the $3^3P_J$ states, ${\cal O}(10^{-5})$.  However, in this case the combined BR's for
some interesting $3P$ decay chains are ${\cal O}(10^{-4})$.  Putting this together we expect
$\sim 0.5$ event/100~fb$^{-1}$ via the $3^3P_2$ intermediate state.  While still below an event 
rate needed to study these states it does offer some hope given the uncertainties in
our assumption for the production cross section and with higher statistics.

Observation of the $n^3D_1$ states in $e^+e^-$ collisions is interesting, even more so
if they can be produced in sufficient numbers to see previously unobserved excited bottomonium
states in their decays.  Unfortunately the production cross section is proportional to the
BR to $e^+e^-$  
which is roughly three orders of magnitude lower than for the $S$-wave states. The
small number of signal events will also make it challenging to see the $n^3D_1$
states over backgrounds.  
Given these caveats we make a rough estimate of the  
number of $n^3D_1$ produced 
by multiplying the ratio of $nD/2S$ BR's to $e^+e^-$ times the $2^3S_1$ cross section
to obtain the $n^3D_1$ production cross section.  For the $1^3D_1$ this gives  
$\sim 13$~pb. For 100~fb$^{-1}$
of integrated luminosity we expect $\sim 1.3\times 10^6$ $1^3D_1$'s to be produced.  The expected
number of events for several $1^3D_1$ decay chains is given in Table~\ref{tab:1D}.  This 
will yield 50 events in $1^3D_1\to \mu^+\mu^-$ but many more in radiative decay chains
via intermediate $1P$ states.

In the same way as we estimated $\sigma (e^+e^-\to 1^3D_1)$ we obtain
 $\sigma(e^+e^- \to 2^3D_1) \sim 18$~pb.  For 100~fb$^{-1}$
of integrated luminosity we expect $\sim 2\times 10^6$ $2^3D_1$'s produced.  We use
this number to estimate the number of events in the different decay chains 
given in Table~\ref{tab:2Dee}.  With these assumptions it should be possible to observe the
$2^3D_1$.  While the $\mu^+\mu^-$ mode may be the cleanest, other modes proceeding via
radiative transition decay chains offer higher statistics, in particular decay chains
proceeding by
$1P$ and $2P$ intermediate states.  More interesting is the possibility 
of observing the $1^3F_2$ for the first time.  The decay chain with the highest statistics
proceeds via $1^3F_2 \to 1^3D_1 \to 1^3P_1$ which would also yield information on
the $1^3D_1$ state.   One can easily scale our projected events up or down to reflect
actual cross sections and integrated luminosities.

The $3D$ and $4D$ are above 
$B\bar{B}$ threshold so have total widths
${\cal O}(100 \hbox{ MeV})$ resulting in BR $\sim 10^{-8}$ and a cross section $\sim 8$~fb.
For reasonable integrated luminosities and using the BR's for the $3^3D_1$ and $4^3D_1$ 
given in Tables \ref{tab:sum_3Dwave} and \ref{tab:sum_4Dwave2} we do not expect that the
$3^3D_1$ and $4^3D_1$ states can be observed in $e^+e^-$ collisions, 
as suggested by the results given in Tables \ref{tab:3Dee} and \ref{tab:4Dee}.

\begin{table*}[tp]
\caption{The $2^3D_1$ Decay chains, branching ratios and event estimates 
for Belle II.  
The event
numbers are based on $2\times 10^6$ $\Upsilon_{3}(2D)$'s produced assuming $\sigma =18$~pb
and 100~fb$^{-1}$ integrated luminosity.
\label{tab:2Dee}}
\begin{tabular}{lllc} \hline \hline
Parent & Decay chain & Combined BR & Events  \\
\hline 
$2^3D_1$	&	$\stackrel{5.28\times 10^{-3}\%}{\longrightarrow} \mu^+\mu^-$
			& $5.3\times 10^{-5}$ & 106 \\
		&  $\stackrel{1\%}{\longrightarrow} 2^3P_2 \gamma \; (170.9)
		\stackrel{10.6\%}{\longrightarrow} 2^3S_1 \gamma \; (242.5)
		\stackrel{1.93\%}{\longrightarrow} \mu^+\mu^-$ & $2.0 \times 10^{-5}$ & 40 \\
		&  $\stackrel{1\%}{\longrightarrow} 2^3P_2 \gamma \; (170.9)
		\stackrel{7.0\%}{\longrightarrow} 1^3S_1 \gamma \; (776.5)
		\stackrel{2.48\%}{\longrightarrow} \mu^+\mu^-$ & $1.7 \times 10^{-5}$ & 34 \\
		&  $\stackrel{0.05\%}{\longrightarrow} 1^3P_2 \gamma \; (515.4)
		\stackrel{19.1\%}{\longrightarrow} 1^3S_1 \gamma \; (441.6)
		\stackrel{2.48\%}{\longrightarrow} \mu^+\mu^-$ & $2.4 \times 10^{-6}$ & 5 \\
		&  $\stackrel{17.2\%}{\longrightarrow} 2^3P_1 \gamma \; (183.9)
		\stackrel{19.9\%}{\longrightarrow} 2^3S_1 \gamma \; (229.6)
		\stackrel{1.93\%}{\longrightarrow} \mu^+\mu^-$ & $6.6 \times 10^{-4}$ & 1320 \\
		&  $\stackrel{17.2\%}{\longrightarrow} 2^3P_1 \gamma \; (183.9)
		\stackrel{9.2\%}{\longrightarrow} 1^3S_1 \gamma \; (764.3)
		\stackrel{2.48\%}{\longrightarrow} \mu^+\mu^-$ & $3.9 \times 10^{-4}$ & 780 \\
		&  $\stackrel{2.4\%}{\longrightarrow} 1^3P_1 \gamma \; (533.8)
		\stackrel{33.9\%}{\longrightarrow} 1^3S_1 \gamma \; (423.0)
		\stackrel{2.48\%}{\longrightarrow} \mu^+\mu^-$ & $2.0 \times 10^{-4}$ & 400 \\
		& $\stackrel{28.1\%}{\longrightarrow} 2^3P_0 \gamma \; (206.4)
		\stackrel{4.6\%}{\longrightarrow} 2^3S_1 \gamma \; (207.1)
		\stackrel{1.93\%}{\longrightarrow} \mu^+\mu^-$ & $2.5 \times 10^{-4}$ & 500 \\
		& $\stackrel{28.1\%}{\longrightarrow} 2^3P_0 \gamma \; (206.4)
		\stackrel{0.9\%}{\longrightarrow} 1^3S_1 \gamma \; (743.1)
		\stackrel{2.48\%}{\longrightarrow} \mu^+\mu^-$ & $6.3 \times 10^{-5}$ & 126 \\
		& $\stackrel{7.7\%}{\longrightarrow} 1^3P_0 \gamma \; 565.4)
		\stackrel{1.76\%}{\longrightarrow} 1^3S_1 \gamma \; (391.1)
		\stackrel{2.48\%}{\longrightarrow} \mu^+\mu^-$ & $3.4 \times 10^{-5}$ & 68 \\
		& $\stackrel{4.2\%}{\longrightarrow} 1^3F_2 \gamma \; (90.5)
		\stackrel{14\%}{\longrightarrow} 1^3D_2 \gamma \; (184.3)
		\stackrel{74.7\%}{\longrightarrow} 1^3P_1 \gamma \; (267.3)
		\stackrel{33.9\%}{\longrightarrow} 1^3S_1 \gamma \; (423.0)
		\stackrel{2.48\%}{\longrightarrow} \mu^+\mu^-$ & $3.7 \times 10^{-5}$ & 74 \\
		& $\stackrel{4.2\%}{\longrightarrow} 1^3F_2 \gamma \; (90.5)
		\stackrel{14\%}{\longrightarrow} 1^3D_2 \gamma \; (184.3)
		\stackrel{22\%}{\longrightarrow} 1^3P_2 \gamma \; (248.4)
		\stackrel{19.1\%}{\longrightarrow} 1^3S_1 \gamma \; (441.6)
		\stackrel{2.48\%}{\longrightarrow} \mu^+\mu^-$ & $6.1 \times 10^{-6}$ & 12 \\
		& $\stackrel{4.2\%}{\longrightarrow} 1^3F_2 \gamma \; (90.5)
		\stackrel{0.35\%}{\longrightarrow} 1^3D_3 \gamma \; (176.5)
		\stackrel{91.0\%}{\longrightarrow} 1^3P_2 \gamma \; (256.0)
		\stackrel{19.1\%}{\longrightarrow} 1^3S_1 \gamma \; (441.6)
		\stackrel{2.48\%}{\longrightarrow} \mu^+\mu^-$ & $6.3 \times 10^{-7}$ & 1 \\
		& $\stackrel{4.2\%}{\longrightarrow} 1^3F_2 \gamma \; (90.5)
		\stackrel{82.4\%}{\longrightarrow} 1^3D_1 \gamma \; (194.1)
		\stackrel{1.6\%}{\longrightarrow} 1^3P_2 \gamma \; (239.1)
		\stackrel{19.1\%}{\longrightarrow} 1^3S_1 \gamma \; (441.6)
		\stackrel{2.48\%}{\longrightarrow} \mu^+\mu^-$ & $2.6 \times 10^{-6}$ & 5 \\
		& $\stackrel{4.2\%}{\longrightarrow} 1^3F_2 \gamma \; (90.5)
		\stackrel{82.4\%}{\longrightarrow} 1^3D_1 \gamma \; (194.1)
		\stackrel{28\%}{\longrightarrow} 1^3P_1 \gamma \; (258.0)
		\stackrel{33.9\%}{\longrightarrow} 1^3S_1 \gamma \; (423.0)
		\stackrel{2.48\%}{\longrightarrow} \mu^+\mu^-$ & $8.1 \times 10^{-5}$ & 162 \\
		& $\stackrel{4.2\%}{\longrightarrow} 1^3F_2 \gamma \; (90.5)
		\stackrel{82.4\%}{\longrightarrow} 1^3D_1 \gamma \; (194.1)
		\stackrel{47.1\%}{\longrightarrow} 1^3P_0 \gamma \; (290.5)
		\stackrel{1.76\%}{\longrightarrow} 1^3S_1 \gamma \; (391.1)
		\stackrel{2.48\%}{\longrightarrow} \mu^+\mu^-$ & $7.1 \times 10^{-6}$ & 14 \\
\hline \hline
\end{tabular}
\end{table*}

\begin{table*}[tp]
\caption{The $3^3D_1$ Decay chains and combined branching ratios.  These are small enough that we do not expect enough events at Belle II for this state to be observed.
\label{tab:3Dee}}
\begin{tabular}{lll} \hline \hline
Parent & Decay chain & Combined BR  \\
\hline 
$3^3D_1$	&	$\stackrel{2.30\times 10^{-6}\%}{\longrightarrow} \mu^+\mu^-$
			& $2.3\times 10^{-8}$  \\
		&  $\stackrel{5.6\times 10^{-4}\%}{\longrightarrow} 3^3P_2 \gamma \; (168.6)
		\stackrel{3.8\%}{\longrightarrow} 3^3S_1 \gamma \; (171.6)
		\stackrel{2.18\%}{\longrightarrow} \mu^+\mu^-$ & $4.6 \times 10^{-9}$  \\
		&  $\stackrel{5.6\times 10^{-4}\%}{\longrightarrow} 3^3P_2 \gamma \; (168.6)
		\stackrel{1.8\%}{\longrightarrow} 2^3S_1 \gamma \; (492.9)
		\stackrel{1.93\%}{\longrightarrow} \mu^+\mu^-$ & $1.9 \times 10^{-9}$  \\
		&  $\stackrel{9.2\times 10^{-3}\%}{\longrightarrow} 3^3P_1 \gamma \; (180.4)
		\stackrel{7.2\%}{\longrightarrow} 3^3S_1 \gamma \; (159.8)
		\stackrel{2.18\%}{\longrightarrow} \mu^+\mu^-$ & $1.4 \times 10^{-7}$  \\
		&  $\stackrel{9.2\times 10^{-3}\%}{\longrightarrow} 3^3P_1 \gamma \; (180.4)
		\stackrel{2.6\%}{\longrightarrow} 2^3S_1 \gamma \; (481.4)
		\stackrel{1.93\%}{\longrightarrow} \mu^+\mu^-$ & $4.6 \times 10^{-8}$  \\
		&  $\stackrel{9.2\times 10^{-3}\%}{\longrightarrow} 3^3P_1 \gamma \; (180.4)
		\stackrel{1.1\%}{\longrightarrow} 1^3S_1 \gamma \; (1003.0)
		\stackrel{2.48\%}{\longrightarrow} \mu^+\mu^-$ & $2.5 \times 10^{-8}$  \\
		&  $\stackrel{1.35\times 10^{-2}\%}{\longrightarrow} 3^3P_0 \gamma \; (196.2)
		\stackrel{0.31\%}{\longrightarrow} 3^3S_1 \gamma \; (144.0)
		\stackrel{2.18\%}{\longrightarrow} \mu^+\mu^-$ & $9.1 \times 10^{-9}$  \\
		&  $\stackrel{1.35\times 10^{-2}\%}{\longrightarrow} 3^3P_0 \gamma \; (196.2)
		\stackrel{0.077\%}{\longrightarrow} 2^3S_1 \gamma \; (466.2)
		\stackrel{1.93\%}{\longrightarrow} \mu^+\mu^-$ & $2.0 \times 10^{-9}$  \\
		&  $\stackrel{1.35\times 10^{-2}\%}{\longrightarrow} 3^3P_0 \gamma \; (196.2)
		\stackrel{0.01\%}{\longrightarrow} 1^3S_1 \gamma \; (988.5)
		\stackrel{2.48\%}{\longrightarrow} \mu^+\mu^-$ & $3.3 \times 10^{-10}$  \\
		&  $\stackrel{2\times 10^{-5}\%}{\longrightarrow} 2^3P_2 \gamma \; (420.4)
		\stackrel{10.6\%}{\longrightarrow} 2^3S_1 \gamma \; (242.5)
		\stackrel{1.93\%}{\longrightarrow} \mu^+\mu^-$ & $4.1 \times 10^{-10}$  \\
		&  $\stackrel{2\times 10^{-5}\%}{\longrightarrow} 2^3P_2 \gamma \; (196.2)
		\stackrel{7.0\%}{\longrightarrow} 1^3S_1 \gamma \; (776.5)
		\stackrel{2.48\%}{\longrightarrow} \mu^+\mu^-$ & $3.5 \times 10^{-10}$  \\
		&  $\stackrel{9.3\times 10^{-4}\%}{\longrightarrow} 2^3P_1 \gamma \; (433.8)
		\stackrel{19.9\%}{\longrightarrow} 2^3S_1 \gamma \; (229.6)
		\stackrel{1.93\%}{\longrightarrow} \mu^+\mu^-$ & $3.6 \times 10^{-8}$ \\
		&  $\stackrel{9.3\times 10^{-4}\%}{\longrightarrow} 2^3P_1 \gamma \; (433.8)
		\stackrel{9.2\%}{\longrightarrow} 1^3S_1 \gamma \; (764.3)
		\stackrel{2.48\%}{\longrightarrow} \mu^+\mu^-$ & $2.1 \times 10^{-8}$  \\
		&  $\stackrel{1.3\times 10^{-4}\%}{\longrightarrow} 1^3P_1 \gamma \; (533.8)
		\stackrel{33.9\%}{\longrightarrow} 1^3S_1 \gamma \; (423.0)
		\stackrel{2.48\%}{\longrightarrow} \mu^+\mu^-$ & $1.1 \times 10^{-8}$  \\
		& $\stackrel{2.7\times 10^{-3}\%}{\longrightarrow} 2^3P_0 \gamma \; (454.9)
		\stackrel{4.6\%}{\longrightarrow} 2^3S_1 \gamma \; (207.1)
		\stackrel{1.93\%}{\longrightarrow} \mu^+\mu^-$ & $2.4 \times 10^{-8}$  \\
		& $\stackrel{2.7\times 10^{-3}\%}{\longrightarrow} 2^3P_0 \gamma \; (454.9)
		\stackrel{0.9\%}{\longrightarrow} 1^3S_1 \gamma \; (743.1)
		\stackrel{2.48\%}{\longrightarrow} \mu^+\mu^-$ & $6.0 \times 10^{-9}$  \\
		& $\stackrel{5.7\times 10^{-4}\%}{\longrightarrow} 1^3P_0 \gamma \; (806.1)
		\stackrel{1.76\%}{\longrightarrow} 1^3S_1 \gamma \; (391.1)
		\stackrel{2.48\%}{\longrightarrow} \mu^+\mu^-$ & $2.5 \times 10^{-9}$  \\
		& $\stackrel{2.6\times 10^{-3}\%}{\longrightarrow} 2^3F_2 \gamma \; (75.7)
		\stackrel{1.98\times 10^{-2}\%}{\longrightarrow} 2^3D_1 \gamma \; (172.6)
		\stackrel{5.28\times 10^{-3}\%}{\longrightarrow} \mu^+\mu^-$ & $2.7 \times 10^{-13}$  \\
\hline \hline
\end{tabular}
\end{table*}

\begin{table*}[tp]
\caption{The $4^3D_1$ Decay chains and combined branching ratios.  These are small enough that we do not expect enough events at Belle II for this state to be observed.
\label{tab:4Dee}}
\begin{tabular}{lll} \hline \hline
Parent & Decay chain & Combined BR  \\
\hline 
$4^3D_1$	&	$\stackrel{3.04\times 10^{-6}\%}{\longrightarrow} \mu^+\mu^-$
			& $3.0\times 10^{-8}$ \\
		&  $\stackrel{4\times 10^{-5}\%}{\longrightarrow} 3^3P_2 \gamma \; (392.7)
		\stackrel{3.8\%}{\longrightarrow} 3^3S_1 \gamma \; (171.6)
		\stackrel{2.18\%}{\longrightarrow} \mu^+\mu^-$ & $3.3 \times 10^{-10}$  \\
		&  $\stackrel{4\times 10^{-5}\%}{\longrightarrow} 3^3P_2 \gamma \; (392.7)
		\stackrel{1.8\%}{\longrightarrow} 2^3S_1 \gamma \; (492.9)
		\stackrel{1.93\%}{\longrightarrow} \mu^+\mu^-$ & $1.4 \times 10^{-10}$  \\
		&  $\stackrel{4\times 10^{-5}\%}{\longrightarrow} 3^3P_2 \gamma \; (392.7)
		\stackrel{1.1\%}{\longrightarrow} 1^3S_1 \gamma \; (1003.0)
		\stackrel{2.48\%}{\longrightarrow} \mu^+\mu^-$ & $1.1 \times 10^{-10}$ \\
		&  $\stackrel{1.8\times 10^{-3}\%}{\longrightarrow} 3^3P_1 \gamma \; (404.2)
		\stackrel{7.2\%}{\longrightarrow} 3^3S_1 \gamma \; (159.8)
		\stackrel{2.18\%}{\longrightarrow} \mu^+\mu^-$ & $2.8\times 10^{-8}$  \\
		&  $\stackrel{1.8\times 10^{-3}\%}{\longrightarrow} 3^3P_1 \gamma \; (404.2)
		\stackrel{2.6\%}{\longrightarrow} 2^3S_1 \gamma \; (481.4)
		\stackrel{1.93\%}{\longrightarrow} \mu^+\mu^-$ & $9.0 \times 10^{-9}$  \\
		&  $\stackrel{1.8\times 10^{-3}\%}{\longrightarrow} 3^3P_1 \gamma \; (404.2)
		\stackrel{1.1\%}{\longrightarrow} 1^3S_1 \gamma \; (1003.0)
		\stackrel{2.48\%}{\longrightarrow} \mu^+\mu^-$ & $4.9 \times 10^{-9}$  \\
		&  $\stackrel{5.6\times 10^{-3}\%}{\longrightarrow} 3^3P_0 \gamma \; (419.6)
		\stackrel{0.31\%}{\longrightarrow} 3^3S_1 \gamma \; (144.0)
		\stackrel{2.18\%}{\longrightarrow} \mu^+\mu^-$ & $3.8 \times 10^{-9}$  \\
		&  $\stackrel{5.6\times 10^{-3}\%}{\longrightarrow} 3^3P_0 \gamma \; (419.6)
		\stackrel{0.077\%}{\longrightarrow} 2^3S_1 \gamma \; (466.2)
		\stackrel{1.93\%}{\longrightarrow} \mu^+\mu^-$ & $8.3 \times 10^{-10}$  \\
		&  $\stackrel{5.6\times 10^{-3}\%}{\longrightarrow} 3^3P_0 \gamma \; (419.6)
		\stackrel{0.01\%}{\longrightarrow} 1^3S_1 \gamma \; (988.5)
		\stackrel{2.48\%}{\longrightarrow} \mu^+\mu^-$ & $1.4 \times 10^{-10}$  \\
		&  $\stackrel{2.1\times 10^{-6}\%}{\longrightarrow} 2^3P_2 \gamma \; (639.1)
		\stackrel{10.6\%}{\longrightarrow} 2^3S_1 \gamma \; (242.5)
		\stackrel{1.93\%}{\longrightarrow} \mu^+\mu^-$ & $4.3 \times 10^{-11}$  \\
		&  $\stackrel{1\times 10^{-4}\%}{\longrightarrow} 2^3P_1 \gamma \; (652.3)
		\stackrel{19.9\%}{\longrightarrow} 2^3S_1 \gamma \; (229.6)
		\stackrel{1.93\%}{\longrightarrow} \mu^+\mu^-$ & $3.8 \times 10^{-9}$  \\
		&  $\stackrel{1\times 10^{-4}\%}{\longrightarrow} 2^3P_1 \gamma \; (652.3)
		\stackrel{9.2\%}{\longrightarrow} 1^3S_1 \gamma \; (764.3)
		\stackrel{2.48\%}{\longrightarrow} \mu^+\mu^-$ & $2.3 \times 10^{-9}$  \\
		& $\stackrel{3.6\times 10^{-4}\%}{\longrightarrow} 2^3P_0 \gamma \; (672.9)
		\stackrel{4.6\%}{\longrightarrow} 2^3S_1 \gamma \; (207.1)
		\stackrel{1.93\%}{\longrightarrow} \mu^+\mu^-$ & $3.2 \times 10^{-9}$  \\
		& $\stackrel{3.6\times 10^{-4}\%}{\longrightarrow} 2^3P_0 \gamma \; (672.9)
		\stackrel{0.9\%}{\longrightarrow} 1^3S_1 \gamma \; (743.1)
		\stackrel{2.48\%}{\longrightarrow} \mu^+\mu^-$ & $8.0 \times 10^{-10}$  \\
		&  $\stackrel{2.6\times 10^{-4}\%}{\longrightarrow} 1^3P_1 \gamma \; (986.0)
		\stackrel{33.9\%}{\longrightarrow} 1^3S_1 \gamma \; (423.0)
		\stackrel{2.48\%}{\longrightarrow} \mu^+\mu^-$ & $2.2 \times 10^{-8}$  \\
		& $\stackrel{1.0\times 10^{-3}\%}{\longrightarrow} 1^3P_0 \gamma \; 1016.7)
		\stackrel{1.76\%}{\longrightarrow} 1^3S_1 \gamma \; (391.1)
		\stackrel{2.48\%}{\longrightarrow} \mu^+\mu^-$ & $4.4 \times 10^{-9}$  \\
\hline \hline
\end{tabular}
\end{table*}

\section{Summary}

In this paper we calculated the properties of bottomonium mesons including masses, 
radiative transitions, annihilation decays, hadronic transitions and strong OZI allowed decays
for states above threshold.  These results were included in extensive tables with estimated 
BR's to different final states.  While we are interested in how these predictions 
fare against experimental measurements as a test of our understanding in the context
of the constituent quark model, the main objective of this work is to make predictions that
can assist experimentalists in finding missing bottomonium states and measuring their
properties.

We estimated the number of events expected in Run II of the LHC in the context of the LHCb 
experiment but they should also be relevant to the ATLAS and CMS experiments.  
We expect that significant numbers of  $\chi_{b2}(3P)$ and $\chi_{b1}(3P)$
will be produced and decay via radiative transitions to the $\Upsilon(1S)$ and 
$\Upsilon(2S)$ which will subsequently decay to $\mu^+\mu^-$.  Likewise we expect a large
number of $\chi_{b2}(4P)$ and $\chi_{b1}(4P)$'s to decay to $\Upsilon(3S)$.  Thus, a 
promising search strategy for excited $P$-wave mesons is to reconstruct $\Upsilon$'s in 
$\mu^+\mu^-$ and look at the invariant mass distributions of the $\Upsilon$'s with one
$\gamma$.   We also expect that $\eta_b(3S)$ can be seen in this final state.  

Turning to
the $D$-waves, the $1^3D_3$ and $1^3D_2$ will undergo radiative transitions to the
$1^3P_2$ and $1^3P_1$ respectively so they might be seen in final states with 
$\gamma \gamma \mu^+ \mu^-$.  Similarly, the $2D$ states will decay to $2^3P_{2,1}$ which 
decay to $2^3S_1$.  Thus,  it might be possible to see most of the $1D$ and $2D$ spin
triplet multiplets in the $\gamma \gamma \mu^+ \mu^-$ final state.  A challenge is that 
the photon energy for $2^3D_3 \to 2^3P_2$ is almost identical to that from the
 $2^3D_1 \to 2^3P_1$ transition so that one would need to be careful in looking at the 
 invariant mass distributions of the final state.
 
 For $e^+e^-$ collisions, large numbers of  $1^3D_J$ states will be produced by
 sitting on the $\Upsilon(3S)$ so it might be possible to resolve the three states and
 determine the splittings between members of the multiplets.  Sitting on the $\Upsilon(4S)$
 will produce $3^3P_{2,1}$ in radiative transitions.  It should be possible to observe 
 the $1^3D_1$ and $2^3D_1$ by an energy scan at the appropriate energy.  Sitting on
 the $2^3D_1$ resonance it might be possible to observe the $1^3F_2$ via radiative
 transitions from the $2^3D_1$.
 
 The LHC experiments and Belle II hold the promise to increase our knowledge of
 bottomonium mesons. This improved knowledge will test the reliability of models of 
 quarkonium physics.  Lattice QCD is making ever more precise calculations of bottomonium
 mesons and it is important that these calculations be held to account by experiment.  
We expect that the phenomenological predictions presented in this paper will be a useful
 tool for experimentalists to do so.

\acknowledgments

SG thanks Bryan Fulsom, Peter Lewis and Hassan Jawahery 
for helpful communications and discussions.  
SG thanks the Department of Physics University
of Toronto, the Department of Physics and Astronomy University of Hawaii and TRIUMF 
for their hospitality where some of this work was done.
This research was supported in part 
the Natural Sciences and Engineering Research Council of Canada under grant number 121209-2009 SAPIN.

\appendix
\section{The $^3P_0$ Model}
\label{app:3P0}

The $^3P_0$ quark pair creation model \cite{Micu:1968mk,Le Yaouanc:1972ae,Ackleh:1996yt,Blundell:1995ev,Barnes:2005pb} is
used to calculate OZI allowed strong bottomonium decays.  Given that details of the
calculations such as phase conventions 
are important and not always clearly stated in the literature we summarize the details of 
the $^3P_0$ model to assist an interested reader in reproducing our results.

In the $^3P_0$ quark pair creation model \cite{Micu:1968mk,Le Yaouanc:1972ae,Ackleh:1996yt,Blundell:1995ev,Barnes:2005pb}, 
a $q\bar{q}$ pair is created from the vacuum in a $^3P_0$ state (the quantum numbers of the vacuum).  
The angular momentum and spin of the created $q\bar{q}$ pair are therefore $L_P = 1$, $S_P = 1$, and $J_P = 0$, 
so that $M_{L_P} = -M_{S_P} \equiv m$.  The transition operator for $q\bar{q}$ pair creation can be written as
\begin{eqnarray}
T&=&-3\gamma  \sum_{m} \langle 1 1; m -m | 0 0\rangle   \sqrt{96\pi} \int d^3p_q d^3p_{\bar{q}} \delta^3(\vec{p}_q+\vec{p}_{\bar{q}})  \cr
&&\times \mathcal{Y}_{1m}\left(\frac{\vec{p}_q-\vec{p}_{\bar{q}}}{2}\right) \chi_{1-m} \phi_0 \omega_0 b_q^\dag(\vec{p}_q) d_{\bar{q}}^\dag(\vec{p}_{\bar{q}}) 
 \label{eq:Toperator}
 \end{eqnarray}
where $b_q^\dag(\vec{p}_q)$ and $d_{\bar{q}}^\dag(\vec{p}_{\bar{q}})$ are the creation operators for the quark and antiquark, respectively.  The momenta of the created quark, $\vec{p}_q$, and the created antiquark, $\vec{p}_{\bar{q}}$, are integrated over all possible values, such that the delta function ensures that their total momentum is zero in their centre-of-mass frame. The spin triplet state of the created $q\bar{q}$ pair is described by its spin wavefunction $\chi_{1 -m}$ and the momentum-space distribution of the created pair is described by the solid harmonic, written in terms of the spherical harmonic as $\mathcal{Y}_{L M_L}(\vec{k}) \equiv |\vec{k}|^LY_{L M_L}(\theta_k,\phi_k)$.   The $SU(3)$ flavour singlet wavefunction of the created pair is $\phi_0 = \frac{1}{\sqrt{3}}(u\bar{u}+d\bar{d}+s\bar{s})$ and its colour singlet wavefunction is $\omega_0$.  The overall factor of 3 in Eq.~\ref{eq:Toperator} will cancel out when evaluating the colour overlap.  The factor of $\sqrt{96\pi}$ arises from the normalization and field theory conventions of Refs. \cite{Ackleh:1996yt, Barnes:2005pb, Segovia:2012}.  The amplitude for quark-pair creation from the vacuum can therefore be described by a single free parameter, $\gamma$.  Any differences in the constant factors that appear in Eq.~\ref{eq:Toperator} simply result in a rescaling of the value of $\gamma$.  For example, the value of $\gamma$ used in Refs. \cite{Blundell:1995ev, Blundell:1996as} is larger than ours by a factor of $\sqrt{96\pi}$ due to the absence of this factor in their $T$ operator.

The $S$-matrix for the meson strong decay $A \to BC$ is defined as 
\begin{equation}
S \equiv I - 2\pi i \delta(E_f-E_i) T 
\end{equation}
so that 
\begin{equation}
\langle BC | T | A \rangle = \delta^3(\vec{p}_{A} - \vec{p}_{B} - \vec{p}_C) \mathcal{M}^{M_{J_A} M_{J_B} M_{J_C}}
\end{equation}
where, using the normalization from Refs. \cite{Ackleh:1996yt, Barnes:2005pb, Segovia:2012}, the helicity amplitude is given by
\begin{widetext}
\begin{eqnarray}
\label{eq:helicityamplitude}
\mathcal{M}^{M_{J_A} M_{J_B} M_{J_C}}(\vec{P}) = \gamma \sum && \langle L_A M_{L_A} S_A M_{S_A} | J_A M_{J_A} \rangle \langle L_B M_{L_B} S_B M_{S_B} | J_B M_{J_B} \rangle \langle L_C M_{L_C} S_C M_{S_C} | J_C M_{J_C} \rangle \\
&& \langle 1 m 1 -m | 0 0 \rangle \langle \chi^{14}_{S_B M_{S_B}} \chi^{32}_{S_C M_{S_C}} |  \chi^{12}_{S_A M_{S_A}} \chi^{34}_{1 -m} \rangle \nonumber \\
&& \left[ \langle \phi^{14}_B \phi^{32}_C | \phi^{12}_A \phi^{34}_0 \rangle I(\vec{P}, m_1, m_2, m_3)  + (-1)^{1+S_A+S_B+S_C}\langle \phi^{32}_B \phi^{14}_C | \phi^{12}_A \phi^{34}_0 \rangle I(-\vec{P}, m_2, m_1, m_3) \right] \nonumber
\end{eqnarray}
\end{widetext}
where the sum is over $M_{L_A}$, $M_{S_A}$, $M_{L_B}$, $M_{S_B}$, $M_{L_C}$, $M_{S_C}$, and $m$.  The two terms in the last factor correspond to the two possible diagrams.  In the first diagram, quark 1 from meson $A$ ends up in meson $B$ and antiquark 2 from meson $A$ ends up in meson $C$.  In the second diagram, quark 1 from meson $A$ ends up in meson $C$ and antiquark 2 from meson $A$ ends up in meson $B$.  Indices 3 and 4 refer to the created quark and antiquark, respectively.

The momentum space integral for the first diagram is given by 
\begin{widetext}
\begin{eqnarray}
I(\vec{P}, m_1, m_2, m_3) = \sqrt{96\pi} \int d^3p   \mathcal{Y}_{1m}(\vec{p})  \psi_{n_A, L_A, M_{L_A}} \left(\vec{p}+\vec{P}\right) \psi^*_{n_B, L_B, M_{L_B}}\left(\vec{p}+\frac{m_3}{m_1+m_3}\vec{P}\right) && \nonumber \\
\times\psi^*_{n_C, L_C, M_{L_C}}\left(\vec{p}+\frac{m_3}{m_2+m_3}\vec{P}\right) &&
\label{eq:spatialintegral}
\end{eqnarray}
\end{widetext}
where $m_1$, $m_2$ and $m_3=m_4$ are the constituent quark masses and we have taken $\vec{P} \equiv \vec{P}_B = -\vec{P}_C$ in the centre-of-mass frame of $A$.  To evaluate the spatial integral for the second diagram, we simply interchange $B\leftrightarrow C$, which amounts to making the replacement $m_1 \leftrightarrow m_2$  
and $\vec{P}\to -\vec{P}$ in Eq.~\ref{eq:spatialintegral}, leading to the second term of Eq.~\ref{eq:helicityamplitude}.

The techniques found in Appendix A of \cite{Roberts:1992} were useful in simplifying and evaluating the spatial integrals.
For the meson space wavefunctions, we use the momentum-space simple harmonic oscillator (SHO) wavefunctions, given by
\begin{equation}
\psi^{SHO}_{nLM_L}(\vec{p})=R^{SHO}_{nL}(p)Y_{L M_L}(\theta_p,\phi_p)
\end{equation}
where the radial wavefunctions are given by
\begin{eqnarray}
R^{SHO}_{nL}(p) &=& \frac{(-1)^n (-i)^L}{\beta^{\frac{3}{2}}}\sqrt{\frac{2 n\!}{\Gamma(n+L+\frac{3}{2})}}  \cr
&& \times  \left( \frac{p}{\beta} \right)^L  L^{L+\frac{1}{2}}_{n}\left( \frac{p^2}{\beta^2} \right)  e^{-p^2/(2\beta^2)}
\end{eqnarray}
and $L^{L+\frac{1}{2}}_{n}(p^2/\beta^2)$ is an associated Laguerre polynomial.  
We use the SHO wavefunctions such that a meson with quantum numbers $n^{2S+1}L_J$ in spectroscopic notation uses $\psi^{SHO}_{n-1,LM_L}$ for its momentum-space wavefunction.  The values we use for the effective harmonic oscillator parameter, $\beta$, are listed in Tables~\ref{tab:Upsilonparams1}-\ref{tab:Bparams}.

The colour matrix element 
\begin{equation}
\langle \omega_B^{14} \omega_C^{32} |  \omega_A^{12} \omega_0^{34} \rangle = \langle \omega_B^{32} \omega_C^{14} |  \omega_A^{12} \omega_0^{34} \rangle = \frac{1}{3}
\end{equation}
does not explicitly appear in Eq.~\ref{eq:helicityamplitude} since it cancels the overall factor of 3 from Eq.~\ref{eq:Toperator}.

The flavour matrix element can be easily found by writing the flavour wavefunctions of mesons $A$, $B$ and $C$ and that of the created quark pair as $5\times5$ matrices with rows indicating the quark flavour ($u, d, s, c, b$) and columns indicating the anti-quark flavour ($\bar{u}, \bar{d}, \bar{s}, \bar{c},\bar{b}$).  For example, the flavour wavefunction for the created $q\bar{q}$ pair is 
\begin{equation}
\phi_0 = \frac{1}{\sqrt{3}}(u\bar{u}+d\bar{d}+s\bar{s}) = \frac{1}{\sqrt{3}} \left( \begin{array}{ccccc}  1 & 0 & 0 & 0 & 0 \\ 0 & 1 & 0 & 0 & 0 \\ 0 & 0 & 1 & 0 & 0 \\ 0 & 0 & 0 & 0 & 0 \\ 0 & 0 & 0 & 0 & 0 \end{array} \right)
\end{equation}
The flavour overlaps for each of the two terms in Eq.~\ref{eq:helicityamplitude} are therefore given by
\begin{eqnarray}
\langle \phi^{14}_B \phi^{32}_C | \phi^{12}_A \phi^{34}_0 \rangle = Tr[\phi_A^\top \phi_B \phi_0^\top \phi_C] \\
\langle \phi^{32}_B \phi^{14}_C | \phi^{12}_A \phi^{34}_0 \rangle = Tr[\phi_A^\top \phi_C \phi_0^\top \phi_B]
\end{eqnarray}

The spin matrix elements for the first and second diagrams are written in terms of the Wigner $9j$ symbols \cite{de-Shalit} as 
\begin{widetext}
\begin{eqnarray}
\label{spinoverlap1}
\langle \chi^{14}_{S_B M_{S_B}} \chi^{32}_{S_C M_{S_C}} |  \chi^{12}_{S_A M_{S_A}} \chi^{34}_{1 -m} \rangle &=& (-1)^{1+S_C}\sqrt{3(2S_A+1)(2S_B+1)(2S_C+1)} \cr&&\sum_{S,M_S} \langle S_B M_{S_B} S_C M_{S_C} | S M_S \rangle \langle S_A M_{S_A} 1 -m | S M_S \rangle \left\{ \begin{array}{ccc} \frac{1}{2} & \frac{1}{2} & S_A \\  \frac{1}{2} &  \frac{1}{2} & 1 \\ S_B & S_C & S \end{array} \right\} \\
\label{spinoverlap2}
\langle \chi^{32}_{S_B M_{S_B}} \chi^{14}_{S_C M_{S_C}} |  \chi^{12}_{S_A M_{S_A}} \chi^{34}_{1 -m} \rangle &=& (-1)^{1+S_A+S_B+S_C} \langle \chi^{14}_{S_B M_{S_B}} \chi^{32}_{S_C M_{S_C}} |  \chi^{12}_{S_A M_{S_A}} \chi^{34}_{1 -m} \rangle
\end{eqnarray}
\end{widetext}
where the spin matrix element for the second diagram was obtained using an alternative definition for the $9j$ symbols that couple the quarks differently \cite{de-Shalit}.  This expression, given in Eq.~\ref{spinoverlap2}, was used to simplify Eq.~\ref{eq:helicityamplitude}.

Using the Jacob-Wick formula \cite{Jacob:1959, Richman:1984}, the helicity amplitudes $\mathcal{M}^{M_{J_A} M_{J_B} M_{J_C}}$, given by Eq.~\ref{eq:helicityamplitude}, are converted to partial wave amplitudes $\mathcal{M}^{LS}$ via
\begin{widetext}
\begin{eqnarray}
\mathcal{M}^{LS}(P) = \frac{\sqrt{4\pi(2L+1)}}{2J_A+1} \sum_{M_{J_B}, M_{J_C}} \left. \langle L 0 S M_{J_A} | J_A M_{J_A} \rangle \langle J_B M_{J_B} J_C M_{J_C} | S M_{J_A} \rangle \mathcal{M}^{M_{J_A} M_{J_B} M_{J_C}}(P \hat{z})\right|_{M_{J_A} = M_{J_B}+M_{J_C}} \cr
\end{eqnarray}
\end{widetext}
where $\vec{S} = \vec{J}_B + \vec{J}_C$ and $\vec{J}_A = \vec{L} + \vec{S}$ such that
\begin{eqnarray}
| J_B - J_C | \leq S \leq J_B + J_C \\
| J_A - S | \leq L \leq J_A + S
\end{eqnarray}
and the outgoing momentum of meson $B$, $\vec{P} \equiv P\hat{z}$, is chosen to lie along the $\hat{z}$-axis in the centre-of-mass frame of meson $A$ so that the helicities and angular momentum projections are related by $M_{J_B} = \lambda_B$, $M_{J_C} = -\lambda_C$ and $M_{J_A}=M_{J_B}+M_{J_C}=\lambda_B-\lambda_C$.  This on-shell momentum is conveniently written in terms of the masses of mesons $A$, $B$ and $C$ as
\begin{equation}
P=\frac{\sqrt{[M_A^2-(M_B+M_C)^2][M_A^2-(M_B-M_C)^2]}}{2M_A}
\end{equation}

Using relativistic phase space, as described in Ref.~\cite{Blundell:1995ev,Ackleh:1996yt}, the partial width for a given partial wave amplitude is given by
\begin{equation}
\Gamma^{LS}=2\pi P \mathcal{S} \frac{E_B E_C}{M_A}|\mathcal{M}^{LS}|^2
\end{equation}
where $E_B=\sqrt{M_B^2+P^2}$, $E_C=\sqrt{M_C^2+P^2}$, and $\mathcal{S}$ is a symmetry factor given by
\begin{equation}
\mathcal{S}=\frac{1}{1+\delta_{BC}} = \left\{ \begin{array}{ll}  \frac{1}{2} & \textrm{if $B$ and $C$ are identical} \\ 1 & \textrm{otherwise} \end{array}  \right.
\end{equation}
Finally, the strong decay width for a given decay mode of meson $A$ is just the sum of its partial widths:
\begin{equation}
\Gamma = \sum_{L,S}\Gamma^{LS}.
\end{equation}
The calculations of the strong decay widths, as outlined in this section, were performed using the Mathematica software package, version 7.0 \cite{Wolfram}.



\begin{thebibliography}{99}


\bibitem{Aad:2011ih} 
  G.~Aad {\it et al.}  [ATLAS Collaboration],
  ``Observation of a new $\chi_b$ state in radiative transitions to $\Upsilon(1S)$ and $\Upsilon(2S)$ at ATLAS,''
  Phys.\ Rev.\ Lett.\  {\bf 108}, 152001 (2012)
  [arXiv:1112.5154 [hep-ex]].
  
\bibitem{Chisholm:2014sca} 
  A.~Chisholm,
  ``Measurements of the $\chi_c$ and $\chi_b$ quarkonium states in $pp$ collisions with the ATLAS experiment,''
  CERN-THESIS-2014-071.

\bibitem{Drutskoy:2012gt} 
  A.~G.~Drutskoy, F.~-K.~Guo, F.~J.~Llanes-Estrada, A.~V.~Nefediev and J.~M.~Torres-Rincon,
  ``Hadron physics potential of future high-luminosity $B$-factories at the $\Upsilon(5S)$ and above,''
  Eur.\ Phys.\ J.\ A {\bf 49}, 7 (2013)
  [arXiv:1210.6623 [hep-ph]].

\bibitem{Lewis:2012ir} 
  R.~Lewis and R.~M.~Woloshyn,
  ``Higher angular momentum states of bottomonium in lattice NRQCD,''
  Phys.\ Rev.\ D {\bf 85}, 114509 (2012)
  [arXiv:1204.4675 [hep-lat]].

\bibitem{Lewis:2011ti} 
  R.~Lewis and R.~M.~Woloshyn,
  ``Excited Upsilon Radiative Decays,''
  Phys.\ Rev.\ D {\bf 84}, 094501 (2011)
  [arXiv:1108.1137 [hep-lat]].

\bibitem{Lewis:2012bh} 
  R.~Lewis and R.~M.~Woloshyn,
  ``More about excited bottomonium radiative decays,''
  Phys.\ Rev.\ D {\bf 86}, 057501 (2012)
  [arXiv:1207.3825 [hep-lat]].


\bibitem{Dowdall:2013jqa} 
  R.~J.~Dowdall {\it et al.}  [HPQCD Collaboration],
  Phys.\ Rev.\ D {\bf 89}, no. 3, 031502 (2014)
  [arXiv:1309.5797 [hep-lat]].


\bibitem{Brambilla:2010cs} 
  N.~Brambilla, S.~Eidelman, B.~K.~Heltsley, R.~Vogt, G.~T.~Bodwin, E.~Eichten, A.~D.~Frawley and A.~B.~Meyer {\it et al.},
  ``Heavy quarkonium: progress, puzzles, and opportunities,''
  Eur.\ Phys.\ J.\ C {\bf 71}, 1534 (2011)
  [arXiv:1010.5827 [hep-ph]].
  
\bibitem{Patrignani:2012an} 
  C.~Patrignani, T.~K.~Pedlar and J.~L.~Rosner,
  ``Recent Results in Bottomonium,''
    Ann.\ Rev.\ Nucl.\ Part.\ Sci.\  {\bf 63}, 21 (2013)
  [arXiv:1212.6552].
  
  
\bibitem{Eichten:2007qx} 
  E.~Eichten, S.~Godfrey, H.~Mahlke and J.~L.~Rosner,
  ``Quarkonia and their transitions,''
  Rev.\ Mod.\ Phys.\  {\bf 80}, 1161 (2008)
  [hep-ph/0701208].


\bibitem{godfrey85xj}
S. Godfrey and N. Isgur,
``Mesons In A Relativized Quark Model With Chromodynamics,''
Phys. Rev. D32, 189 (1985).


\bibitem{Micu:1968mk} 
  L.~Micu,
  ``Decay rates of meson resonances in a quark model,''
  Nucl.\ Phys.\ B {\bf 10}, 521 (1969).
  
\bibitem{Le Yaouanc:1972ae} 
  A.~Le Yaouanc, L.~Oliver, O.~Pene and J.~C.~Raynal,
  ``Naive quark pair creation model of strong interaction vertices,''
  Phys.\ Rev.\ D {\bf 8}, 2223 (1973).


\bibitem{Barnes:2003vb} 
  T.~Barnes and S.~Godfrey,
  ``Charmonium options for the X(3872),''
  Phys.\ Rev.\ D {\bf 69}, 054008 (2004)
  [hep-ph/0311162].

\bibitem{Godfrey:2003kg} 
  S.~Godfrey,
  ``Testing the nature of the $D_{sJ}^*(2317)^+$ and $D_{sJ}(2463)^+$ states using radiative transitions,''
  Phys.\ Lett.\ B {\bf 568}, 254 (2003)
  [hep-ph/0305122].


\bibitem{Blundell:1995ev} 
  H.~G.~Blundell and S.~Godfrey,
  ``The $\xi(2220)$ reexamined: Strong decays of the $1^3F_2$ and $1^3F_4$ $s\bar{s}$ mesons,''
  Phys.\ Rev.\ D {\bf 53}, 3700 (1996)
  [hep-ph/9508264].

  \bibitem{Barnes:2005pb} 
  T.~Barnes, S.~Godfrey and E.~S.~Swanson,
  ``Higher charmonia,''
  Phys.\ Rev.\ D {\bf 72}, 054026 (2005)
  [hep-ph/0505002].
    
    
     \bibitem{Godfrey:2014} 
  S.~Godfrey and K.~Moats,
  ``$D_{sJ}^*(2860)$ mesons as excited $D$-wave $c\bar{s}$ states,''
  Phys.\ Rev.\ D {\bf 90}, 117501 (2014)
  [arXiv:1409.0874 [hep-ph]].
 
 

\bibitem{Godfrey:1985by} 
  S.~Godfrey,
  ``High Spin Mesons in the Quark Model,''
  Phys.\ Rev.\ D {\bf 31}, 2375 (1985).

\bibitem{godfrey85b}
S.~Godfrey and N.~Isgur,
``Isospin Violation In Mesons And The Constituent Quark Masses,''
Phys.\ Rev.\ D {\bf 34}, 899 (1986).


\bibitem{Godfrey:2004ya} 
  S.~Godfrey,
  ``Spectroscopy of $B_c$ mesons in the relativized quark model,''
  Phys.\ Rev.\ D {\bf 70}, 054017 (2004)
  [hep-ph/0406228].

\bibitem{Godfrey:2005ww} 
  S.~Godfrey,
  ``Properties of the charmed P-wave mesons,''
  Phys.\ Rev.\ D {\bf 72}, 054029 (2005)
  [hep-ph/0508078].
  



\bibitem{Aubert:2003fg} 
  B.~Aubert {\it et al.}  [BaBar Collaboration],
  ``Observation of a narrow meson decaying to $D_s^+ \pi^0$ at a mass of 2.32-GeV/c$^2$,''
  Phys.\ Rev.\ Lett.\  {\bf 90}, 242001 (2003)
  [hep-ex/0304021].

\bibitem{Besson:2003cp} 
  D.~Besson {\it et al.}  [CLEO Collaboration],
  ``Observation of a Narrow Resonance of Mass 2.46 GeV/c$^2$ Decaying to $D^{*+}_s \pi^0$ and Confirmation of the $D^*_{sJ}(2317)$ State,''
  Phys.\ Rev.\ D {\bf 68}, 032002 (2003)
  [Erratum-ibid.\ D {\bf 75}, 119908 (2007)]
  [hep-ex/0305100].

\bibitem{Krokovny:2003zq} 
  P.~Krokovny {\it et al.}  [Belle Collaboration],
  ``Observation of the $D_{sJ}(2317)$ and $D_{sJ}(2457)$ in $B$ decays,''
  Phys.\ Rev.\ Lett.\  {\bf 91}, 262002 (2003)
  [hep-ex/0308019].
  
  \bibitem{Choi:2003ue}
   S.-K.~Choi {\it et al.}  [Belle Collaboration],
  ``Observation of a Narrow Charmoniumlike State in Exclusive $B^\pm \rightarrow K^\pm \pi^+ \pi^- J/\psi$ Decays,"
    Phys.\ Rev.\ Lett.\  {\bf 91}, 262001 (2003)
  [hep-ex/0309032].

\bibitem{Godfrey:2008nc} 
  S.~Godfrey and S.~L.~Olsen,
  ``The Exotic XYZ Charmonium-like Mesons,''
  Ann.\ Rev.\ Nucl.\ Part.\ Sci.\  {\bf 58}, 51 (2008)
  [arXiv:0801.3867 [hep-ph]].

\bibitem{Godfrey:2009qe} 
  S.~Godfrey,
  ``Topics in Hadron Spectroscopy in 2009,''
  arXiv:0910.3409 [hep-ph].

\bibitem{Braaten:2013oba} 
  E.~Braaten,
  ``Theoretical Interpretations of the XYZ Mesons,''
  arXiv:1310.1636 [hep-ph].

\bibitem{DeFazio:2012sg} 
  F.~De Fazio,
  ``New Spectroscopy of Heavy Mesons,''
  PoS HQL {\bf 2012}, 001 (2012)
  [arXiv:1208.4206 [hep-ph]].

\bibitem{Eichten:2004uh} 
  E.~J.~Eichten, K.~Lane and C.~Quigg,
  ``Charmonium levels near threshold and the narrow state $X(3872) \to \pi^{+}\pi^{-}J/\psi$,''
  Phys.\ Rev.\ D {\bf 69}, 094019 (2004)
  [hep-ph/0401210].
  
\bibitem{Olive:2014kda} 
  K.~A.~Olive {\it et al.}  [Particle Data Group Collaboration],
  ``Review of Particle Physics,''
  Chin.\ Phys.\ C {\bf 38}, 090001 (2014).
    
\bibitem{Aaij:2014hla} 
  R.~Aaij {\it et al.}  [LHCb Collaboration],
  ``Measurement of the $\chi_b(3P)$ mass and of the relative rate of $\chi_{b1}(1P)$ and $\chi_{b2}(1P)$ production,''
  JHEP {\bf 1410}, 88 (2014)
  [arXiv:1409.1408 [hep-ex]].
  
\bibitem{Aaij:2014caa} 
  R.~Aaij {\it et al.}  [LHCb Collaboration],
  ``Study of $\chi _{{\mathrm {b}}}$ meson production in $\mathrm {p} $ $\mathrm {p} $ collisions at $\sqrt{s}=7$ and $8{\mathrm {\,TeV}} $ and observation of the decay $\chi _{{\mathrm {b}}}\mathrm {(3P)} \rightarrow \Upsilon \mathrm {(3S)} {\gamma } $,''
  Eur.\ Phys.\ J.\ C {\bf 74}, no. 10, 3092 (2014)
  [arXiv:1407.7734 [hep-ex]].


  
  \bibitem{Kwo88a}
W.~Kwong and J.~L.~Rosner,
``D Wave Quarkonium Levels of the Upsilon Family,''
Phys. Rev. D38, 279 (1988).


\bibitem{Skwarnicki:2005pq} 
  T.~Skwarnicki,
  ``CLEO results on transitions in heavy quarkonia,''
  hep-ex/0505050.
  


  
\bibitem{Grant:1992fi} 
  A.~Grant and J.~L.~Rosner,
  ``Dipole transition matrix elements for systems with power law potentials,''
  Phys.\ Rev.\ D {\bf 46}, 3862 (1992).
  
  
  
\bibitem{Godfrey:2001eb} 
  S.~Godfrey and J.~L.~Rosner,
  Phys.\ Rev.\ D {\bf 64}, 074011 (2001)
  [Erratum-ibid.\ D {\bf 65}, 039901 (2002)]
  [hep-ph/0104253].

\bibitem{Aubert:2008ba} 
  B.~Aubert {\it et al.}  [BaBar Collaboration],
  Phys.\ Rev.\ Lett.\  {\bf 101}, 071801 (2008)
  [Erratum-ibid.\  {\bf 102}, 029901 (2009)]
  [arXiv:0807.1086 [hep-ex]].
  
  
  \bibitem{Nov78}
V.A.Novikov, L.B.Okun, M.A.Shifman, A.I.Vainshtein, M.B.Voloshin, 
and V.I.Zakharov, Phys. Rept. C41, 1 (1978).


\bibitem{Daghighian:1987ru} 
  F.~Daghighian and D.~Silverman,
  Phys.\ Rev.\ D {\bf 36}, 3401 (1987).

\bibitem{Ferretti:2014xqa} 
  J.~Ferretti, G.~Galatà and E.~Santopinto,
  Phys.\ Rev.\ D {\bf 90}, no. 5, 054010 (2014)
  [arXiv:1401.4431 [nucl-th]].
  
\bibitem{Ebert:2002pp} 
  D.~Ebert, R.~N.~Faustov and V.~O.~Galkin,
  Phys.\ Rev.\ D {\bf 67}, 014027 (2003)
  [hep-ph/0210381].
  
\bibitem{Wei-Zhao:2013sta} 
  T.~Wei-Zhao, C.~Lu, Y.~You-Chang and C.~Hong,
  Chin.\ Phys.\ C {\bf 37}, 083101 (2013)
  [arXiv:1308.0960 [hep-ph]].
  
\bibitem{Pineda:2013lta} 
  A.~Pineda and J.~Segovia,
  Phys.\ Rev.\ D {\bf 87}, no. 7, 074024 (2013)
  [arXiv:1302.3528 [hep-ph]].

  
  

\bibitem{App75}
T.Appelquist and H.D.Politzer, Phys. Rev. Lett. 34, 43 (1975);

\bibitem{DeR75}
A.DeRujula and S.L.Glashow, Phys. Rev. Lett. 34, 46 (1975).

\bibitem{Cha75}
M.Chanowitz, Phys. Rev. D12, 918 (1975).

\bibitem{Bar76a}
R.Barbieri, R.Gatto and R. K\"ogerler, 
Phys. Lett. B60, 183 (1976).

\bibitem{Bar76b}
R.Barbieri, R.Gatto and E.Remiddi, 
Phys. Lett. B61, 465 (1976).



\bibitem{Bar79}
R.Barbieri, G.Curci, E.d'Emilio and E.Remiddi, 
Nucl. Phys. B154, 535 (1979).

\bibitem{Kwo88b} 
W.Kwong, P.B.Mackenzie, R.Rosenfeld, and J.L.Rosner,
Phys. Rev. D37, 3210 (1988).

\bibitem{Ack92a}
E.S.Ackleh and T.Barnes,
Phys. Rev. D45, 232 (1992). 

\bibitem{Ack92b}
E.S.Ackleh, T.Barnes and F.E.Close,
Phys. Rev. D46, 2257 (1992).

\bibitem{Belanger:1987cg} 
  G.~Belanger and P.~Moxhay,
  ``Three Gluon Annihilation of $D$ Wave Quarkonium,''
  Phys.\ Lett.\ B {\bf 199}, 575 (1987).
  
\bibitem{Ber91}
L.Bergstr\"om and P.Ernstr\"om, 
Phys. Lett. B267, 111 (1991).

\bibitem{Robinett:1992px} 
  R.~W.~Robinett and L.~Weinkauf,
  ``Covariant formalism for $F$-wave quarkonium production and annihilation: Application to $^3F_J \to g g$ decays,''
  Phys.\ Rev.\ D {\bf 46}, 3832 (1992).
  
\bibitem{Bradley:1980eh} 
  A.~Bradley and A.~Khare,
  ``{QCD} Correction to the Leptonic Decay Rate of $D$ Wave Vector Mesons,''
  Z.\ Phys.\ C {\bf 8}, 131 (1981).
  

  
  
  

\bibitem{Yan80}
T.M.Yan, Phys. Rev. D22, 1652 (1980).

\bibitem{Kua81}
Y.P.Kuang and T.M.Yan, 
Phys. Rev. D24, 2874 (1981).

\bibitem{Kua88}
Y.P.Kuang, S.F.Tuan and T.M.Yan, 
Phys. Rev. D37, 1210 (1988).

\bibitem{Kua90}
Y.P.Kuang and T.M.Yan,
 ``Hadronic Transitions of $D$ Wave Quarkonium and $\Psi(3770) \to J/\Psi \pi \pi$,''
Phys. Rev. D41, 155 (1990).

\bibitem{Rosner03}
J.~L.~Rosner,
 ``Prospects for detection of $\Upsilon(1D) \to \Upsilon(1S) \pi \pi$ via $\Upsilon(3S) \to \Upsilon(1D) + X$,''
Phys.\ Rev.\ D {\bf 67}, 097504 (2003)
[arXiv:hep-ph/0302122].

\bibitem{Vol80} 
M.B.Voloshin and V.I.Zakharov, 
Phys. Rev. Lett. 45, 688 (1980).

\bibitem{Nov81} 
V.A.Novikov and M.A.Shifman, 
Z. Phys. C8, 43 (1981).

\bibitem{Iof80} 
B.L.Ioffe and M.A.Shifman, 
Phys. Lett. 95B, 99 (1980).

\bibitem{Vol86} 
M.B.Voloshin, 
Sov. J. Nucl. Phys. 43, 1011 (1986).

\bibitem{Vol03a} 
M.B.Voloshin 
Phys. Lett. B562, 68 (2003)
[hep-ph/0302261].

\bibitem{Ko94}
P.~Ko,
 ``Search for $\eta_c^\prime$ and $h_c(^1P_1)$ states in the $e^+ e^-$ annihilations,''
Phys.\ Rev.\ D {\bf 52}, 1710 (1995)
[arXiv:hep-ph/9505299].

\bibitem{Mox88}
P.Moxhay, 
Phys. Rev. 37, 2557 (1988).

\bibitem{Ko93}
P.~Ko,
``On The Branching Ratio Of $\Upsilon (1D) \to \Upsilon (1S) \pi \pi$,''
Phys.\ Rev.\ D {\bf 47}, 208 (1993).

\bibitem{Kuang:2006me} 
  Y.~P.~Kuang,
  Front.\ Phys.\ China {\bf 1}, 19 (2006)
  [hep-ph/0601044].



\bibitem{Godfrey:2002rp} 
  S.~Godfrey and J.~L.~Rosner,
  ``Production of singlet $P$ wave $c\bar{c}$ and $b\bar{b}$ states,''
  Phys.\ Rev.\ D {\bf 66}, 014012 (2002)
  [hep-ph/0205255].

\bibitem{Rosner:2005ry} 
  J.~L.~Rosner {\it et al.}  [CLEO Collaboration],
  Phys.\ Rev.\ Lett.\  {\bf 95}, 102003 (2005)
  [hep-ex/0505073].
  
\bibitem{Rubin:2005px} 
  P.~Rubin {\it et al.}  [CLEO Collaboration],
  Phys.\ Rev.\ D {\bf 72}, 092004 (2005)
  [hep-ex/0508037].
  
\bibitem{Lees:2011zp} 
  J.~P.~Lees {\it et al.}  [BaBar Collaboration],
  Phys.\ Rev.\ D {\bf 84}, 091101 (2011)
  [arXiv:1102.4565 [hep-ex]].
  
\bibitem{Severini:2003qw} 
  H.~Severini {\it et al.}  [CLEO Collaboration],
  Phys.\ Rev.\ Lett.\  {\bf 92}, 222002 (2004)
  [hep-ex/0307034].
  
\bibitem{Eichten:1994gt} 
  E.~J.~Eichten and C.~Quigg,
  Phys.\ Rev.\ D {\bf 49}, 5845 (1994)
  [hep-ph/9402210].


\bibitem{Chen:2008}
  K.~Chen {\it et al.}  [Belle Collaboration],
  ``Observation of Anomalous $\Upsilon(1S)\pi^+\pi^-$ and $\Upsilon(2S)\pi^+\pi^-$ Production near the $\Upsilon(5S)$ Resonance,''
  Phys.\ Rev.\ Lett. {\bf 100}, 112001 (2008)
  [arXiv:0710.2577 [hep-ex]].


\bibitem{Segovia:2014mca} 
  J.~Segovia, D.~R.~Entem and F.~Fernandez,
  ``Puzzles in hadronic transitions of heavy quarkonium with two pion emission,''
  arXiv:1409.7079 [hep-ph].
  
  
    
  
  
  
  

\bibitem{Ali:2009es} 
  A.~Ali, C.~Hambrock and M.~J.~Aslam,
  Phys.\ Rev.\ Lett.\  {\bf 104}, 162001 (2010)
  [Phys.\ Rev.\ Lett.\  {\bf 107}, 049903 (2011)]
  [arXiv:0912.5016 [hep-ph]].
  
\bibitem{Ali:2010pq} 
  A.~Ali, C.~Hambrock and S.~Mishima,
  Phys.\ Rev.\ Lett.\  {\bf 106}, 092002 (2011)
  [arXiv:1011.4856 [hep-ph]].


\bibitem{Chen:2011qx} 
  D.~Y.~Chen, J.~He, X.~Q.~Li and X.~Liu,
  Phys.\ Rev.\ D {\bf 84}, 074006 (2011)
  [arXiv:1105.1672 [hep-ph]].
  
  
  
  
  
\bibitem{delAmoSanchez:2010kz} 
  P.~del Amo Sanchez {\it et al.}  [BaBar Collaboration],
  ``Observation of the $\Upsilon(1^3D_J)$ Bottomonium State through Decays to $\pi^+\pi^-\Upsilon(1S)$,''
  Phys.\ Rev.\ D {\bf 82}, 111102 (2010)
  [arXiv:1004.0175 [hep-ex]].
  



  

  


  
  


  
\bibitem{Ackleh:1996yt} 
  E.~S.~Ackleh, T.~Barnes and E.~S.~Swanson,
  ``On the mechanism of open-flavor strong decays,''
  Phys.\ Rev.\ D {\bf 54}, 6811 (1996)
  [hep-ph/9604355].
  

  
\bibitem{Ferretti:2014} 
  J.~Ferretti and E.~Santopinto,
  ``Higher Mass Bottomonia,''
  Phys.\ Rev.\ D {\bf 90}, 094022 (2014)
  [arXiv:1306.2874 [hep-ph]].
  
\bibitem{Segovia:2012} 
  J.~Segovia, D.~R.~Entem and E.~Fernandez,
  ``Scaling of the $^3P_0$ strength in heavy meson decays,''
  Phys.\ Lett.\ B {\bf 714}, 322 (2012).
  
\bibitem{Ebert:2014} 
  D.~Ebert, R.~N.~Faustov and V.O.~Galkin
  ``Strong decays of vector mesons to pseudoscalar mesons in the relativistic quark model,''
  [arXiv:1412.4534 [hep-ph]].

\bibitem{Blundell:1996as} 
  H.~G.~Blundell,
  ``Meson properties in the quark model: A look at some outstanding problems,''
  hep-ph/9608473.

\bibitem{Close:2005se} 
  F.~E.~Close and E.~S.~Swanson,
  ``Dynamics and decay of heavy-light hadrons,''
  Phys.\ Rev.\ D {\bf 72}, 094004 (2005)
  [hep-ph/0505206].



  





   
\bibitem{Bodwin:1994jh} 
  G.~T.~Bodwin, E.~Braaten and G.~P.~Lepage,
  ``Rigorous QCD analysis of inclusive annihilation and production of heavy quarkonium,''
  Phys.\ Rev.\ D {\bf 51}, 1125 (1995)
  [Erratum-ibid.\ D {\bf 55}, 5853 (1997)]
  [hep-ph/9407339].
  
  
\bibitem{Likhoded:2012hw} 
  A.~K.~Likhoded, A.~V.~Luchinsky and S.~V.~Poslavsky,
  ``Production of $\chi_{b}$-mesons at LHC,''
  Phys.\ Rev.\ D {\bf 86}, 074027 (2012)
  [arXiv:1203.4893 [hep-ph]].
  
\bibitem{Han:2014kxa} 
  H.~Han, Y.~Q.~Ma, C.~Meng, H.~S.~Shao, Y.~J.~Zhang and K.~T.~Chao,
  ``$\Upsilon(nS)$ and $\chi_b(nP)$ production at hadron colliders in nonrelativistic QCD,''
  arXiv:1410.8537 [hep-ph].
  
\bibitem{Ali:2013xba} 
  A.~Ali, C.~Hambrock and W.~Wang,
  ``Hadroproduction of $\Upsilon(nS)$ above $B\bar B$ Thresholds and Implications for $Y_b(10890)$,''
  Phys.\ Rev.\ D {\bf 88}, no. 5, 054026 (2013)
  [arXiv:1306.4470 [hep-ph]].

  
\bibitem{Aad:2012dlq} 
  G.~Aad {\it et al.}  [ATLAS Collaboration],
  Phys.\ Rev.\ D {\bf 87}, no. 5, 052004 (2013)
  [arXiv:1211.7255 [hep-ex]].
  
\bibitem{Chatrchyan:2013yna} 
  S.~Chatrchyan {\it et al.}  [CMS Collaboration],
  Phys.\ Lett.\ B {\bf 727}, 101 (2013)
  [arXiv:1303.5900 [hep-ex]].
 
    
\bibitem{Adachi:2011ji} 
  I.~Adachi {\it et al.}  [Belle Collaboration],
  ``First observation of the $P$-wave spin-singlet bottomonium states $h_b(1P)$ and $h_b(2P)$,''
  Phys.\ Rev.\ Lett.\  {\bf 108}, 032001 (2012)
  [arXiv:1103.3419 [hep-ex]].
    
\bibitem{Bonvicini:2004yj} 
  G.~Bonvicini {\it et al.}  [CLEO Collaboration],
  ``First observation of a Upsilon(1D) state,''
  Phys.\ Rev.\ D {\bf 70}, 032001 (2004)
  [hep-ex/0404021].
  
\bibitem{Godfrey:2001vc} 
  S.~Godfrey and J.~L.~Rosner,
  ``Production of the $D$ wave $b\bar{b}$ states,''
  Phys.\ Rev.\ D {\bf 64}, 097501 (2001)
  [Erratum-ibid.\ D {\bf 66}, 059902 (2002)]
  [hep-ph/0105273].
  
\bibitem{Lees:2014qea} 
  J.~P.~Lees {\it et al.}  [BaBar Collaboration],
  ``Bottomonium spectroscopy and radiative transitions involving the $\chi_{bJ}(1P,2P)$ states at BABAR,''
  Phys.\ Rev.\ D {\bf 90}, no. 11, 112010 (2014)
  [arXiv:1410.3902 [hep-ex]].

\bibitem{Abazov:2012gh} 
  V.~M.~Abazov {\it et al.}  [D0 Collaboration],
  ``Observation of a narrow mass state decaying into $\Upsilon(1S) + \gamma$ in $p\bar{p}$ collisions at $\sqrt{s} = 1.96$ TeV,''
  Phys.\ Rev.\ D {\bf 86}, 031103 (2012)
  [arXiv:1203.6034 [hep-ex]].
  
  
  

\bibitem{Bevan:2014iga} 
  A.~J.~Bevan {\it et al.}  [BaBar and Belle Collaborations],
  Eur.\ Phys.\ J.\ C {\bf 74}, no. 11, 3026 (2014)
  [arXiv:1406.6311 [hep-ex]].


\bibitem{Roberts:1992} 
  W.~Roberts and B.~Silvestre-Brac
  ``General method of calculation of any hadronic decay in the $^3P_0$ model,''
  Few-Body Systems, {\bf 11}, 177 (1992)
  
\bibitem{de-Shalit} 
A.~de-Shalit and I.~Talmi,
  ``Nuclear Shell Theory,''
Academic Press, New York (1963)
  
\bibitem{Jacob:1959} 
  M.~Jacob, and G.~C.~Wick,
  ``On the General Theory of Collisions for Particles with Spin,''
  Ann.\ Phys.\ {\bf 7}, 404 (1959).

\bibitem{Richman:1984} 
  J.~D.~Richman,
  ``An Experimenter's Guide to the Helicity Formalism,''
 Technical Note CALT-68-1148, California Institute of Technology (1984).
   
\bibitem{Wolfram} 
  Wolfram Research, Inc.,
  ``Mathematica, Version 7.0,''
 Wolfram Research, Inc., Champaign, Illinois (2007)

  
  

\end{thebibliography}
\end{document}